\documentclass[twocolumn]{aastex631}

\extrafloats{100}
\usepackage{rotating}

\shorttitle{The massive cluster IRS 13}
\shortauthors{Pei{\ss}ker et al.}

\begin{document}

\title{The Evaporating Massive Embedded Stellar Cluster IRS 13 Close to Sgr~A*. I. Detection of a rich population of dusty objects in the IRS 13 cluster}

\correspondingauthor{Florian Pei{\ss}ker}
\email{peissker@ph1.uni-koeln.de}

\author[0000-0002-9850-2708]{Florian Pei$\beta$ker}
\affil{I.Physikalisches Institut der Universit\"at zu K\"oln, Z\"ulpicher Str. 77, 50937 K\"oln, Germany}

\author[0000-0001-6450-1187]{Michal Zaja\v{c}ek}
\affil{Department of Theoretical Physics and Astrophysics, Faculty of Science, Masaryk University, Kotl\'a\v{r}sk\'a 2, 611 37 Brno, Czech Republic}

\author{Lauritz Thomkins}
\affil{I.Physikalisches Institut der Universit\"at zu K\"oln, Z\"ulpicher Str. 77, 50937 K\"oln, Germany}

\author[0000-0001-6049-3132]{Andreas Eckart}
\affil{I.Physikalisches Institut der Universit\"at zu K\"oln, Z\"ulpicher Str. 77, 50937 K\"oln, Germany}
\affil{Max-Plank-Institut f\"ur Radioastronomie, Auf dem H\"ugel 69, 53121 Bonn, Germany}

\author[0000-0001-5342-5713]{Lucas Labadie}
\affil{I.Physikalisches Institut der Universit\"at zu K\"oln, Z\"ulpicher Str. 77, 50937 K\"oln, Germany}

\author[0000-0002-5760-0459]{Vladim\'ir Karas}
\affil{Astronomical Institute, Czech Academy of Sciences, Bo\v{c}n\'{i} II 1401, CZ-14100 Prague, Czech Republic}

\author[0000-0001-7134-9005]{Nadeen B. Sabha}
\affil{Institut f\"ur Astro- und Teilchenphysik, Universit\"at Innsbruck, Technikerstr. 25, 6020 Innsbruck, Austria}

\author[0000-0002-5859-3932]{Lukas Steiniger}
\affil{I.Physikalisches Institut der Universit\"at zu K\"oln, Z\"ulpicher Str. 77, 50937 K\"oln, Germany}

\author[0000-0002-4902-3794]{Maria Melamed}
\affil{I.Physikalisches Institut der Universit\"at zu K\"oln, Z\"ulpicher Str. 77, 50937 K\"oln, Germany}

\begin{abstract}

A detailed analysis of the Nuclear Stellar Cluster (NSC) concedes not only the existence of the S-cluster with its fast-moving stars and the supermassive black hole (SMBH) Sgr~A*. It also reveals an embedded region of gas and dust with an exceptionally high stellar density called IRS 13. The IRS 13 cluster can be divided into the northern and the eastern counterparts, called IRS 13N and IRS 13E, respectively. This work will focus on both regions and study their most prominent members using rich infrared and radio/submm data baselines. Applying a multiwavelength analysis enables us to determine a comprehensive photometric footprint of the investigated cluster sample. Using the raytracing-based radiative transfer model \texttt{HYPERION}, the spectral energy distribution of the IRS 13 members suggests a stellar nature of the dusty sources. These putative Young Stellar Objects (YSOs) have a comparable spectroscopic identification to the D and G sources in or near the S cluster. Furthermore, we report the existence of a population of dusty sources in IRS 13 that can be mostly identified in the H-, K-, and L-band. Together with the objects reported in literature, we propose that this population is the outcome of a recent star formation process. Furthermore, we report that these presumably young objects are arranged in a disk structure. Although it cannot be excluded that the intrinsic arrangement of IRS 13 does show a disk structure, we find indications that the investigated cluster sample might be related to the counterclockwise disk.
\end{abstract}


\section{Introduction} \label{sec:1}

The bright and variable radio source Sgr~A*, identified as the supermassive black hole (SMBH), is located at the center of the Nuclear Star Cluster \citep[NSC;][]{Menten1997,Eckart2017,Tursunov2020,2022RvMP...94b0501G}. Sheltering a rich depot of various types of stars, the NSC in the Galactic center (GC) enables detailed studies of its structure and components \citep[][]{Schoedel2009, Baumgardt2018, Shahzamanian2022}. As a prominent sub-structure of the NSC, the IRS 13 cluster {has drawn} attention because of the possibility of {hosting} an intermediate-mass black hole (IMBH) with several $\sim\,10^4\,M_{\odot}$ \citep[see][]{zwart2002, Maillard2004, Schoedel2005}. Although every attempt to find such an IMBH resulted in a dead end \citep[see the X-ray observations in][]{Zhu2020, Wang2020}, the question remains why the embedded cluster IRS 13 seems to resist the gravitational, and consequently disruptive, influence of Sgr~A* \citep{muzic2008}. With this correlation in mind, \cite{Tsuboi2017b} analyzed the ionized gas associated with IRS 13E showing velocities of up to several hundred km/s on a highly eccentric orbit around a central component of the cluster, namely E3 \citep{Fritz2010}. Tsuboi et al. suggested that the blue and redshifted velocities might indicate the presence of an IMBH responsible for the circular motion of the ionized gas. Although the existence of an IMBH is disputed \citep{Zhu2020}, the IRS 13 cluster features various additional fruitful scientific topics \citep{Paumard2006}. For example, \cite{Eckart2004a} investigated the population of presumably Young Stellar Objects (YSOs) in IRS 13N. Photometric analysis of these dusty IRS 13N objects showed similarities with the D-sources \citep[also donated as G-sources, see][]{Peissker2020b, Ciurlo2020} in the S-cluster \citep[][]{Eckart2004a, Eckart2013} suggesting a common nature.\newline 
On larger scales, {and in comparison with the observations of single objects such as the mentioned D-sources}, \citet{Lutz1993} analyzed the forbidden iron line emission in the {\it inner parsec} showing a bow-shock-like distribution. The strongest [FeIII] emission was located at the position of the IRS 13 cluster, which also {included} the region of the prominent early-type star IRS 2L \citep{Buchholz2013, Roche2018}. As we found in \cite{Peissker2020b}, all the dusty objects located to the west of the Br$\gamma$-bar \citep[][]{Schoedel2011, Peissker2020c} do exhibit prominent [FeIII] lines, while the spectra of all the sources, which are located in {projection} to the east of the bar, do not {exhibit [FeIII] emission}. The basic mechanism {behind this dichotomy} is still under debate and may be part of a larger scientific frame \citep[][]{Jalali2014, peissker2021c} that will be discussed in the upcoming publications.\newline
A different debate accompanies the analysis of the dusty sources of the Galactic center \citep[for an overview, see][]{Peissker2020b, Ciurlo2020}. This discussion started with the observation of {the fast-moving} G2 a decade ago \citep{Gillessen2012}. While the authors of Gillessen et al. proposed a {coreless} cloud nature of the object, several follow-up studies questioned this classification and suggested a stellar origin to explain the emission of G2 \citep[][]{Murray-Clay2012, Scoville2013, Eckart2013,Zajacek2014,Shahzamanian2016,Zajacek2017}. {Currently, numerous authors are in favor of} a stellar nature of G2, especially because of the missing flare activity of Sgr~A* that was proposed for the periapse. For example, \cite{Witzel2014} showed a point-like L-band source close to Sgr~A* with no elongation. Recently, we underlined the classification {of G2} as a low-mass star {embedded in a dusty envelope} by analyzing a large data baseline covering the epochs between 2005-2019 \citep{peissker2021c}. Observations of other objects, such as X7 \citep{Clenet2003a, Clenet2005a, muzic2010}, revealed a similar nature compared to G2 \citep{peissker2021}. The data suggest that {these} dusty sources belong to a stellar subclass which shows {characteristics similar to} YSOs \citep{Lada1987}. Based on observed colors of dusty sources found in IRS 13, \cite{Eckart2004a} classified the investigated objects as YSOs. In this work, we will focus on the dusty sources of IRS 13, which seem to follow the same morphology as G2, X7, and the D-sources \citep{Peissker2020b, peissker2021c}.\newline
Based on a multi-wavelength photometric analysis, we use the radiative transfer {code} HYPERION \citep{Robitaille2011} to investigate the stellar type of the dusty sources. Furthermore, we test the validity of HYPERION by analyzing the flux density distribution of IRS 3. In addition to the dusty sources found in \cite{Eckart2004a}, we identify 33 newly discovered objects that can be observed in the H-, K-, and L-band. For this new population of objects, we find a similar photometric footprint compared to the literature known dusty sources suggesting a similar nature. Compared with a uniform cluster, IRS 13 seems to show an underlying pattern regarding the normalized angular momentum vector, which advocates a counterclockwise disk membership. However, this particular point will be investigated in detail in Paper II. In this paper, we focus on the detection and analysis of the newly discovered sources and the classification of the dusty objects.
This work is structured as follows. In Sec. \ref{sec:analysis}, we will list the used instruments and the related telescopes. We also give an overview of the used methods and tools for the analysis. {Section \ref{sec:analysis}} is followed by the results in Sec. \ref{sec:results} where we present the identification and the analysis of the dusty sources of the IRS 13N cluster.
The results of Sec. \ref{sec:results} are discussed in Sec. \ref{sec:discuss}. Subsequent to Sec. \ref{sec:discuss}, we conclude the discussion in Sec. \ref{sec:conclusion}. In Appendix \ref{ref:data_appendix}, we list the used data {and show supporting results of our analysis presented}.

\section{Data and Tools} 
\label{sec:analysis}
In this section, we will introduce the instruments that were used to observe the GC and describe the applied techniques for the analysis. The {public available} archival data is listed in Appendix \ref{ref:data_appendix}.

\subsection{SINFONI and NACO}

The Spectrograph for INtegral Field Observations in the Near Infrared \citep[SINFONI,][]{Eisenhauer2003, Bonnet2004} and Nasmyth Adaptive Optics System (NAOS) – Near-Infrared Imager and Spectrograph (CONICA), abbreviated as NACO \citep{Lenzen2003, Rousset2003}, were mounted at the Very Large Telescope on top of Cerro Paranal (Chile).
The near-infrared imager NACO operates in the H-, K-, L-, and M-band and provides a set of narrow filters. NACO is equipped with an S13, S27, and S54 camera with a related spatial pixel scale of 13.3 mas, 27.0 mas, and 54.3 mas, respectively.\newline
Furthermore, SINFONI is capable of providing a spectrum along with the produced image due to its Integrated Field Unit (IFU). Hence, every pixel shows a related spectrum, resulting in a 3d data cube (two spatial dimensions and one spectral dimension). The SINFONI data used here were observed in the H+K band \normalfont{($1.4\,-\,2.4\,\mu m$) with a related pixel scale of 0.1" and a spectral resolution of 1500.} Adaptive optics is enabled for both instruments. We apply common reduction steps, like the LINEARITY/DARK correction resulting in FLAT FIELDING. The pre-mentioned reduction steps are applied to {the data of} both instruments. Because of the spectroscopic characteristics of SINFONI, we also use a WAVELENGTH and DISTORTION calibration. {It should be noted that} NACO and SINFONI are decommissioned {since 2019}. As a successor, ERIS \citep{Davies2018} combines the capabilities of NACO and SINFONI.

\subsection{ALMA}

{The Atacama Large (Sub)Millimeter Array (ALMA) is located on the Chajnantor plateau (Chile).} The radio and submm observations can be executed between 31 and 1000 GHz. The majority of the ALMA CO data used in this work (Prog. ID: 2015.1.01080.S) is reduced with Common Astronomy Software Applications \citep[CASA, ][]{Casa2022} and was analyzed and discussed in \cite{Tsuboi2017, tsuboi2019, Tsuboi2020a, Tsuboi2020b}. In addition, we use scientific-ready data from the ALMA archive related to the Prog. ID 2012.1.00543.S \citep{Martin2012, Moser2017}. The ALMA data discussed and analyzed in this work shows CO v=0 (transition 3-2) at 343 GHz.

\subsection{Radiative transfer model}

\begin{figure}[htbp]
	\centering
	\includegraphics[width=.5\textwidth]{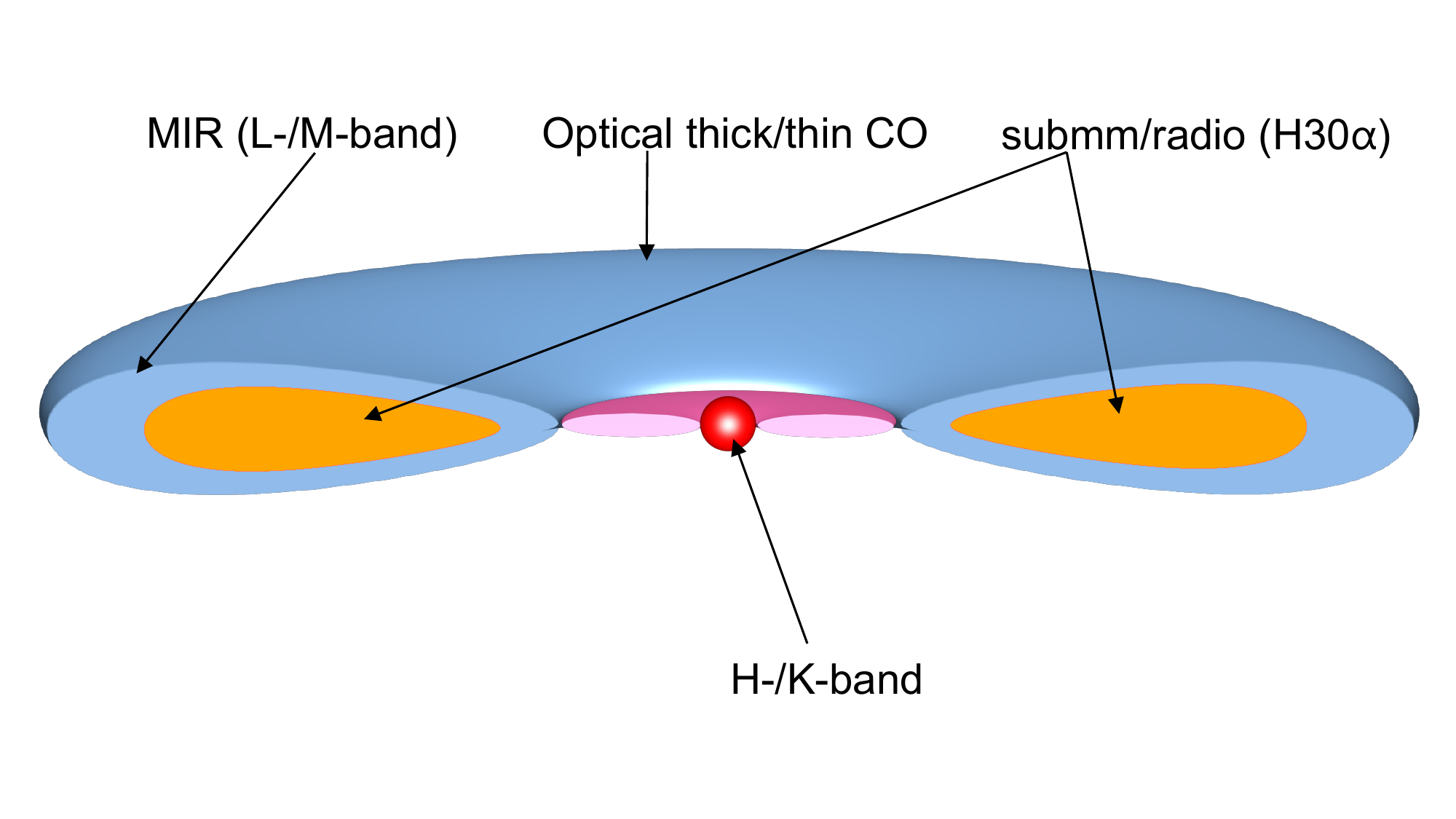}
	\caption{Sketch of a class I YSO. As indicated, various components are only detectable at specific wavelengths/bands. Using this model of a class I YSO, we expect higher luminous mid-infrared (L-/M-band) emission compared to the near-infrared (H-/K-band). This sketch is inspired by a similar figure shown in \cite{SiciliaAguilar2016}.}
\label{fig:yso_sketch}
\end{figure}

For the flux analysis based on the presented multi-wavelength observations, we use the radiative transfer {code} HYPERION\footnote{HYPERION: an open-source parallelized three-dimensional dust continuum radiative transfer code.} using dust grains as ray-tracing sources \citep{Robitaille2011, Robitaille2017}. {The spectrum that serves as an input quantity for HYPERION, we use flux density values estimated from the magnitude of the related source. Consequently,} we measure the peak counts of the object of interest and compare it with a reference source with known properties. For this, we use
\begin{equation}
    {\rm mag}_{\rm obj}\,=\,{\rm mag}_{\rm ref}-2.5 \log{\left(\frac{\text{counts}_{\rm obj}}{\text{counts}_{\rm ref}}\right)}
    \label{eq:mag_count}
\end{equation}
where mag$_{\rm ref}$ and counts$_{\rm ref}$ refer to the reference source, mag$_{\rm obj}$ and counts$_{\rm obj}$ to the analyzed object. {We estimate the source flux with} 
\begin{equation}
    f_{\rm obj}\,=\,f_{\rm ref}\,\times\,10^{[-0.4(\text{mag}_{\rm obj}-\text{mag}_{\rm ref})]}
    \label{eq:flux}
\end{equation}
where $f_{\rm ref}$ donates the flux (called zero flux) of the reference source. The basic settings for the code are listed in Table \ref{tab:sed_values}.
\begin{table}[hbt!]
    \centering
    \begin{tabular}{|cc|}
         \hline
           Properties & Setting\\
           \hline
           Number of Photons & 10$^4$ \\
           Raytracing sources & 10$^2$ \\
           Number of Iterations & 10 \\
           \hline
           \hline
    \end{tabular}
    \caption{Basic input parameters for HYPERION. Additional parameters for the model are listed in Sec. \ref{sec:results}. To optimize the computational time for the investigated cluster members, we reduce the number of photons and raytracing sources by two orders of magnitude compared to the model results discussed in \cite{peissker2023b}. Due to the low coverage of the spectral distribution because of the deficient amount of flux density values, the reduced number of radiators is not reflected in a significant qualitative difference.} 
    \label{tab:sed_values}
\end{table}
The code allows modeling various components of YSOs, such as the gaseous accretion disk or bipolar cavities \citep[see Fig. \ref{fig:yso_sketch} and][]{SiciliaAguilar2016}. {For the code used in this analysis, we model a flared disk with increasing height. The shape of the flared disk is described with
\begin{equation}
    h_{(R)}\,=\,h_0 \left(\frac{R}{R_0}\right)^{\beta}
\end{equation}
where we set $\beta=1.25$ following the settings used in \cite{Robitaille2011, Robitaille2017}}.
Furthermore, a flattened rotational and infalling dust envelope \citep[Ulrich type, see][]{Ulrich1976} can be modeled depending on the flux density.
Since HYPERION uses grains as emitters, the properties of dust are directly related to the outcome of the model. Therefore, in the following section, we will outline the dust model used for the radiative transfer analysis.
For the assumed model that is used in HYPERION, we construct a dusty envelope and a gaseous accretion disk that are arranged around a stellar core. Due to the lack of high-resolution \normalfont{(spectral/spatial)} IFU data covering the IRS 13 region, we cannot validate or exclude the presence of bipolar cavities \citep[an example of these cavities is displayed in][]{Peissker2019}. In summary, our assumed model resembles the composition of a class I YSO (Fig. \ref{fig:yso_sketch}).

\subsubsection{Dust models}

The composition of dust has a {particular} impact on photometric studies in the GC. It is, therefore, obvious that the evolution of dust models is coupled to a precise knowledge of the different spectral species. For example, \cite{Weingartner2001} limit their extinction law to the presence of carbonaceous and silicate grains. In contrast, the authors of \cite{Fritz2011} investigated the incorporation of many more infrared emission lines such as CO, CO$_2$, aliphates, and silicates. In addition, Fritz et al. consider the presence of ice particles (H$_2$O and CO ice) in agreement with the studies by \citet{Moneti2001} and \citet{moultaka2015}. \citet{Fritz2011} conclude that the model of \cite{Zubko2004} is the best-fitting model to describe the extinction toward the GC. However, Zubko et al. uses R$_V\,=\,A_V / E_{B-V}\,=\,3.1$ which is consistent with the work by \cite{Draine2003}. Therefore, we use \cite{Draine2003} to model the dust grains used in this work to perform a spectral analysis.

\subsection{High-pass filter}
\label{sec:high-pass}

The high-pass filtering technique is a common tool for deblurring imaging data by minimizing the influence of the PSF wings of a bright star. While the Lucy Richardson (LR) algorithm \citep[][]{Lucy1974} offers a variety of setup parameters, the smooth-subtract algorithm is a robust approach to analyzing the data. \normalfont{Critically, the LR algorithm tends to transform elongated structures into point sources. While this does not necessarily exclude the usage of the algorithm on regions with extended structures, IRS 13 shows elongated and compact objects with unknown nature\footnote{Please see DS28 and DS33 shown in Fig. \ref{fig:finding_chart} which form an elongated structure with $\alpha$. Using the LR algorithm without the knowledge of the nature of these sources might bias the interpretation.}.} 
The accessible implementation into the process of analyzing the data is done by smoothing the original image I$_{\rm orig}$ with a Gaussian kernel with a size that should be in the range of the PSF {measured in the data}. The resulting smoothed image I$_{\rm smo}$ describes a low-pass filtered version of I$_{\rm orig}$. With
\begin{equation}
    I_{{\rm orig}}\,-\,I_{\rm smo}\,=\,I_{\rm high}
\end{equation}
we {acquire} the high-pass filtered version I$_{\rm high}$ of I$_{\rm orig}$. To enhance the image quality, one can apply a Gaussian smoothing filter smaller than the PSF to I$_{\rm high}$. A comparable description of this process is outlined in \cite{Peissker2022} as well. A rather qualitative comparison is presented in the following, where we investigate the astrometric and photometric imprint of the high-pass filter on the data.
In general, we find no significant difference between a stellar position determined in I$_{orig}$ or I$_{high}$. 
\begin{figure*}[t!]
	\centering
	\includegraphics[width=.8\textwidth]{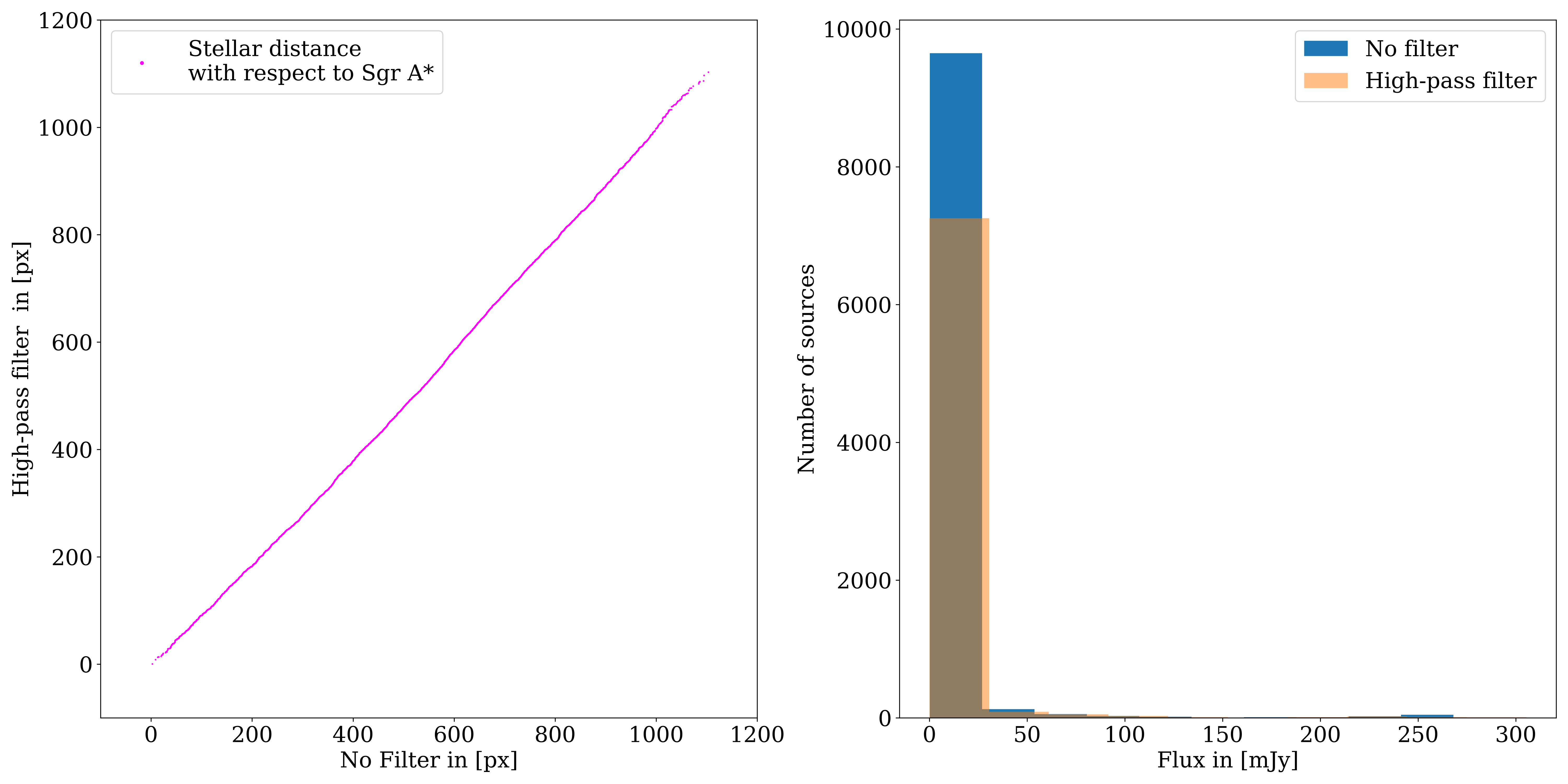}
	\caption{Comparison of raw data with processed data treated with a high-pass filter. The plot on the left shows the distance of several thousand stars with respect to Sgr~A*. This plot underlines the robustness of the used high-pass filter regarding an astrometric analysis \normalfont{since every individual star shows the same distance to Sgr~A* independent of the used approach (filter/no filter)}. On the right side, we show the K-band flux distribution of about 10000 stars and estimate a difference between filtered and nonfiltered data of about 20$\%$. Please see the text for details.}
\label{fig:filter_comp}
\end{figure*}
For the flux shown in Fig. \ref{fig:filter_comp}, we find an uncertainty of about 20$\%$ between filtered and non-filtered data. Taking into account the usual flux density uncertainties shown, for example, in \cite{peissker2023b} and this work, the value distribution shown in Fig. \ref{fig:filter_comp} is well inside the expected range. We emphasize that we expect a flux difference between high-pass filtered and non-filtered data due to the presence of elongated structures, such as the mini-spiral (see Fig. \ref{fig:finding_chart}). High-pass filtering tends to convert elongated structures into point sources, resulting in a broader flux distribution as shown in Fig. \ref{fig:filter_comp}. Therefore, the analysis of individual sources with the high-pass filter presented here should be carried out with caution.

\section{Results}
\label{sec:results}
\begin{figure*}[htbp!]
	\centering
	\includegraphics[width=1.\textwidth]{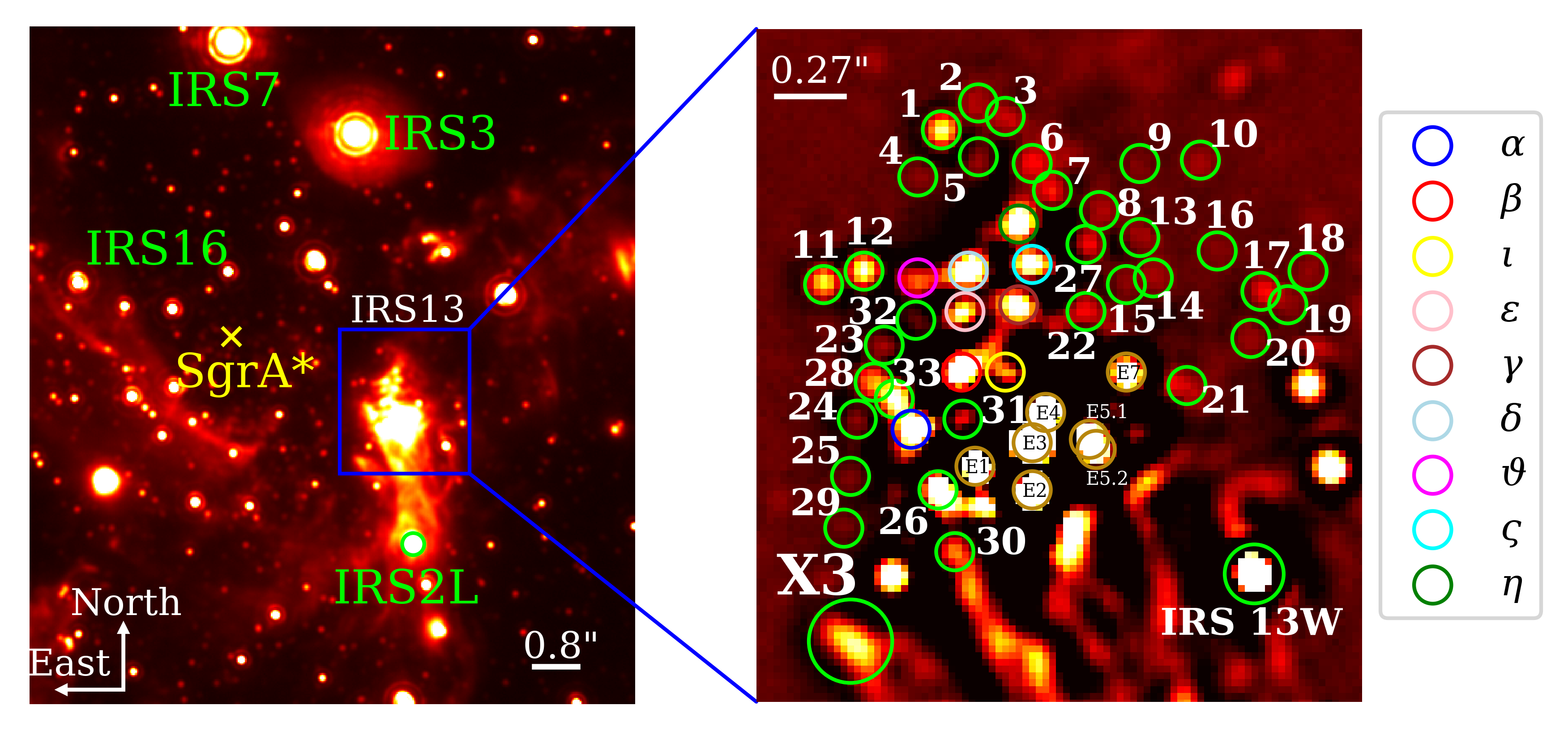}
	\caption{Finding chart for the dusty sources associated with a zoomed view toward the IRS 13 cluster. While the Greek-named sources are previously analyzed in the literature, we find additional sources that we enumerate to avoid confusion. The sources E1-E7 belong to IRS 13E and are marked for clarity. The bow-shock source X3 is analyzed in \cite{peissker2023b}. In addition, IRS 13W is associated with a M3 giant star \citep[][]{Maillard2004}. The data shown was observed with NACO in 2004. Whereas the left image displays a continuum overview of the direct vicinity of Sgr~A* (about $0.45\times 0.50$ pc), the right represents a zoomed-in high-pass filtered view towards IRS13. Please consult Appendix \ref{sec:kband-counter-app} and Appendix \ref{sec:hband-counter-app} for the H- and K-band counterparts of the sources marked. In addition, Sec. \ref{sec:light-curves-app} displays all newly identified dusty sources in the H-, K-, and L-band, including their related light curve. \normalfont{Every source not marked in this finding chart is not considered to be a cluster member due to its proper motion or previous studies \citep{Pfuhl2014, Gautam2019}.}}
\label{fig:finding_chart}
\end{figure*}

In the following, we present the results of the multi-wavelength analysis of IRS 13. From the proper motion analysis, we derive the cluster membership of the individual sources. Using photometric measurements in various bands, we classify the observed objects and estimate the related stellar mass. In Fig. \ref{fig:finding_chart}, we show a NACO L-band overview of the region of interest. 
The region displayed in Fig. \ref{fig:finding_chart} is called IRS 13 and can be (historically) {divided up into} IRS 13N and IRS 13E. The North-East nomenclature may result in confusion due to the proper motion of the IRS 13E-related sources and the coinciding IRS 13N objects. For ease of confusion, we will only use the term IRS 13 here when referring to the sources of IRS 13N and IRS 13E.

\subsection{Photometric analysis}
\label{sec:photo_results}
The analysis of about two decades of NIR and MIR NACO data revealed 33 sources that can be observed in {various} bands in addition to previously known dust-enshrouded objects. {In Fig. \ref{fig:finding_chart}, we show the L-band detection of all the investigated sources in this work. In addition, Appendix \ref{sec:kband-counter-app} and Appendix \ref{sec:hband-counter-app} reveal the related H- and K-band identification of the dusty sources (DS). The rich data set permits us to produce light curves and individual detections of all new DS objects (Appendix \ref{sec:light-curves-app}). Furthermore, Table \ref{tab:ds_sources_ident} lists the magnitudes of the newly discovered DS objects.} We compare all the known IRS 13 objects with the literature and list the new sources identified in Table \ref{tab:ID}, Appendix \ref{ref:dusty_objects_ident}.
\begin{table*}
\centering
\begin{tabular}{|cccccccc|}
\hline
\hline
ID & H-band & K-band & L-band & K-L & $\Delta$K-L & H-K & $\Delta$H-K\\ 
\hline
DS1 & 14.40 $\pm$ 0.15 & 12.37 $\pm$ 0.50 & 10.13 $\pm$ 0.33& 2.24 & 0.59 & 2.03&0.52\\
DS2 & - & - & 12.60 $\pm$ 0.24 & - & - & & \\
DS3 & - & - & 12.30 $\pm$ 0.30  & - & - & & \\
DS4 & 18.27 $\pm$ 0.59 & 15.71 $\pm$ 0.35 & 14.03 $\pm$ 1.41 & 1.68 & 1.45 & 2.56 & 0.68 \\
DS5 & - & - & 14.37 $\pm$ 1.68  & - & - & & \\
DS6 & 18.91 $\pm$ 0.2 & 17.62 $\pm$ 0.44 & 11.43 $\pm$ 0.54 & 6.19 & 0.69 & 1.29 & 0.48\\
DS7 & - & - & 11.54 $\pm$ 0.48  & - & - & - & -\\
DS8 & 19.17 $\pm$ 0.2 & 17.57 $\pm$ 0.44 & 12.56 $\pm$ 0.76 & 5.01 & 0.87 & 1.60 & 0.48\\
DS9 & 17.11 $\pm$ 0.45 & 14.61 $\pm$ 0.76 & 13.91 $\pm$ 1.09 & 0.7 & 1.32 & 2.50 & 0.88\\
DS10 & 16.50 $\pm$ 0.31 & 14.19 $\pm$ 0.69 & 12.82 $\pm$ 0.46 & 1.37 & 0.82 & 2.31 & 0.75\\
DS11 & 14.22 $\pm$ 0.35 & 12.10 $\pm$ 0.70 & 10.46 $\pm$ 0.64 & 1.64 & 0.94 & 2.12 & 0.78\\
DS12 & 14.04 $\pm$ 0.36 & 11.86 $\pm$ 0.72 & 10.23 $\pm$ 0.60 & 1.63 & 0.93 & 2.18 & 0.80\\
DS13 & 18.55 $\pm$ 0.46 & 16.51 $\pm$ 0.78 & 12.80 $\pm$ 0.54 & 3.71 & 0.94 & 2.04 & 0.90\\
DS14 & 18.84 $\pm$ 0.94 & 17.33 $\pm$ 0.92 & 12.76 $\pm$ 0.39 & 4.57 & 0.99 & 1.51 & 1.31\\
DS15 & 18.77 $\pm$ 0.37 & 17.28 $\pm$ 0.53 & 13.65 $\pm$ 0.56 & 3.73 & 0.77 & 1.49 & 0.64\\
DS16 & 18.61 $\pm$ 0.71 & 17.39 $\pm$ 1.05 & 13.75 $\pm$ 0.83 & 3.64 & 1.33 & 1.22 & 1.26\\
DS17 & 15.68 $\pm$ 0.35 & 13.28 $\pm$ 0.87 & 11.76 $\pm$ 0.64 & 1.52 & 1.08 & 2.40 & 0.93\\
DS18 & 19.75 $\pm$ 0.95 & 17.49 $\pm$ 0.40 & 13.39 $\pm$ 0.54 & 4.10 & 0.67 & 2.26 & 1.03\\
DS19 & 19.55 $\pm$ 0.01 & 18.23 $\pm$ 0.64 & 13.95 $\pm$ 0.75 & 4.28 & 0.98 & 1.32 & 0.64\\
DS20 & 18.67 $\pm$ 0.15 & 15.81 $\pm$ 0.85 & 13.58 $\pm$ 0.72 & 2.23 & 1.11 & 2.86 & 0.86\\
DS21 & 20.23 $\pm$ 1.17 & 17.83 $\pm$ 1.01 & 12.12 $\pm$ 0.59 & 5.71 & 1.16 & 2.40 & 1.54\\
DS22 & 19.90 $\pm$ 1.21 & 17.40 $\pm$ 0.35 & 11.91 $\pm$ 0.54 & 5.49 & 0.65 & 2.57 & 1.25\\
DS23 & 18.65 $\pm$ 0.37 & 17.33 $\pm$ 0.59 & 14.73 $\pm$ 1.39 & 2.60 & 1.51 & 1.32 & 0.69\\
DS24 & 15.07 $\pm$ 0.40 & 13.05 $\pm$ 0.74 & 13.71 $\pm$ 1.65 & -0.66 & 1.80 & 2.02 &0.84\\
DS25 & 16.28 $\pm$ 0.30 & 14.49 $\pm$ 0.75 & 15.46 $\pm$ 1.34 & -0.97 & 1.53 & 1.79 &0.80\\
DS26 & 17.20 $\pm$ 0.62 & 15.64 $\pm$ 1.16 &  9.13 $\pm$ 0.59 & 6.51 & 1.30 & 1.56 & 1.31\\
DS27 & 14.83 $\pm$ 0.29 & 12.77 $\pm$ 0.61 & 12.08 $\pm$ 0.66 & 0.69 & 0.89 & 2.06 & 0.67\\
DS28 & 16.14 $\pm$ 0.27 & 15.80 $\pm$ 0.92 & 10.76 $\pm$ 0.42 & 5.04 & 1.01 & 0.34 & 0.95\\
DS29 & 16.17 $\pm$ 0.36 & 14.42 $\pm$ 0.86 & 16.29 $\pm$ 1.33 & -1.87 & 1.58 & 1.75 &0.93\\
DS30 & 15.67 $\pm$ 0.47 & 13.83 $\pm$ 0.74 & 10.92 $\pm$ 0.51 & 2.91 & 0.89 & 1.84 & 0.87\\
DS31 & 14.84 $\pm$ 0.29 & 12.97 $\pm$ 1.13 & 13.08 $\pm$ 2.06 & -0.11 & 2.34 & 1.87 &1.16\\
DS32 & 16.74 $\pm$ 0.32 & 15.02 $\pm$ 0.80 & 14.71 $\pm$ 0.93 & 0.31 & 1.22 & 1.72 & 0.86\\
DS33 & 18.39 $\pm$ 1.11 & 16.16 $\pm$ 0.62 &  9.92 $\pm$ 0.57 & 6.24 & 0.84 & 2.23 & 1.27\\
\hline
\end{tabular}
\caption{{Mean dereddened magnitudes of the DS sources analyzed in this work. We list the mean magnitude (see also Appendix \ref{ref:dusty_objects_ident}) and calculate the variance of the individual standard deviation. Hence, the uncertainty of the K-L and H-K colors is given by the square root of the variance and represent the total standard deviation.}}
\label{tab:ds_sources_ident}
\end{table*}
Consistent with the literature, we identified all known sources in the L-, K-, H-, and M-band. Because every previous analysis of the cluster covered only a fraction of the objects investigated here, our objective was to provide a complete list of all sources with a consistent nomenclature. {To avoid confusion with existing studies of the region}, we adapt the nomenclature for the brightest sources (E1-E7 and $\alpha$-$\iota$) of the IRS 13 cluster.

For the analysis of the data, we applied the introduced image sharpener and extracted the positions with a Gaussian fit with dimensions that correspond to the PSF of the data. The FWHM is about 5 to 6 pixels. With a spatial pixel scale for the L-band data of 27 mas, the dimensions {of the corresponding PSF} are about 1.3"-1.6". For the K-band data and a {related} spatial pixel scale of 13 mas, the dimensions of the PSF transfer to 0.6"-0.7". Since the sources studied {in this work} have dominant MIR emissions {suffering from reduced confusion and crowding}, we {focus on} the L-band and M-band whenever the detection of the objects in the NIR is blended. \normalfont{Due to the prominent and variable background of the crowded and dense cluster, we did not apply a local background subtraction since we consider confusion as the dominant source of uncertainty. Especially DS4 and DS5 suffer from confusion and blending effects that are confronted by the usage of the mean covering almost two decades of observations (Table \ref{tab:ds_sources_ident}).} 

\subsection{Proper motion}
\label{ref:proper_motion}

\begin{figure}[htbp!]
	\centering
	\includegraphics[width=.5\textwidth]{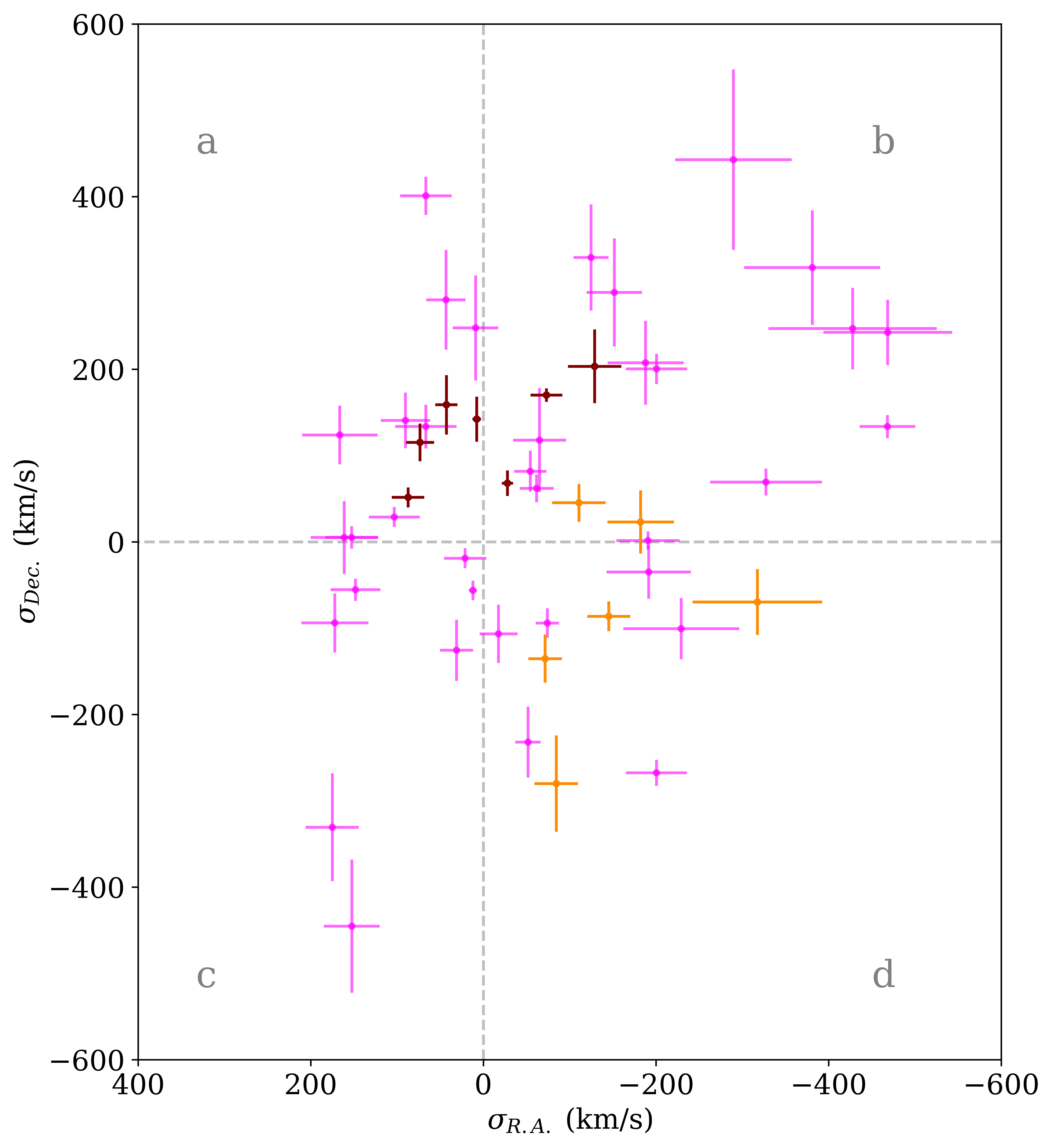}
	\caption{Proper motion of all observed IRS 13 sources (see Fig. \ref{fig:finding_chart}). The E-stars (E1-E5) are {orange} colored, the {brown} data points represent the bright dusty sources ($\alpha$-$\eta$). {We find a slight over and under density regarding the shown data points in the corresponding quadrant. Please see the text for details.}}
\label{fig:pm_membership}
\end{figure}

\begin{figure}[htbp!]
	\centering
	\includegraphics[width=.5\textwidth]{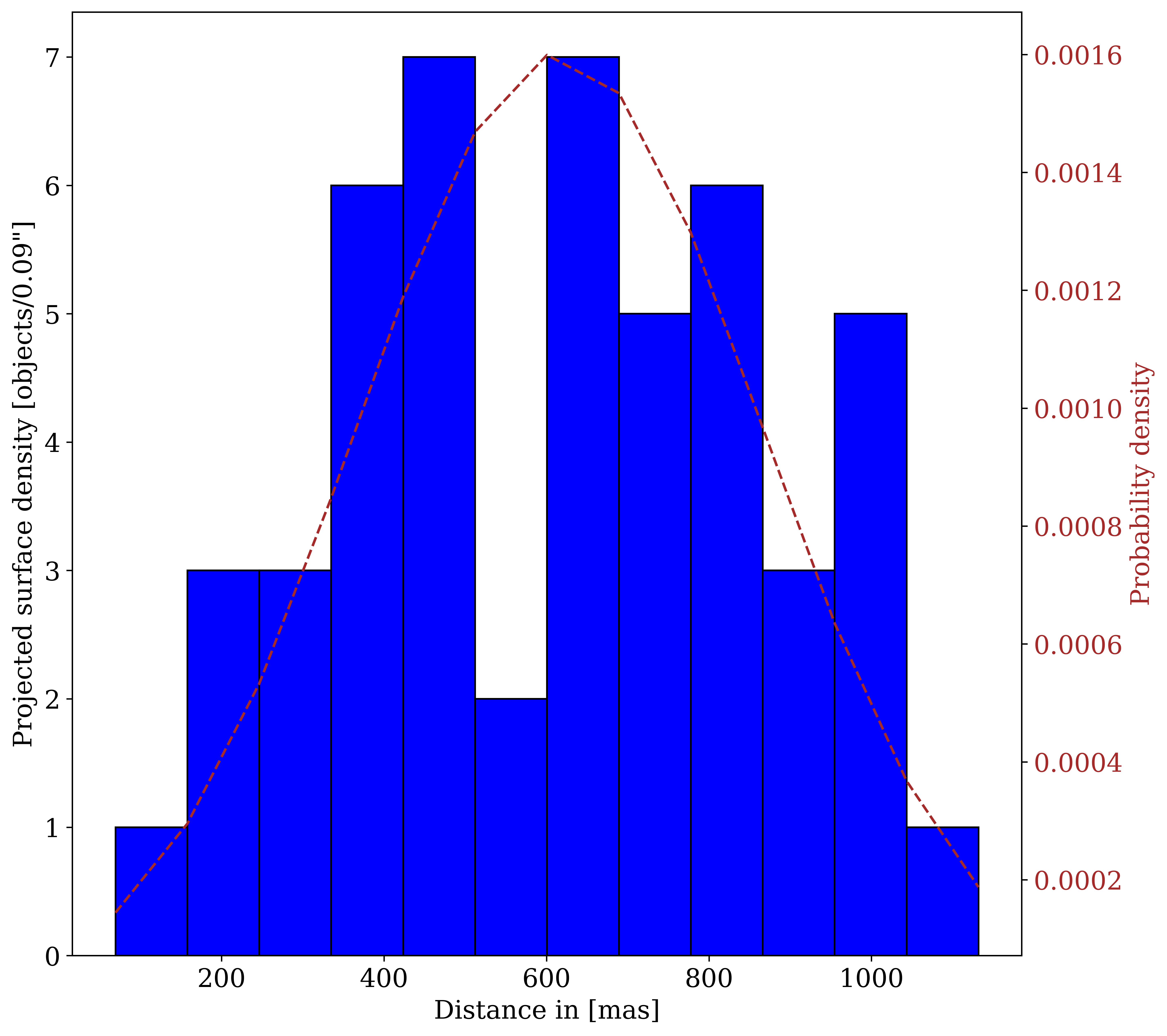}
	\caption{Gaussian distribution of the DS sources around the geometrical center of IRS 13. The blue-colored histogram shows the number of sources inside the bin of the size of 0.09". The brown dashed line represents a Gaussian fit of the blue-colored projected surface density. Since the geometrical center of the cluster is located in an empty region, the Gaussian probability density drops towards origin of the stellar distribution.}
\label{fig:density_propability}
\end{figure}

\begin{figure*}[htbp!]
	\centering
	\includegraphics[width=1.\textwidth]{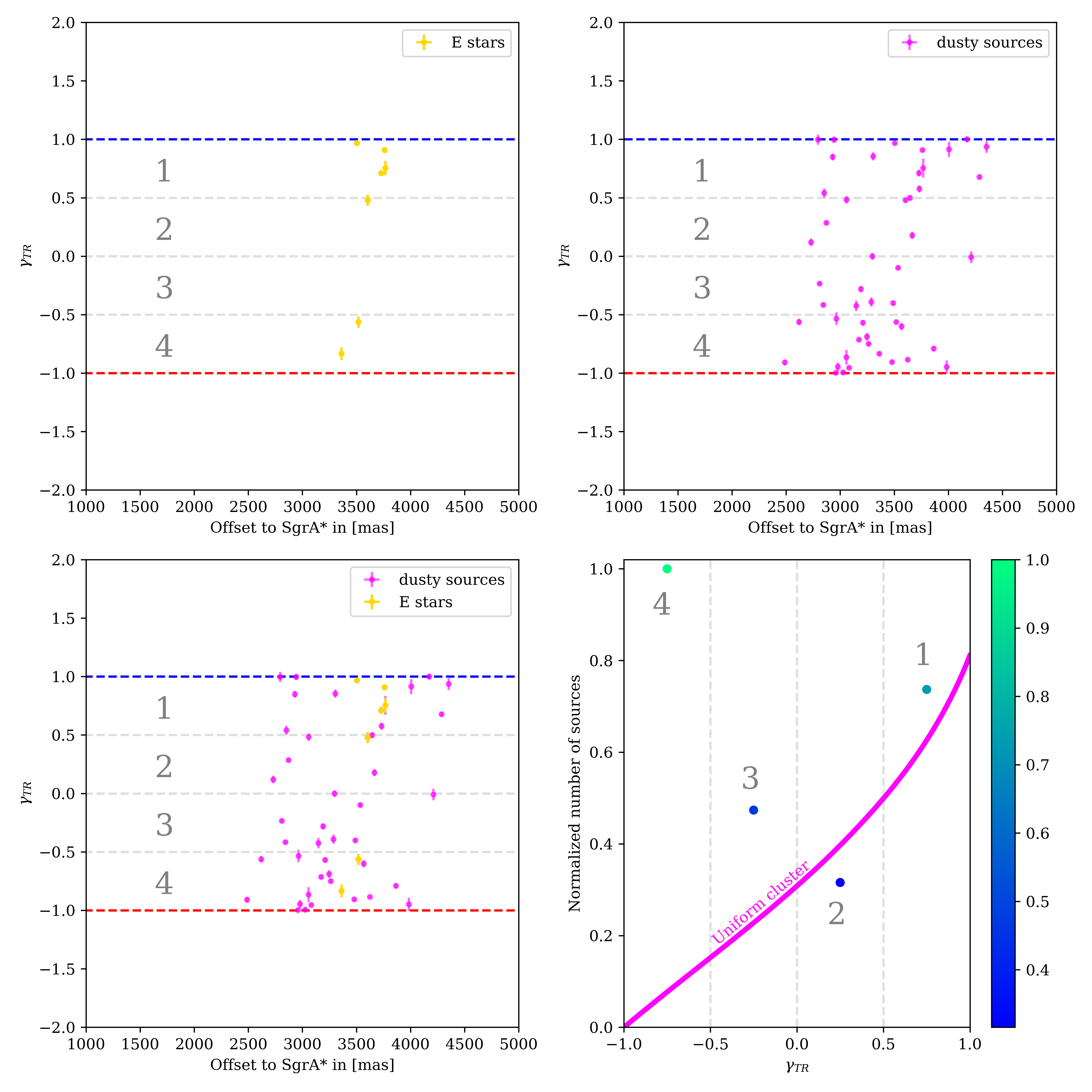}
	\caption{Anisotropy parameter $\gamma_{TR}$ for the investigated sources in IRS 13. In the upper row, we illustrate $\gamma_{TR}$ for the E-stars (left) and the dusty sources (r). Dividing the anisotropy parameter into four bins reveals an overdensity for some of the sources close to $\gamma_{TR}\approx\pm 1$ implying the existence of substructures. In the lower-left plot, we show all investigated sources of the IRS 13 cluster. The uncertainties of the anisotropy parameters are determined with error propagation of the related standard deviation of the proper motion values (Appendix \ref{ref:proper_motion}). The grey numbers and dashed vertical lines represent the related bin. The lower right plot shows the normalized number of sources for each bin as a function of $\gamma_{TR}$. The magenta-colored line represents the normalized theoretical probability distribution that one would expect for a uniform cluster. The comparison of the observed structure of IRS 13 with a uniform cluster suggests anomalies that are responsible for a non-uniform distribution.} 
\label{fig:anisotropy_parameter}
\end{figure*}

\begin{figure}[htbp!]
	\centering
	\includegraphics[width=.5\textwidth]{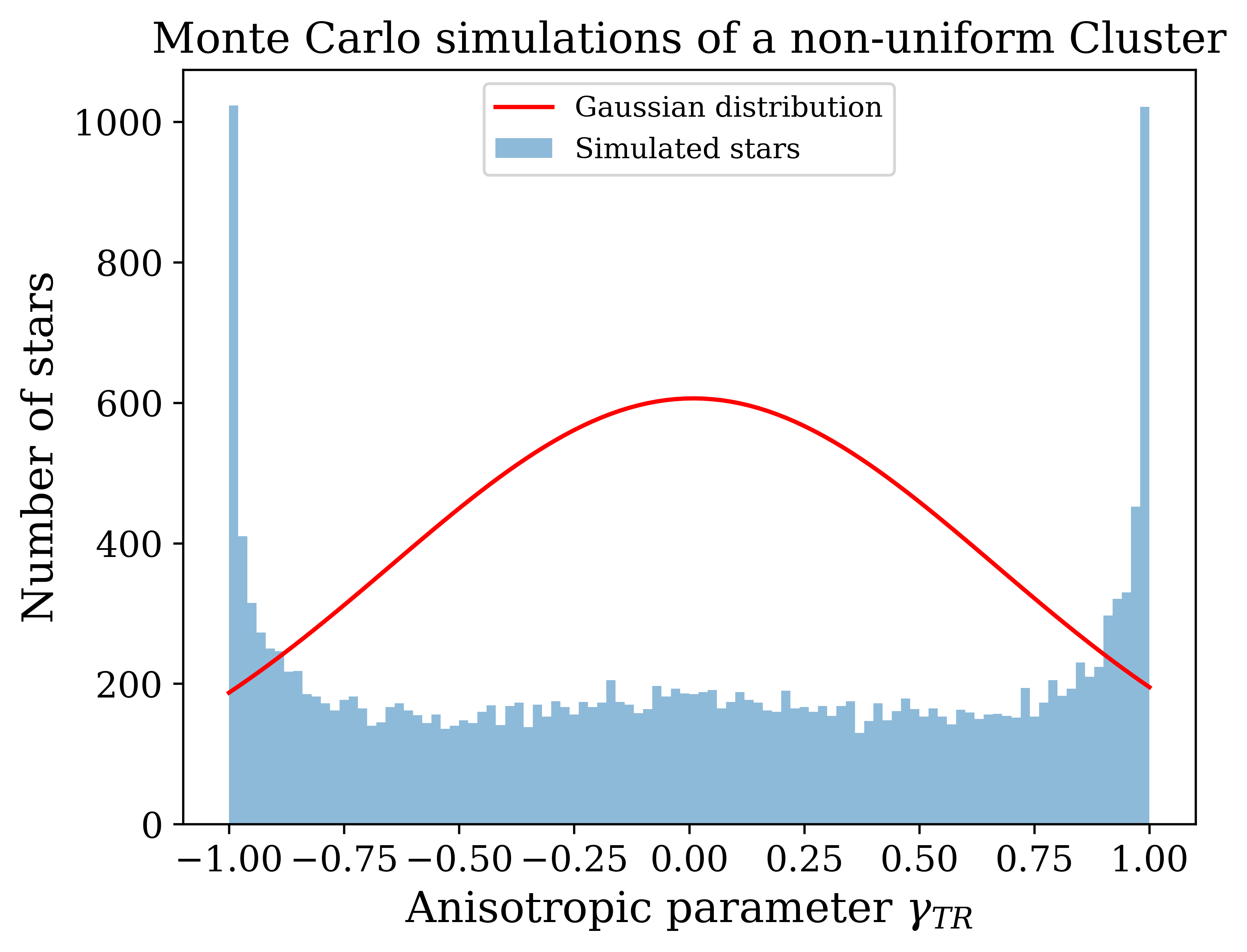}
	\caption{Monte Carlo simulations of $\gamma_{TR}$ for 10000 stars. As we have shown in Fig. \ref{fig:anisotropy_parameter}, the anisotropy parameter peaks at $\pm 1$, which indicates a non-uniform cluster. In comparison, we show a Gaussian distribution.}
\label{fig:monte_carlo}
\end{figure}

Simultaneously to the photometric analysis presented in Sec. \ref{sec:photo_results}, we estimate the proper motion of the investigated cluster members. Due to the chance of confusion regarding the detectability of the DS sources, we use K- and L-band observations whenever possible carried out with NACO between 2002 and 2019. Except for 2014 and 2015, we trace the objects listed in Table \ref{tab:ID} in the majority of available observations. We fit a PSF-sized Gaussian to the individual sources to extract its position (Table \ref{tab:prop}). The origin of our reference frame coincides with the position of Sgr~A*. For this, we identify the position of the B2V star S2 and use its well-known and observed orbital solution. From the orbital solution and the position of S2 \citep{Do2019S2}, we derive the location of Sgr~A*. We refer to Appendix \ref{sec:individual_pos_app}, which lists all positions of S2 and the dusty sources investigated in this work.
Since the IRS 13 cluster is about 0.12 parsec away from the location of the SMBH, we assume an approximately vanishing velocity v$_{Sgr~A*}$ of Sgr~A*. Even for objects close to Sgr~A*, the velocity effect caused by v$_{Sgr~A*}$ is in the subpixel regime \citep[][]{Parsa2017}.
We list the resulting proper motion of the DS sources and all other objects investigated in this work in Table \ref{tab:prop}. Due to the high degree of crowding, the standard deviation-based uncertainties may not cover the full set of entities.
\begin{table*}[htbp!]
\centering
\setlength{\tabcolsep}{0.5pt}
\begin{tabular}{@{}lrrrrrrrr@{}}
\toprule
ID &
  \multicolumn{1}{c}{\begin{tabular}[c]{@{}c@{}}$r_{0,R.A.}$\\ (mas)\end{tabular}} &
  \multicolumn{1}{c}{\begin{tabular}[c]{@{}c@{}}$r_{0,Dec.}$\\ (mas)\end{tabular}} &
  \multicolumn{1}{c}{\begin{tabular}[c]{@{}c@{}}$v_{R.A.}$\\ (km/s)\end{tabular}} &
  \multicolumn{1}{c}{\begin{tabular}[c]{@{}c@{}}$v_{Dec.}$\\ (km/s)\end{tabular}} &
  \multicolumn{1}{c}{\begin{tabular}[c]{@{}c@{}}$\Delta r_{0,R.A.}$ \\ (mas)\end{tabular}} &
  \multicolumn{1}{c}{\begin{tabular}[c]{@{}c@{}}$\Delta r_{0,Dec.}$\\ (mas)\end{tabular}} &
  \multicolumn{1}{c}{\begin{tabular}[c]{@{}c@{}}$\Delta v_{R.A.}$ \\ (km/s)\end{tabular}} &
  \multicolumn{1}{c}{\begin{tabular}[c]{@{}c@{}}$\Delta v_{Dec.}$\\ (km/s)\end{tabular}} \\ 
  \hline
$\alpha$    & -2677.0 & -1479.1 & 87.4   & 51.6   & 3.2  & 1.9  & 18.9 & 11.5  \\
$\beta$     & -2890.0 & -1246.0 & 73.4   & 115.4  & 3.0  & 4.1  & 16.2 & 21.8  \\
$\gamma$    & -3088.0 & -1003.0 & -73.0  & 170.2  & 3.4  & 3.2  & 18.5 & 7.8   \\
$\delta$    & -2903.6 & -854.3  & 7.9    & 142.3  & 1.0  & 4.3  & 5.0  & 26.1  \\
$\epsilon$  & -2883.0 & -1015.0 & 42.9   & 158.9  & 2.3  & 6.7  & 12.9 & 34.5  \\
$\zeta$     & -3150.8 & -826.5  & -128.9 & 203.5  & 5.5  & 7.7  & 31.0 & 42.7  \\
$\eta$      & -3104.0 & -654.9  & -27.8  & 68.0   & 1.2  & 2.7  & 6.6  & 14.8  \\
$\vartheta$ & -2700.0 & -892.0  & 90.4   & 140.8  & 4.8  & 5.5  & 28.7 & 32.1  \\
$\iota$     & -3043.4 & -1240.4 & -54.2  & 81.9   & 3.4  & 4.3  & 18.7 & 23.8  \\
1           & -2795.0 & -276.4  & -74.0  & -94.0  & 2.0  & 2.6  & 13.6 & 17.1  \\
2           & -2954.4 & -175.8  & 9.2    & 248.0  & 4.6  & 10.5 & 26.4 & 60.9  \\
3           & -3070.9 & -251.8  & 43.5   & 280.6  & 4.4  & 10.1 & 22.8 & 57.8  \\
4           & -2687.0 & -480.3  & 21.2   & -18.8  & 5.1  & 2.0  & 24.6 & 11.5  \\
5           & -2935.5 & -408.9  & -65.0  & 117.9  & 4.6  & 9.0  & 30.8 & 60.3  \\
6           & -3182.0 & -421.7  & -151.6 & 289.2  & 6.0  & 12.8 & 32.1 & 62.6  \\
7           & -3217.6 & -531.3  & -124.7 & 329.7  & 5.2  & 11.9 & 20.2 & 61.5  \\
8           & -3430.4 & -636.4  & -289.6 & 443.0  & 11.7 & 18.6 & 67.6 & 104.6 \\
9           & -3597.6 & -435.1  & 31.1   & -125.5 & 3.2  & 5.6  & 19.3 & 35.3  \\
10          & -3844.4 & -387.9  & 152.5  & -445.1 & 4.4  & 10.3 & 32.3 & 77.1  \\
11          & -2323.4 & -891.0  & 12.3   & -56.2  & 0.9  & 2.1  & 4.6  & 11.3  \\
12          & -2484.7 & -830.9  & 175.1  & -330.6 & 5.7  & 11.4 & 30.8 & 62.4  \\
13          & -3593.2 & -723.4  & -381.0 & 317.9  & 17.1 & 14.1 & 78.8 & 66.4  \\
14          & -3629.7 & -867.4  & -468.4 & 242.8  & 14.3 & 7.9  & 74.6 & 37.9  \\
15          & -3526.0 & -917.0  & -427.7 & 247.2  & 20.3 & 10.4 & 97.6 & 47.1  \\
16          & -3909.4 & -767.9  & -17.5  & -106.5 & 3.3  & 5.5  & 21.9 & 33.8  \\
17          & -4106.0 & -930.8  & -61.6  & 62.1   & 3.9  & 2.8  & 19.6 & 16.2  \\
18          & -4267.0 & -858.9  & -191.4 & -34.7  & 8.7  & 5.2  & 48.9 & 31.1  \\
19          & -4178.5 & -956.4  & -229.2 & -100.4 & 13.4 & 7.2  & 67.2 & 35.4  \\
20          & -4017.9 & -1125.5 & -190.6 & 1.7    & 7.5  & 2.1  & 36.6 & 10.6  \\
21          & -3781.6 & -1318.7 & -327.4 & 69.3   & 12.9 & 2.8  & 64.7 & 15.6  \\
22          & -3383.1 & -1021.1 & -187.9 & 207.5  & 8.5  & 9.8  & 44.0 & 48.6  \\
23          & -2543.5 & -1155.7 & 161.4  & 5.1    & 6.8  & 7.3  & 39.0 & 42.2  \\
24          & -2461.0 & -1441.3 & 172.1  & -93.8  & 7.8  & 6.4  & 38.8 & 34.1  \\
25          & -2410.6 & -1687.7 & 152.6  & 5.2    & 5.9  & 2.6  & 30.5 & 13.0  \\
26          & -2801.4 & -1750.3 & 103.3  & 28.9   & 4.7  & 2.0  & 29.5 & 11.5  \\
27          & -3404.0 & -720.0  & -51.6  & -231.9 & 2.8  & 7.9  & 14.7 & 41.1  \\
28          & -2551.8 & -1318.4 & 166.4  & 124.0  & 7.6  & 6.0  & 43.7 & 34.0  \\
29          & -2429.7 & -1719.9 & 66.85  & 401.1  &  8.9 & 6.6  & 29.9 & 22.1 \\
30          & -2921.1 & -2047.5 & 66.85  & 133.7  & 10.6 & 7.6  & 35.6 & 25.4  \\
31          & -2948.4 & -1474.2 & -200.5 & 200.5  & 10.7 & 5.4  & 35.6 & 17.5  \\
32          & -2730.0 & -1064.7 & -468.0 & 133.7  & 9.6  & 4.0  & 32.2 & 13.3  \\
33          & -2921.1 & -1283.1 & -200.5 & -267.4 & 10.5 & 4.5  & 35.2 & 15.1  \\
E1          & -2935.0 & -1637.2 & -84.3 & -280.0  & 4.9  & 10.9 & 25.4 & 55.9   \\
E2          & -3161.5 & -1729.6 & -145.2 & -86.2  & 3.8  & 2.4  & 24.9 & 17.3  \\
E3          & -3171.0 & -1521.0 & -71.5  & -135.2 & 3.5  & 5.4  & 19.4 & 27.9  \\
E4          & -3201.0 & -1424.0 & -182.2 & 23.2   & 8.4  & 6.8  & 38.5 & 36.7  \\
E5.0        & -3401.0 & -1528.0 & -110.6 & 45.4   & 5.2  & 3.9  & 31.1 & 22.0  \\
E5.1        & -3413.0 & -1579.3 & -317.4 & -69.6  & 15.2 & 8.6  & 75.1 & 38.2  \\
E7          & -3549.0 & -1260.0 & 148.3  & -55.4  & 6.3  & 2.5  & 28.9 & 12.9  \\ 
\hline
\end{tabular}%
\caption{Proper motions of the sources investigated in this work based on $L$-band observations carried out with NACO between 2002 and 2018. The uncertainties represent the standard deviation. The distance of the DS sources indicates the related position in 2002.}
\label{tab:prop}
\end{table*}
With upcoming JWST observations, we expect decreased astrometric uncertainties from MIRI IFU data due to unique Doppler-shifted emission lines. 
However, we use the results from the presented astrometric analysis to investigate the cluster and the sources for any anisotropy. We will use the approach of \cite{Eckart1997} and \cite{Genzel2000}, where the authors used the anisotropy parameter to analyse the stellar content of the S-cluster. The anisotropy parameter is defined as 
\begin{equation}
    \gamma _{TR}\,=\,\frac{v_T^2 - v_R^2}{v_T^2 + v_R^2}
    \label{eq:anisotropy}
\end{equation}
where v$_T$ and v$_R$ refer to the proper motion components perpendicular and parallel to the projected radius vector in the sky, respectively. The anisotropy parameter $\gamma_{TR}$ provides an accessible numerical approach to investigate IRS 13 related objects for abnormalities. These abnormalities would imply a tendency for a specific stellar type or substructures. A uniform data point distribution indicates a continuous and randomized structure of the cluster. For example, \cite{Ali2020} expected sine-like distribution for the inclination angle. While this can not be transferred to the anisotropy parameter, the Gaussion-shaped and uniform cluster discussed in \cite{Fellenberg2022} can be used for the expected distribution of the dusty sources. Therefore, we will investigate the cluster sample for Gaussian-like structures to investigate the presence of substructures.\newline
In Table \ref{tab:prop}, we list the resulting proper motions of all investigated objects (Table \ref{tab:ID}) with the related distances to Sgr~A*. Using the proper motions listed in Table \ref{tab:prop}, we derive the velocity-velocity diagram (Fig. \ref{fig:pm_membership}).
From the fit and the data points displayed in Fig. \ref{fig:pm_membership}, it is evident that the proper motion of the investigated sources shows {an asymmetric distribution around the geometrical center, implying the presence of a trend. Translating this finding to a density as a function of quadrant yields:}
\begin{itemize}
    \item[1:] 13 sources in a (26.5$\%$)
    \item[2:] 19 sources in b (38.8$\%$)
    \item[3:]  7 sources in c (14.3$\%$)
    \item[4:] 10 sources in d (20.4$\%$) 
\end{itemize}
Based on this analysis, the significance of a trend for the proper motion of the DS sources lacks a reasonable level of explicitness. From the analysis of the proper motion distribution, it is implied that the dusty sources follow a rather uniform arrangement, which is furthermore reflected in the Gaussian-like density probability displayed in Fig. \ref{fig:density_propability}. This figure shows the Gaussian distribution of the dusty sources around the geometrical center of the cluster as a function of the distance d, which is estimated with $p_{\rm (geo)}\,=\,p_{(x,y)} - p_{aver(x,y)}$. In the given relation, $p_{\rm ( geo)}$ is the geometrical center of the cluster, $p_{aver(x,y)}$ the average distance of all cluster sources to Sgr~A*, and $p_{(x,y)}$ the averaged distance of a single source between 2002 and 2018. With this, the presentation of the distribution of the DS sources shown in Fig. \ref{fig:density_propability} does not reflect possible existing anomalies of the cluster.
Despite the decreased surface density at about 0.6 arcsec indicated in Fig. \ref{fig:density_propability}, we find a probability density consistent that resembles a Gaussian function as we would expect for a uniform cluster \citep{Genzel2000}. Because $p_{\rm ( geo)}$ does not necessarily have to be located at the highest stellar density, the Gaussian fit exhibits an offset from $\rm d\,=\,0\,mas$ as illustrated Fig. \ref{fig:density_propability}.
Therefore, we aim to expand the search for substructures and translate the estimated values in Table~\ref{tab:prop} into the anisotropy parameter $\gamma_{TR}$ indicated by Eq. \ref{eq:anisotropy}. In the first three plots displayed in Fig. \ref{fig:anisotropy_parameter}, we show $\gamma_{TR}$ as a function of distance from Sgr~A* for the E-stars, dusty sources, and all combined sources. 
The apparent continuous distribution of the data points for all sources illustrated in Fig. \ref{fig:pm_membership} and Fig. \ref{fig:anisotropy_parameter} is expected due to the shared parameter. However, if we separate the distribution of the anisotropy parameter into four bins with a corresponding size of 0.5 each, we find indications for a slight overdensity close to $\pm 1$. In particular, we find:
\begin{itemize}
    \item[1:] 14 sources between 1.0 and 0.5
    \item[2:]  6 sources between 0.5 and 0.0
    \item[3:]  9 sources between 0.0 and -0.5
    \item[4:] 19 sources between -0.5 and -1.0   
\end{itemize}
We normalize the distribution to the total number of sources and estimate that about 30$\%$ of the sources are located in the 0.5 to 1.0 bin, while almost 40$\%$ can be found in the -0.5 to -1.0 bin, suggesting an overdensity of sources in bin 1 and bin 4. In particular, this overdensity is reflected in the lower right plot of Fig. \ref{fig:anisotropy_parameter}. There, we show the normalized number of sources as a function of the anisotropy parameter $\gamma_{TR}$. In the same plot, we incorporate the theoretical probability distribution (PDF) with a constant anisotropy where adapt the corresponding normalized function 
\begin{equation}
    \rm PDF(\gamma_{TR})d\gamma_{TR}\,=\,\frac{n!(\sqrt{1+\gamma_{TR}})^{2n-1}}{\pi(2n-1)!!\sqrt{1-\gamma_{TR}}}d\gamma_{TR}
    \label{eq:pdf}
\end{equation}
from \cite{Genzel2000}. In the above equation, n refers to a power-law distribution index with $\beta = 1/2-n$. We can now assume numerical values representing a constant anisotropy that classifies uniform clusters. For example, the magenta PDF in the lower right plot of Fig. \ref{fig:anisotropy_parameter} is calculated with Eq. \ref{eq:pdf} and a constant anisotropy of $\beta=-3/2$ resembling the results of \cite{Genzel2000}\footnote{Please refer to Fig. 9 in \cite{Genzel2000}.}. From the data points representing the sources in IRS 13 shown in Fig. \ref{fig:anisotropy_parameter}, it becomes directly obvious that the cluster is not uniform and shows anisotropy that peaks at $\pm 1$ in strong agreement with \cite{Genzel2000}. However, for the results displayed in Fig. \ref{fig:anisotropy_parameter}, we picked only one numerical value for $\beta$ to maintain clarity. To inspect the expected distribution of an anisotropic cluster, we use Monte Carlo simulations shown in Fig. \ref{fig:monte_carlo}.
This figure strengthens our results which show a peak of $\gamma_{TR}$ at $\pm 1$ as well. In Figure \ref{fig:monte_carlo}, we simulate 10000 stars and find an overdensity at $\gamma_{TR}\pm 1$ as for IRS13. Therefore, the investigated cluster that harbors the dusty sources is not uniform.
In addition, we estimate a velocity dispersion from the data listed in Table \ref{tab:prop} and illustrated in Fig. \ref{fig:pm_membership} of {$128.86 \pm 0.14$ km/s} for the cluster. From this we can directly derive the mass that is needed to bind the stars to the cluster. With $M_{\rm IRS13}\,=\,\langle v^2 \rangle \cdot R / G$ where $R=0.01$pc donates the approximate size of the cluster and G the gravitational constant, we get $\sim (3.9\,\pm\,0.1)\times 10^4 M_{\odot}$ in agreement with independently calculated literature values \citep[see][]{Schoedel2005, Paumard2006, Tsuboi2017, Tsuboi2020b}. This enclosed mass estimate can be used to calculate the Hill radius $r_{\rm Hill}$ to inspect the gravitational bounds of the IRS 13 cluster. We use
\begin{equation}
    r_{\rm Hill}\,=\,D(M_{\rm IRS13}/3M_{SgrA*})^{1/3},
    \label{eq:hill}
\end{equation}
where D donates the distance to Sgr~A* and M$_{SgrA*}$ the related mass of the SMBH. We use D = 0.15 pc and M$_{SgrA*}\,=\,4\times 10^6\,M_{\odot}$ \citep{Peissker2022, eht2022} and get $r_{\rm Hill}\,=\,0.022\,pc\,=\,22\,mpc$. Since we investigate the complete IRS 13 region, including the E-stars and the dusty objects (Fig. \ref{fig:finding_chart}), $r_{\rm Hill}$ is in remarkable agreement with the \normalfont{measured} diameter of the cluster core region ($\approx 45 mpc$\normalfont{, see Sec. \ref{sec:discuss}}).

\subsection{Photometric analysis}

\begin{figure*}[htbp!]
	\centering
	\includegraphics[width=1.\textwidth]{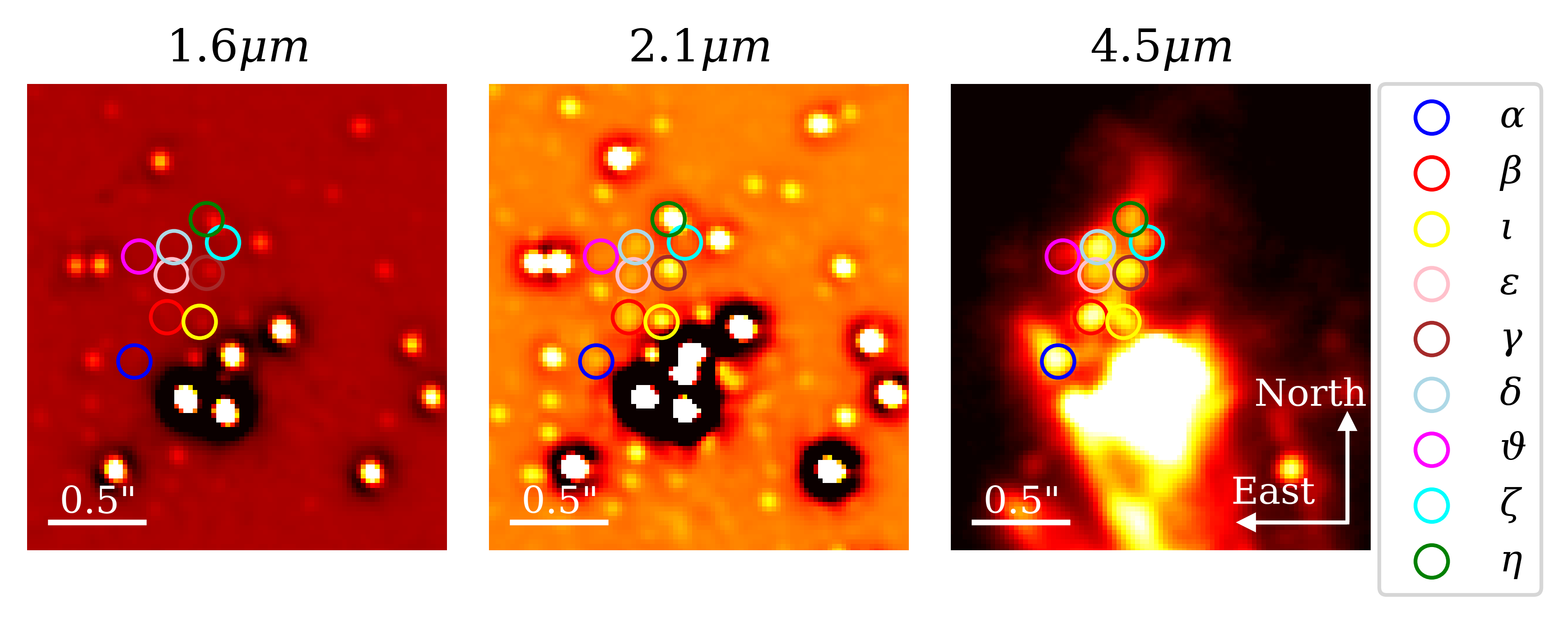}
	\caption{Multi-wavelength view toward IRS 13 observed with NACO. To minimize the influence of dominating PSF wings, we apply an image {sharpener} to the H- and K-band observed at $1.6\,\mu m$ and $2.1\,\mu m$, respectively. The prominent dust features are revealed in the MIR (here: $4.5\mu m$) and are not treated with any filter. Every image was normalized to its peak emission flux. The contrast was adjusted to visualize the presence of the dusty sources in the related band. However, the lower cutoff resulted in regions around bright stars with apparent missing flux. We note that sources such as E3 close to E4 (see Fig. \ref{fig:finding_chart}) have an H-band emission but are suppressed due to the contrast settings.}
\label{fig:finding_chart_mband}
\end{figure*}

\begin{figure*}[htbp!]
	\centering
	\includegraphics[width=1.\textwidth]{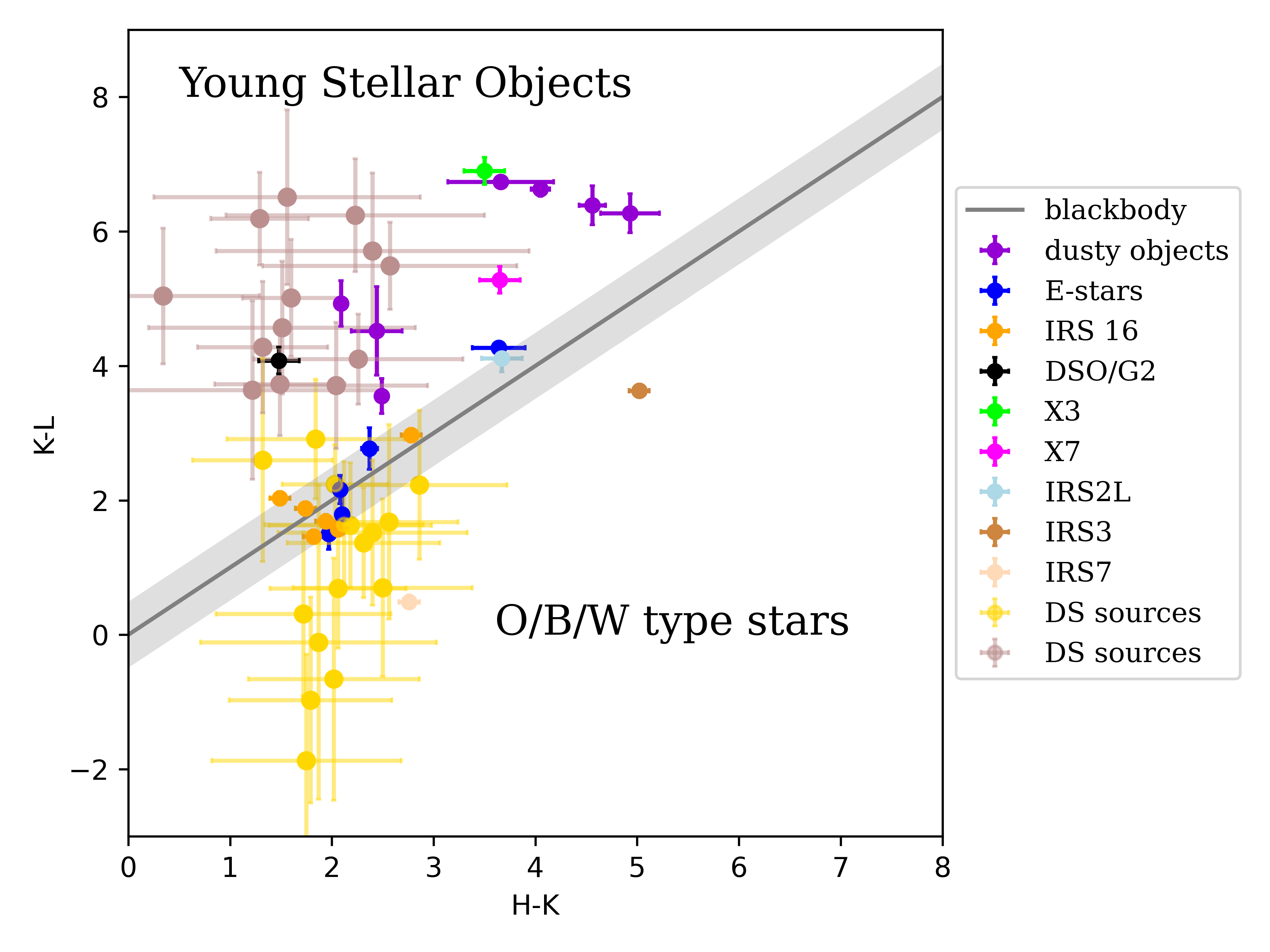}
	\caption{Color-color diagram for some prominent stellar objects in the {\it inner parsec}. The linear gray 
         line represents a one-component black body with increasing temperature and separates known evolved and embedded early-type stars from candidate YSOs \citep[see also][]{Ishii1998, Eckart2004a}. \normalfont{Based on this classification, the photometric data implies two generations of dusty sources, namely YSOs (brown) and main-sequence stars (yellow). Please note that the photometric uncertainty of two DS sources, DS23 and DS30, forbids a strong statement about their exact nature. Overall, the} uncertainties represent the standard deviation of the estimated colors listed in Table \ref{tab:mag_e_sources}. Here, IRS 16 refers to the stars indicated in Fig. \ref{fig:finding_chart}.}
\label{fig:color_color_diagram}
\end{figure*}

We use a multiwavelength approach to investigate the nature of the brightest M-band dust objects in the IRS 13 cluster (Fig. \ref{fig:finding_chart_mband}). Starting from Fig. \ref{fig:finding_chart_mband} in the M-band, we analyze the emission of the dust objects in the H-, K-, and L-band (see also Fig. \ref{fig:finding_chart}). 
Since \cite{Viehmann2006} analyzed an extensive amount of stellar sources in the environment of Sgr~A* in various bands \citep[see also][]{Bhat2022}, we use the close-by star IRS 2L as a reference source (Table \ref{tab:mag_fux_reference_values}). Because Viehmann et al. used part of the here investigated data set, the choice of the reference star ensures a consistent photometric approach. 
\begin{table}[hbt!]
    \centering
    \begin{tabular}{|cccc|}
         \hline 
         \hline
           Filter & Magnitude [mag] & Flux$_{\lambda}$ [Jy] & A$\rm _X$ \\
         \hline
         H-band  &  14.26   & 0.13 & 4.37\\
         K-band  &  10.60   & 0.48 & 2.80\\         
         L-band  &   6.4    & 2.98 & 1.45\\         
         M-band  &   5.5    & 3.98 & 0.58\\     
       \hline
    \end{tabular}
    \caption{Derredened reference values for IRS 2L used in this work based on the analysis of \cite{Viehmann2006}. \normalfont{The reddening vector A$\rm _X$ for the corresponding band is adapted from \cite{Viehmann2007} assuming an optical extinction of A$\rm _V\,=\,25$mag \citep{Scoville2003} using the extinction law from \cite{Rieke1985}. We refer to \cite{Fritz2011} for a detailed discussion of the optical extinction A$\rm _V$.}}
    \label{tab:mag_fux_reference_values}
\end{table}
Due to the dominant {contribution of} Wolf-Rayet and O stars E1, E2, E3, and E4 \citep[][]{Maillard2004} {in all the bands}, we will use a high-pass filter to minimize the PSF wings. {This process used is already described in detail in Sec. \ref{sec:high-pass} and \cite{Peissker2022}. The photometric robustness of high-pass filters compared to the raw data is further investigated in \cite{Ott1999}. To inspect the validity of the proposed photometric robustness discussed in Ott et al., we compare the estimated L-band magnitudes of DS1 (Fig. \ref{fig:finding_chart}) in the raw data with the results of the high-pass filtering. As listed in Table \ref{tab:filter_comparison}, we do not find a significant difference between the filter and non-filtered data in agreement with the analysis of \cite{Ott1999}. We would like to emphasize that the analysis of the investigated dusty objects focuses on the colors defined as the difference between two magnitudes. The colors are not affected by systematic differences potentially induced by the applied high-pass filter because variations would be canceled out.}
\begin{table}[hbt!]
    \centering
    \begin{tabular}{|ccccc|}
         \hline 
         \hline
           Year & No filter & Filter & Mean & $\Delta$ Mean \\
         \hline
         2002  & 10.75 & 10.95 & 10.85 & $\pm$0.10 \\
         2003  & 10.70 & 10.85 & 10.77 & $\pm$0.07 \\         
         2004  & 10.46 & 10.75 & 10.60 & $\pm$0.15 \\         
         2005  & 10.50 & 10.68 & 10.59 & $\pm$0.09 \\     
         2006  & 10.40 & 10.64 & 10.52 & $\pm$0.11 \\  
         2007  & 10.27 & 10.36 & 10.31 & $\pm$0.04  \\  
         2008  & 10.05 & 10.32 & 10.18 & $\pm$0.13   \\  
         2009  & 10.13 & 10.34 & 10.23 & $\pm$0.10   \\  
         2010  & 10.12 & 10.25 & 10.18 & $\pm$0.06   \\  
         2011  & 9.88  & 10.09 & 9.98 & $\pm$0.10   \\  
         2012  & 9.98  & 10.03 & 10.00 & $\pm$0.02  \\  
         2013  & 10.09 & 10.21 & 10.15 & $\pm$0.06    \\  
         2014  &  -   & - & - & -   \\  
         2015  &  -   & - & - & -   \\  
         2016  & 9.42 & 9.58 & 9.50 & $\pm$0.08    \\  
         2017  & 9.88 & 10.19 & 10.03 & $\pm$0.15    \\ 
         2018  & 10.31 & 10.27 & 10.29 & $\pm$0.02   \\ 
         Average &  10.19   & 10.36 & 10.27 & 0.08   \\  
         Median  &  10.13   & 10.32 & 10.27 & 0.08   \\
       \hline
    \end{tabular}
    \caption{Photometric comparison of the applied analysis tools for DS1 in the L-band. For the mean, we average the magnitude derived from the high-pass filtered and non-filtered data. The magnitude differences are marginal and smaller as the usual standard deviation as presented in Table \ref{tab:ds_sources_ident}.}
    \label{tab:filter_comparison}
\end{table}
{However, using Eq. \ref{eq:mag_count} with the magnitudes of reference star IRS2L (Table \ref{tab:mag_fux_reference_values}), we estimate the magnitudes for the dusty sources and the main-sequence stars E1-E7 (Fig. \ref{fig:finding_chart_mband}). Consult Table \ref{tab:mag_e_sources} for the related values, including the standard deviation.}.
From the H-K and K-L colors of the investigated sources, we do find a substantial difference between the two groups (dusty sources - E stars) of cluster members (Fig. \ref{fig:color_color_diagram}). In addition to the sources investigated here, we also include \normalfont{magnitudes from the related publication of} various other objects, such as DSO/G2 \citep[][]{peissker2021c}, X3 \citep{peissker2023b}, and X7 \citep[][]{peissker2021}. {A complete list of all used sources analyzed for Fig. \ref{fig:color_color_diagram} is listed in Table \ref{tab:mag_e_sources}.} 
\begin{table*}
\centering
\setlength{\tabcolsep}{0.999pt}
\begin{tabular}{|c|cccccccccc|}
\hline
\hline
ID & \multicolumn{2}{c}{H-band} & \multicolumn{2}{c}{K-band} & \multicolumn{2}{c}{L-band} & \multicolumn{2}{c}{M-band} & K-L & H-K \\
\hline
   & [mag] &  [mJy] & [mag] &  [mJy] & [mag] &  [Jy] & [mag] &  [Jy] & & \\
\hline
$\alpha$    & 20.68$\pm$0.85 & 0.35$^{+0.41}_{-0.19}$ & 15.75$\pm$0.14 & 4.18  $^{+0.57}_{-0.50}$  & 9.48$\pm$0.30 & 0.17 $^{+0.05}_{-0.04}$ & 7.56$\pm$0.17 &  0.59 $^{+0.10}_{-0.08}$ & 6.27$\pm$0.29 & 4.93$\pm$0.29 \\ 
$\beta$     & 19.74$\pm$0.04 & 0.83$^{+0.03}_{-0.03}$ & 15.18$\pm$0.21 & 7.06  $^{+1.50}_{-1.24}$  & 8.79$\pm$0.78 & 0.32 $^{+0.34}_{-0.16}$ & 7.57$\pm$0.14 &  0.59 $^{+0.08}_{-0.07}$ & 6.39$\pm$0.29 & 4.56$\pm$0.13 \\
$\gamma$    & 16.28$\pm$0.20 &20.22$^{+4.09}_{-3.40}$ & 14.19$\pm$0.30 &17.58  $^{+5.59}_{-4.24}$  & 9.26$\pm$0.98 & 0.21 $^{+0.31}_{-0.12}$ & 7.91$\pm$0.29 &  0.43 $^{+0.13}_{-0.10}$ & 4.93$\pm$0.34 & 2.09$\pm$0.05 \\
$\delta$    & 19.56$\pm$0.48 & 0.98$^{+0.54}_{-0.35}$ & 15.90$\pm$0.28 & 3.64  $^{+1.07}_{-0.82}$  & 9.16$\pm$0.76 & 0.23 $^{+0.23}_{-0.11}$ & 7.70$\pm$0.17 &  0.52 $^{+0.08}_{-0.07}$ & 6.74$\pm$0.09 & 3.66$\pm$0.52\\ 
$\epsilon$  &    -             &     -                  & 16.27$\pm$0.22 & 2.58  $^{+0.58}_{-0.47}$  & 9.50$\pm$1.16 & 0.17 $^{+0.32}_{-0.11}$ & 8.34$\pm$0.52 &  0.29 $^{+0.17}_{-0.11}$ & 6.77$\pm$0.69 &  - \\
$\zeta$     & 20.06$\pm$0.46 & 0.62$^{+0.32}_{-0.21}$ & 16.01$\pm$0.62 & 3.29  $^{+2.53}_{-1.43}$  & 9.38$\pm$0.84 & 0.19 $^{+0.22}_{-0.10}$ & 8.12$\pm$0.29 &  0.35 $^{+0.10}_{-0.08}$ & 6.63$\pm$0.11 & 4.05$\pm$0.09 \\ 
$\eta$      & 15.29$\pm$0.16 &50.34$^{+7.99}_{-6.89}$ & 12.80$\pm$0.19 & 63.27 $^{+12.10}_{-10.15}$& 9.25$\pm$0.71 & 0.21 $^{+0.19}_{-0.10}$ & 8.11$\pm$0.23 &  0.35 $^{+0.08}_{-0.06}$ & 3.55$\pm$0.26 & 2.49$\pm$0.02 \\
$\vartheta$ & 21.04$\pm$0.25 & 0.25$^{+0.06}_{-0.05}$ & -                & -                         &10.29$\pm$1.22 & 0.08 $^{+0.17}_{-0.05}$ & 9.14$\pm$0.39 &  0.13 $^{+0.06}_{-0.04}$ & - &  - \\ 
$\iota$     & 16.67$\pm$0.21 &14.12$^{+3.01}_{-2.48}$ & 14.23$\pm$0.29 & 16.95 $^{+5.19}_{-3.97}$  & 9.71$\pm$1.62 & 0.14 $^{+0.48}_{-0.10}$ & 8.12$\pm$0.45 &  0.35 $^{+0.18}_{-0.12}$ & 4.52$\pm$0.66 & 2.44$\pm$0.25\\
\hline
   & [mag] &  [Jy] & [mag] &  [Jy] & [mag] &  [Jy] & [mag] &  [Jy] & & \\
\hline
E1   & 11.16$\pm$0.06 & 2.25 $^{+0.12}_{-0.12}$ & 9.19$\pm$0.17 & 1.75 $^{+0.29}_{-0.25}$ & 7.69$\pm$0.65 & 0.90 $^{+0.74}_{-0.40}$ & 6.96$\pm$0.33 & 1.03 $^{+0.36}_{-0.27}$ & 1.50$\pm$0.23 & 1.97$\pm$0.03 \\ 
E2   & 11.31$\pm$0.08 & 1.96 $^{+0.15}_{-0.13}$ & 9.23$\pm$0.21 & 1.69 $^{+0.36}_{-0.29}$ & 7.07$\pm$0.64 & 1.60 $^{+1.29}_{-0.71}$ & 6.17$\pm$0.27 & 2.14 $^{+0.60}_{-0.47}$ & 2.16$\pm$0.21 & 2.08$\pm$0.02 \\ 
E3   & 14.43$\pm$0.59 & 0.11 $^{+0.08}_{-0.04}$ &10.79$\pm$0.67 & 2.66 $^{+0.34}_{-0.18}$ & 6.52$\pm$0.51 & 2.66 $^{+1.59}_{-1.00}$ & 5.17$\pm$0.01 & 5.39 $^{+0.04}_{-0.04}$ & 4.27$\pm$0.08 & 3.64$\pm$0.26 \\ 
E4   & 12.62$\pm$0.15 & 0.58 $^{+0.08}_{-0.07}$ &10.25$\pm$0.33 & 1.45 $^{+0.23}_{-0.17}$ & 7.18$\pm$0.65 & 1.45 $^{+1.19}_{-0.65}$ & 6.15$\pm$0.15 & 2.18 $^{+0.32}_{-0.28}$ & 2.77$\pm$0.31 & 2.37$\pm$0.08 \\ 
E5.0 &   -              & -                       &14.24$\pm$0.41 & 0.65 $^{+0.01}_{-0.01}$ & 8.04$\pm$0.99 & 0.65 $^{+0.97}_{-0.39}$ & 6.68$\pm$0.31 & 1.34 $^{+0.44}_{-0.33}$ & 6.20$\pm$0.70 & - \\ 
E5.1 &  -               & -                       &14.53$\pm$0.09 & 0.61 $^{+0.01}_{-0.01}$ & 8.11$\pm$0.87 & 0.61 $^{+0.75}_{-0.34}$ & 6.95$\pm$0.29 & 1.04 $^{+0.32}_{-0.24}$ & 6.42$\pm$0.48 & - \\ 
E7   & 13.21$\pm$0.11 & 0.54 $^{+0.02}_{-0.02}$ &10.61$\pm$0.11 & 0.32 $^{+0.05}_{-0.04}$ & 8.82$\pm$0.71 & 0.32 $^{+0.29}_{-0.15}$ & 8.92$\pm$0.85 & 0.17 $^{+0.20}_{-0.09}$ & 1.79$\pm$0.29 & 2.60$\pm$0.05 \\ 
\hline
   & [mag] &  [Jy] & [mag] &  [Jy] & [mag] &  [Jy] & [mag] &  [Jy] & & \\
\hline
IRS3    & 14.68 $\pm$ 0.12 & 0.08  $^{+0.02}_{-0.01}$   & 9.66 $\pm$ 0.62 & 1.14 $^{+0.88}_{-0.49}$ & 6.03$\pm$0.03 & 4.19 $^{+0.11}_{-0.12}$ & 2.99 $\pm$ 0.14 & 40.16 $^{+5.53}_{-4.86}$ & 3.63$\pm$0.32 & 5.02$\pm$0.35 \\
IRS7    & 9.26 $\pm$ 0.04 & 13.12 $^{+0.36}_{-0.60}$  & 6.50 $\pm$ 0.10 & 20.95 $^{+2.02}_{-1.85}$ & 6.01$\pm$0.05 & 4.26 $^{+0.20}_{-0.19}$ & 3.12 $\pm$ 0.49 & 35.63 $^{+20.33}_{-11.94}$  & 0.49$\pm$0.08 & 2.76$\pm$0.07\\
\hline
\end{tabular}%
\caption{Dereddened magnitude and flux estimates of the E-stars and the brightest dusty sources of the IRS 13 cluster (see Fig. \ref{fig:finding_chart}). {For comparison, we also list the measured values of IRS3 and IRS7. The indicated uncertainties of the magnitudes and fluxes for all objects are based on the standard deviation except for IRS3 and IRS7. The uncertainties for the two bright stars are adapted from published studies, namely \cite{Blum1996}, \cite{Viehmann2006}, and \cite{Pott2008}. The used} reference magnitudes for IRS 2L are listed in Table \ref{tab:mag_fux_reference_values}. For $\epsilon$, E5.0, and E5.1, we do not find H-band emission above the detection limit, which might be due to confusion. {The uncertainties of the magnitudes and fluxes reflect the standard deviation.} As pointed out by \cite{Fritz2010} {and implied by the east-west elongation indicated in \cite{Eckart2004a}}, E3 may be a collection of several less luminous stars. Controversially, \cite{Tsuboi2017} associates E3 with the location of a possible IMBH.}
\label{tab:mag_e_sources}
\end{table*}
The findings {presented in Fig. \ref{fig:color_color_diagram}} are in agreement with the studies and classifications presented in \cite{Peissker2020b} and will be discussed in Sec. \ref{sec:discuss}.
Since we derived the magnitudes of the IRS 13 sources, we will estimate the {related flux density and the corresponding uncertainties with Eq. \ref{eq:flux}}.
The flux density is useful to estimate the SED of the individual sources, which will be {presented} in the next section. {Compared to the literature, we maximize the} spectral coverage and include the radio data observed\footnote{PI: Masato Tsuboi} with ALMA and previously analyzed, e.g., in \cite{Tsuboi2017}. 
\begin{figure}[htbp!]
	\centering
	\includegraphics[width=.5\textwidth]{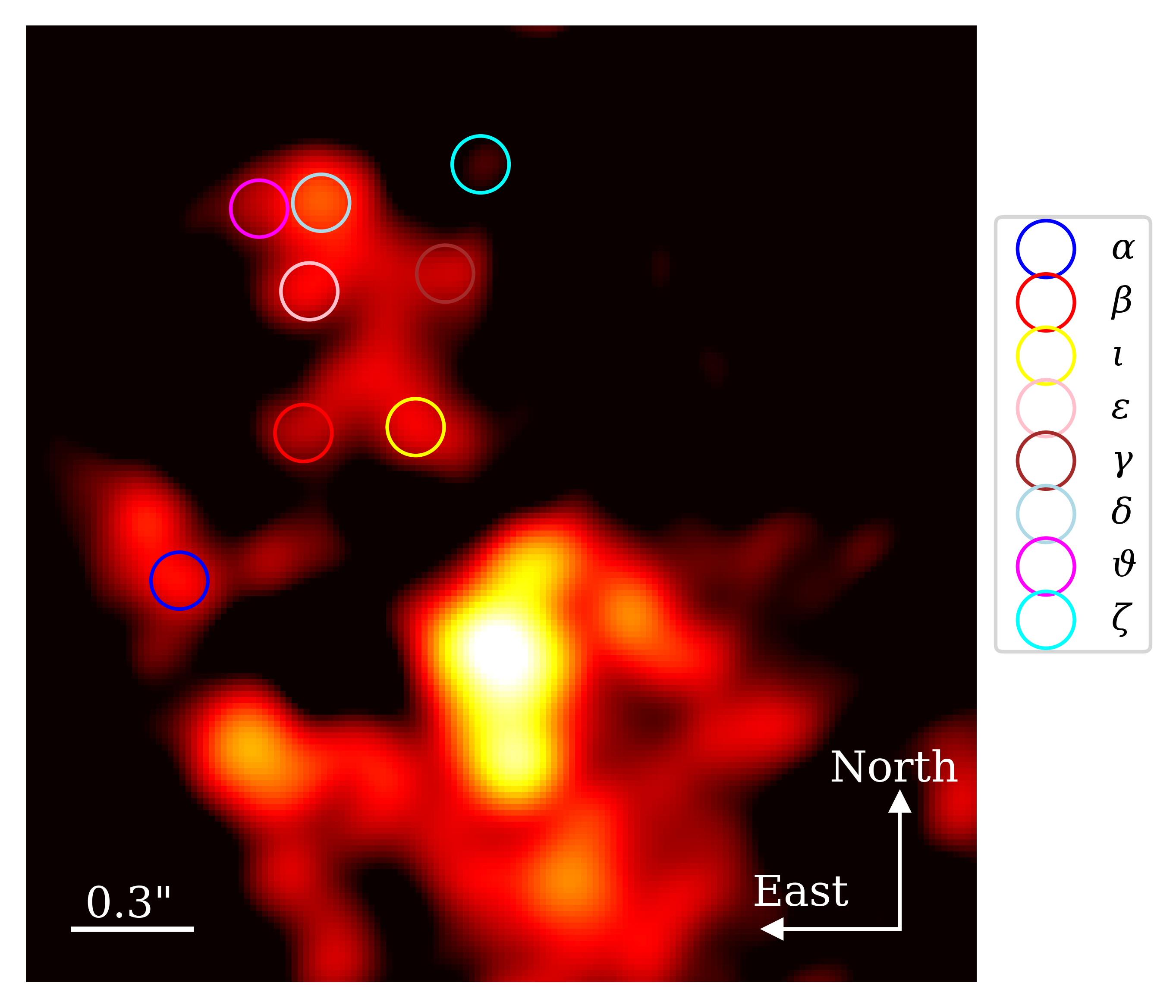}
	\caption{Observation of the IRS 13 cluster with ALMA. The data was observed at 343 GHz and corresponds to CO (v=0). The location of the bright dusty sources that are observed in the infrared is indicated by colored circles. Like the E-stars, the projected location of the dusty sources implies clustering. The related flux of the individual sources in listed in Table \ref{tab:flux_dusty_sources_radio}.}
\label{fig:finding_chart_radio}
\end{figure}
{Due to the science-ready character of the calibrated data, we list the corresponding flux values of the dusty sources in Table \ref{tab:flux_dusty_sources_radio}.}
\begin{table}[htbp!]
\centering
\begin{tabular}{|cc|}
\hline
\hline
ID & CO (v=0), 343 GHz \\ 
\hline
   & [mJy]  \\
\hline
$\alpha$    & 1.43$\pm$0.5 \\
$\beta$     & 1.05$\pm$0.5 \\
$\gamma$    & 1.09$\pm$0.5\\
$\delta$    & 1.91$\pm$0.5\\
$\epsilon$  & 1.38$\pm$0.5\\
$\zeta$     & 0.61$\pm$0.5\\
$\eta$      & (0.19$\pm$0.5)$^{*}$ \\
$\vartheta$ & (1.15$\pm$0.5)$^{**}$\\
$\iota$     & 1.29$\pm$0.5 \\
IRS3        & 129.1$\pm$55.1\\
IRS7        & 34.4$\pm$0.4\\
\hline
\end{tabular}%
\caption{Flux density values for the dusty objects derived from CO ALMA observations. Please see Fig. \ref{fig:finding_chart_radio} for the related source identification. For $\eta$, we only estimate an upper limit, whereas $\vartheta$ seems to be confused with $\delta$. We use IRS 13E3 as a reference source with a corresponding peak flux of 10.5$\pm$0.5 mJy {and adapt the uncertainty as proposed in \cite{Tsuboi2017b}. These submm/radio flux values in combination with the IR values listed in Table \ref{tab:mag_e_sources} are used for the input spectrum of HYPERION.}}
\label{tab:flux_dusty_sources_radio}
\end{table}

\subsection{Spectral Energy Distribution}
\label{sec:sed}
\begin{figure*}[htbp!]
	\centering
	\includegraphics[width=1.\textwidth]{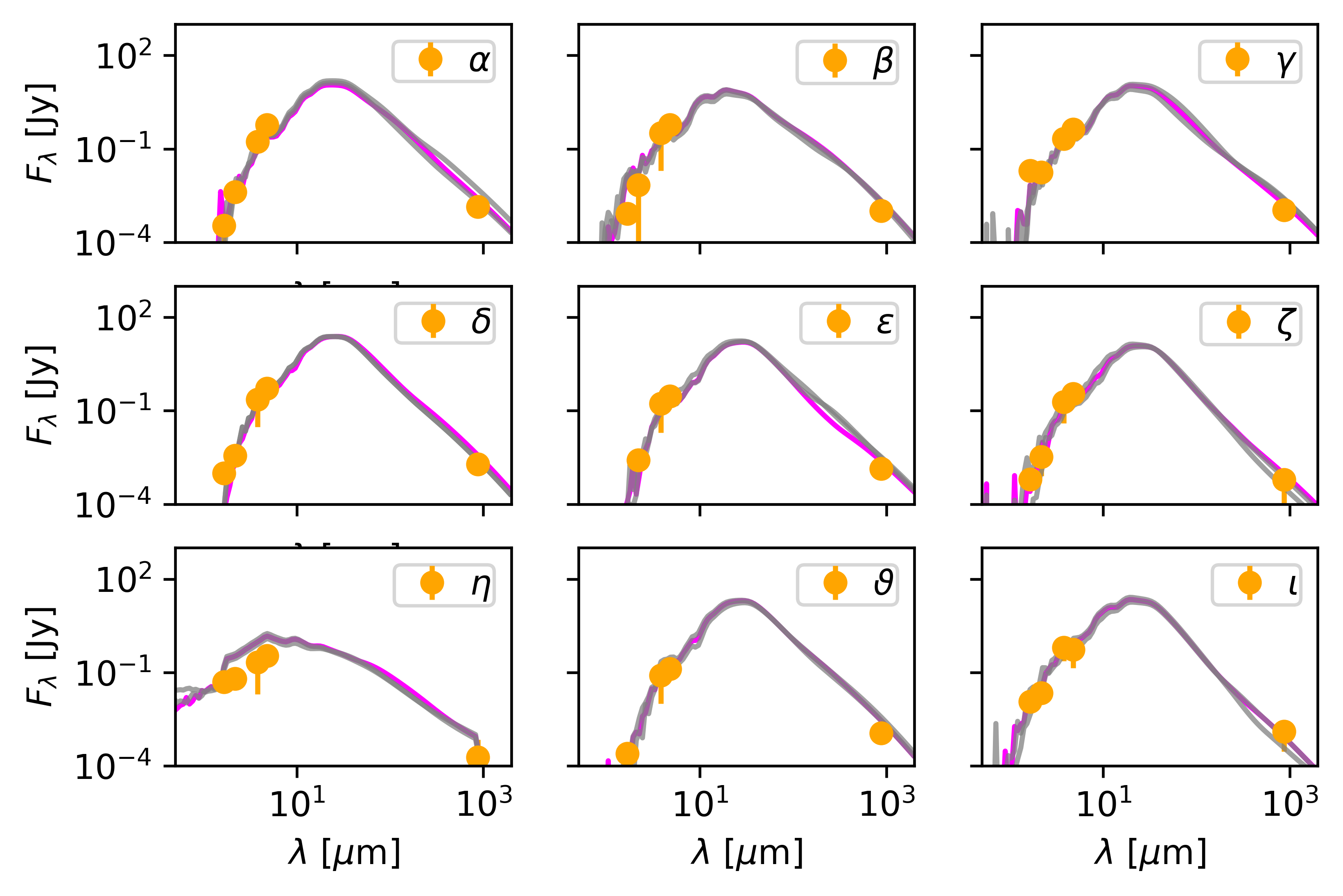}
	\caption{{Best-fit SED} of the brightest dusty sources in the IRS 13 cluster {representing the YSO parameters given in Table \ref{tab:sed_table}. The input spectrum is constructed using the flux density values listed in Table \ref{tab:mag_e_sources} and Table \ref{tab:flux_dusty_sources_radio}. The associated numerical flux density values, including their related uncertainty, are implemented in the SED plots to emphasize the validity of the resulting best-fit parameters.} For {almost} every source, the solution in {magenta} represents the best fit and is associated with an inclination of about 90$^{\circ}$, while the gray SED is related to the maximum and minimum uncertainty indicated in Table \ref{tab:sed_table}. The inclination for the SED, including the related uncertainties of the source $\eta$, equals 70$^{\circ}$.}
\label{fig:sed}
\end{figure*}

The spectral analysis of the O- and W-type stars of the IRS 13 cluster is well covered in the literature \citep[][]{Maillard2004}. Despite bright emission in various bands, the dusty sources analyzed in the literature lack a detailed spectral analysis. We {apply} the 3D radiative transfer model {implemented} in the HYPERION code and incorporate the results listed in {Table \ref{tab:mag_e_sources} and Table \ref{tab:flux_dusty_sources_radio}. These flux density values are used as input parameters from which HYPERION estimates the best-fit SED. {The spectrum is renormalized and ensures a high synergy between the observations and the simulations.}
The uncertainties of the input flux density values (Table \ref{tab:mag_e_sources}) estimated from the standard deviation do account for a variable background, close-by sources, and the stellar density as well as the embedded structure (eminent in the L- and M-band) of the cluster.}\newline
To motivate the usage of HYPERION, which models the emission of YSOs, we refer to the color-color diagram shown in Fig.~\ref{fig:color_color_diagram}, which justifies our approach. Incorporating the derived flux and uncertainty values, we find a best-fit solution for the spectral energy distribution of the dusty sources, as shown in Fig. \ref{fig:sed}.
We list related input parameters, such as, for example, {stellar} mass and luminosity, in Table \ref{tab:sed_table}.
\begin{table*}[]
\setlength{\tabcolsep}{0.99pt}
\centering                                  
\begin{tabular}{|c|ccccccc|}
\toprule
ID     & \multicolumn{1}{c|}{Mass [$M_{\odot}$]} & \multicolumn{1}{c|}{Luminosity [$10^3\times L_{\odot}$]}  & \multicolumn{1}{c|}{Infall rate [$10^{-6}\times \dot{M}_{\odot}$]} & \multicolumn{1}{c|}{Radius [$R_{\odot}$]}& \multicolumn{1}{c|}{Disk mass [$M_{\odot}$]} & \multicolumn{1}{c|}{Disk size [AU]} & Envelope size [AU] \\
\hline
$\alpha$    & 5.0  $\pm$ 1.0 & 9 $\pm$ 1 & 5 $\pm$ 1 & 2 $\pm$ 0.5   & 0.1 $\pm$ 0.01 & 0.06-200 & 0.09-500 \\
\hline
$\beta$     & 8.0  $\pm$ 1.0 & 10 $\pm$  & 0.3 $\pm$ 1 & 3 $\pm$ 0.5 & 0.01 $\pm$ 0.005 & 0.04-200 & 0.09-350 \\
\hline
$\gamma$    & 6.0  $\pm$ 1.0 & 9 $\pm$ 1 & 0.3 $\pm$ 0.1 & 4 $\pm$ 1 & 0.05 $\pm$ 0.01 & 0.04-200 & 0.09-500 \\
\hline
$\delta$    & 10.0 $\pm$ 2.0 & 11 $\pm$ 1 & 0.5 $\pm$ 0.1 & 3 $\pm$ 1 & 0.01 $\pm$ 0.001 & 0.04-200 & 0.09-700 \\
\hline
$\epsilon$  & 7.0  $\pm$ 2.0 & 7  $\pm$ 2 & 0.5 $\pm$ 0.1 & 3 $\pm$ 2 & 0.1 $\pm$ 0.01 & 0.04-200 & 0.09-700 \\
\hline
$\zeta$     & 4.0  $\pm$ 1.0 & 7  $\pm$ 1 & 0.3 $\pm$ 0.1 & 3 $\pm$ 1 & 0.06 $\pm$ 0.01 & 0.04-100 & 0.09-500  \\
\hline
$\eta$      & 0.5  $\pm$ 0.2 & 2  $\pm$ 0.5 & - & 0.8 $\pm$ 0.2 & 0.005 $\pm$ 0.002 & 0.13-50 & -  \\
\hline
$\vartheta$ & 7.0  $\pm$ 1.5 & 8  $\pm$ 1 & 0.5 $\pm$ 0.1 & 3 $\pm$ 1 & 0.01 $\pm$ 0.02 & 0.04-200 & 0.04-700  \\
\hline
$\iota$     & 10   $\pm$ 2.0 & 12 $\pm$ 2 & 0.05 $\pm$ 0.01 & 7 $\pm$ 2 & 0.5 $\pm$ 0.1 & 0.02-100 & 0.04-700  \\
\hline
\end{tabular}
\caption{Best fit parameters describing the flux density distribution of the dusty sources of IRS 13 indicating their stellar nature using the radiative transfer model HYPERION. We motivate the application of the radiative transfer model describing YSOs by the clear color-color classification shown in Fig. \ref{fig:color_color_diagram}. Considering the pronounced slope of the NIR/MIR SED and their related position in the color-color diagram, it is suggested that the nature of these sources allows the classification as candidates YSOs (class I) \citep[][]{Lada1987}. For a demonstration of HYPERION's feedback after incorporating the estimated flux density values for IRS 3, please refer to Section \ref{sec:validity_hyperion}.}
\label{tab:sed_table}
\end{table*}
In agreement with the top-heavy mass function derived by \cite{Paumard2006} or \cite{Lu2013}, we find several massive and high-mass YSOs in the IRS 13 cluster, in line with its exceptional high core density of $\rho_{\rm core}\geq 3\times 10^8 M_{\odot}\,{\rm pc^{-3}}$ \citep{Paumard2006}. Except for $\eta$, all investigated sources exhibit a stellar mass between 4.0 and 10.0 M$_{\odot}$. Taking into account the stellar properties of the dusty sources in combination with the accretion rate ({denoted as the} infall rate in Table \ref{tab:sed_table}), we classify these objects as massive Herbig Ae/Be stars. Due to their nature, these dusty sources have an age of $10^4-10^5$yr and show a strong photometric correlation (Fig. \ref{fig:color_color_diagram}) with the recent discovery of the HMYSO X3 \citep{peissker2023b}. Regarding $\eta$, more data is needed to classify the low-mass source. However, the shape of the related SED implies that $\eta$ could be associated with a {low-mass} T Tauri star \citep{Beckwith1990, Kenyon1995}.

\section{Discussion} 
\label{sec:discuss}
In this Section, we will discuss the results presented above. We will introduce a new substructure of the IRS 13 cluster and motivate detailed upcoming observations in the mid-infrared. Taking into account the results presented, we will further suggest expanding the existing view towards the dimension of the IRS 13 cluster. It is suggested that the cluster shows an elongated tail that is caused by the gravitational interaction of Sgr~A* with IRS 13. 

\subsection{Stellar content of the system}

\begin{figure}[htbp!]
	\centering
	\includegraphics[width=.5\textwidth]{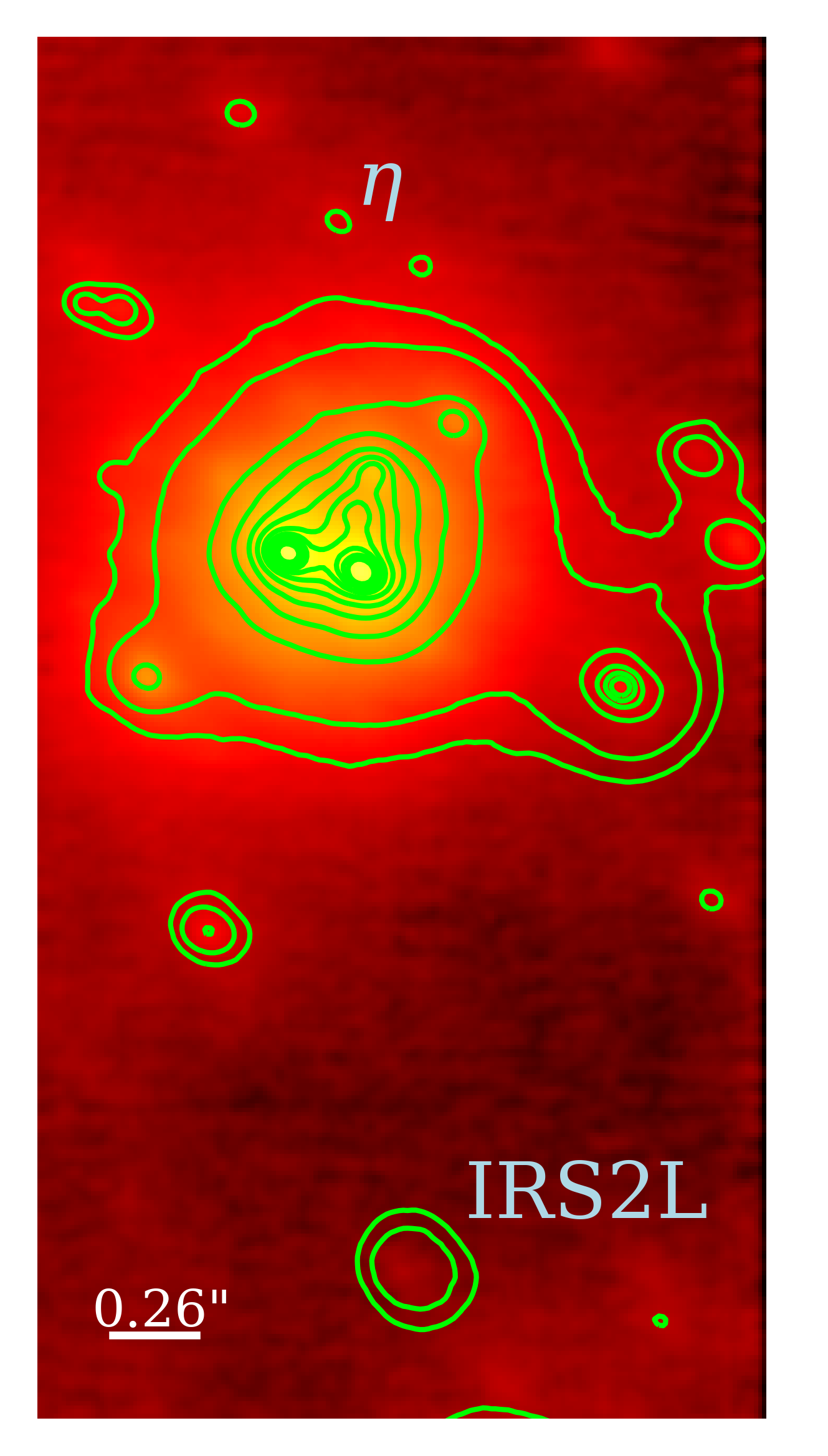}
	\caption{J-band observation of IRS 13 with NACO in 2013. We overlaid lime-colored contour lines adopted from the K-band observations of the same epoch and instrument. The contour levels represent $0.75\%$,$1\%$, $2\%$, $3\%$, $4\%$, $5\%$, $6\%$, $7\%$, $8\%$, $9\%$, and $10\%$ of the normalized K-band NACO data. The faint emission of IRS2L is labeled at the corresponding K- and L-band position whereas the bright emission is associated with IRS 13. We further indicate the position of the low-mass class I YSO $\eta$. Here, North is up, East is to the left.}
\label{fig:irs13_jband}
\end{figure}

NACO L-band observations of IRS 13 revealed 33 unknown objects in addition to previously investigated dust and stellar sources \citep[Table \ref{tab:ID}; see][]{Maillard2004, Eckart2013}. The brightest sources, {donates with greek letters,} can be observed in various bands ranging from the infrared to the radio/submm domain. Since the majority of the studied MIR dust sources exhibit K- and H-band NIR counterparts {(see Appendix \ref{sec:kband-counter-app} and Appendix \ref{sec:hband-counter-app})}, a stellar nature is inevitable. Compared to the main-sequence E-stars (O/WR-type), the K-L and H-K colors of the dusty sources are represented by two to three times higher numerical values. These high infrared H-K and K-L colors {suggest}, together with the survey of \cite{Ishii1998}, a YSO classification for the bright dusty objects of IRS 13. Further studies of YSOs were carried out by \citet{Lada1992}, who used J-H and H-K colors for their classification. Although the geometrical composition of the circumstellar components influences the NIR emission of YSOs, our derived H-K colors are in agreement with studies of intermediate and high-mass YSOs \citep[see also][]{Berrilli1992}. A further indicator that underlines the classification of the dusty sources as YSOs is illustrated in Fig. \ref{fig:sed}. The flux density values, covering a spectral range between the IR and the submm/radio, are fitted with a model representing the typical emission of Class I YSOs. Why the best-fit models presented in Fig. \ref{fig:sed} do not necessarily exclude other interpretations of the flux density distribution of the dusty sources, it is still a strong footprint of YSOs. Observations with the JWST and MIRI will potentially reveal typical emission lines that are associated with YSOs (see Sec. \ref{sec:jwst}).  
We note that there is increased confusion and noise level when investigating J-band NACO observations, which requires a detailed data processing method such as the Lucy-Richardson deconvolution algorithm \citep{Lucy1974}. However, these analysis steps exceed the scope of this work and will be part of a future publication. It should be noted that the classification of the brightest L-band sources in our sample exhibits a flux density distribution that agrees very well with class I YSOs except for $\eta$. The SED of this source shows similarities to a low-mass T Tauri star \citep{Chiang1997, Scoville2013}. Hence, we will focus on the rather ambiguous classification of $\eta$ using the J-H and H-K colors.
Despite the challenges with respect to the J-band analysis of the dusty sources of IRS 13, we identify $\eta$ without confusion about noise (Fig. \ref{fig:irs13_jband}), and infer a related magnitude of mag$_J\,=\,18.8\,\pm\,0.6$ that results in {J-H = 3.5} with H-K= 2.5 (Table \ref{tab:mag_e_sources}). 
Taking into account the color-color analysis of \cite{Lada1992}, {\cite{Ito2008}, and \cite{Ojha2009}}, $\eta$ appears to be a low-mass class I YSO that conflicts with the results of the radiative transfer model presented in Sec. \ref{sec:sed} due to the missing envelope. We can only speculate on possible explanations for the interplay of the missing envelope with the photometric footprint of a Class I YSO candidate. One option could be the intrinsic orientation of the system towards the observer. As implied by the SED results present in Fig. \ref{fig:sed}, the inclination does have a considerably large impact on the shape of the distribution. Another option could be an evolutionary transition to the class II stage or a partial detachment of the envelope, as is already suggested for the class I YSO L1489 IRS \citep{Brinch2007}. Despite the exact classification, the global interpretation as a low-mass YSO is still plausible.
Since DSO/G2 \citep[][]{peissker2021c} is also classified as a low-mass YSO \citep[see][]{Zajacek2017}, it is implied that both sources share a common nature. In summary, the general trend suggests that sources above the solid line illustrated in Fig. \ref{fig:color_color_diagram} can be classified as YSOs\normalfont{, which implies that IRS 13 harbors two generations of stellar objects.} 

\subsection{Validity of the radiative transfer model}
\label{sec:validity_hyperion}

{Here we want to take a critical look at the results regarding the classification of the dusty sources as YSOs. In \cite{Eckart2004a}, the authors proposed for the first time the idea of associating the dusty sources of IRS 13 using a color-color diagram such as the one displayed in Fig. \ref{fig:color_color_diagram}. In agreement with \cite{Eckart2004a} and the analysis of X3 presented in \cite{peissker2023b}, we found distinguishing colors compared to known main-sequence stars such as IRS3 \citep{Pott2008}. To expand the color-color analysis of the dusty sources, we decided to apply the radiative transfer code HYPERION \citep{Robitaille2011, Robitaille2017} to the flux density values listed in Table \ref{tab:mag_e_sources}. Although we find satisfying solutions to the flux density values in combination with the colors of the dusty sources (Fig. \ref{fig:sed}), the outcome of the radiative transfer model could be biased since we already assumed a YSO classification. Therefore, we want to investigate the validity of this approach by using the class I model used for the SED shown in Fig. \ref{fig:sed} with the flux density values of the embedded and cool carbon star IRS 3. This star is most probably in the helium-core burning phase with a related stellar temperature of 3000 K based on the interferometric observations carried out with MIDI/VLTI \citep{Leinert2003, Pott2008}. For the stellar analysis of IRS3 presented in Pott el al., the authors used the one-dimensional radiative transfer code DUSTY \citep{Ivezic1999}, which is developed for AGB stars exhibiting radiatively driven winds. The NIR and MIR flux density values for IRS3 listed in Table \ref{tab:mag_e_sources} are in reasonable agreement with the results presented in Figure 16 in \cite{Pott2008}. Deviations between our estimated flux values and the ones derived by Pott et al. are reflected by the uncertainties given in Table \ref{tab:mag_e_sources} and the error bars shown in Fig. \ref{fig:sed_irs3}.
Using the estimated stellar properties of \cite{Pott2008} results in the magenta-colored SED displayed in Fig. \ref{fig:sed_irs3}. Comparing the magenta-colored result with the measured flux density of IRS3 reveals that HYPERION is not suitable to reflect the SED of the star. Using the upper limit of Pott el al. for a hot C-rich star with an amorphous carbon grain-dominated circumstellar dust distribution produces the brown-colored SED shown in Fig. \ref{fig:sed_irs3}.
\begin{figure}[htbp!]
	\centering
	\includegraphics[width=.5\textwidth]{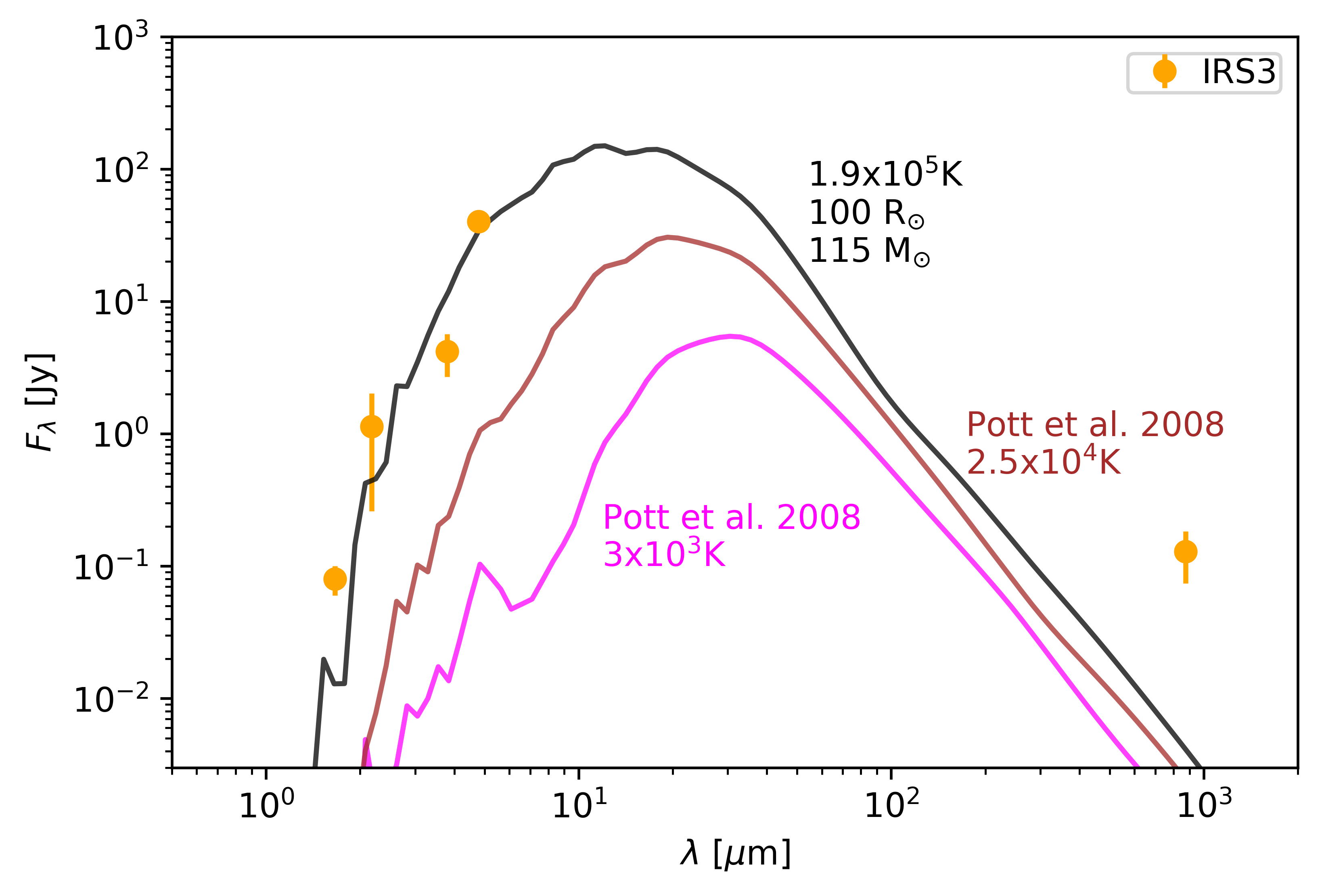}
	\caption{Comparison of different input parameters for the flux density values (orange colored dots) estimated for IRS3 (Table \ref{tab:mag_e_sources}). Please note that the magenta and brown-colored SED is based on the stellar parameters derived by the interferometric broadband analysis by \cite{Pott2008}. Assuming a hypothetical YSO association with IRS3 results in unsatisfying outcomes of the radiative transfer model fit. Only a speculative stellar temperature of $1.9\times 10^5 K$ seems to reproduce the NIR and MIR flux. The submm/radio emission is not fitted by any of the presented SED solutions. Please see the text for details.}
\label{fig:sed_irs3}
\end{figure}
In summary, the authors of Pott et al. conclude that IRS3 is a cool carbon AGB star. However, we speculatively implement a stellar temperature of $1.9\times 10^5 K$ in our radiative transfer model and find that this setting reflects the NIR and MIR emission. Neither of the presented SED solutions for IRS3 using HYPERION does fit the submm/radio flux. Taking into account the silicate absorption feature of IRS3 observed in the N-band \citep{Pott2008}, we can safely conclude that our speculative solution (black SED, Fig. \ref{fig:sed_irs3}) is not valid. It is further well-known that the existence of silicates requires a stellar temperature of a few 1000 K \citep[][]{Kozasa1999,Tsuchikawa2021} in line with the established results for IRS3 of the literature.
Therefore, we conclude that the SED solution displayed in Fig. \ref{fig:sed} is a strong indication for the classification of the dusty sources as YSOs. The radiative transfer model HYPERION is not suitable for embedded main sequence stars, which emphasizes the color-color results presented in Fig. \ref{fig:color_color_diagram}.}

\subsection{Formation scenarios for the IRS13 cluster}

\begin{figure}[htbp!]
	\centering
	\includegraphics[width=.5\textwidth]{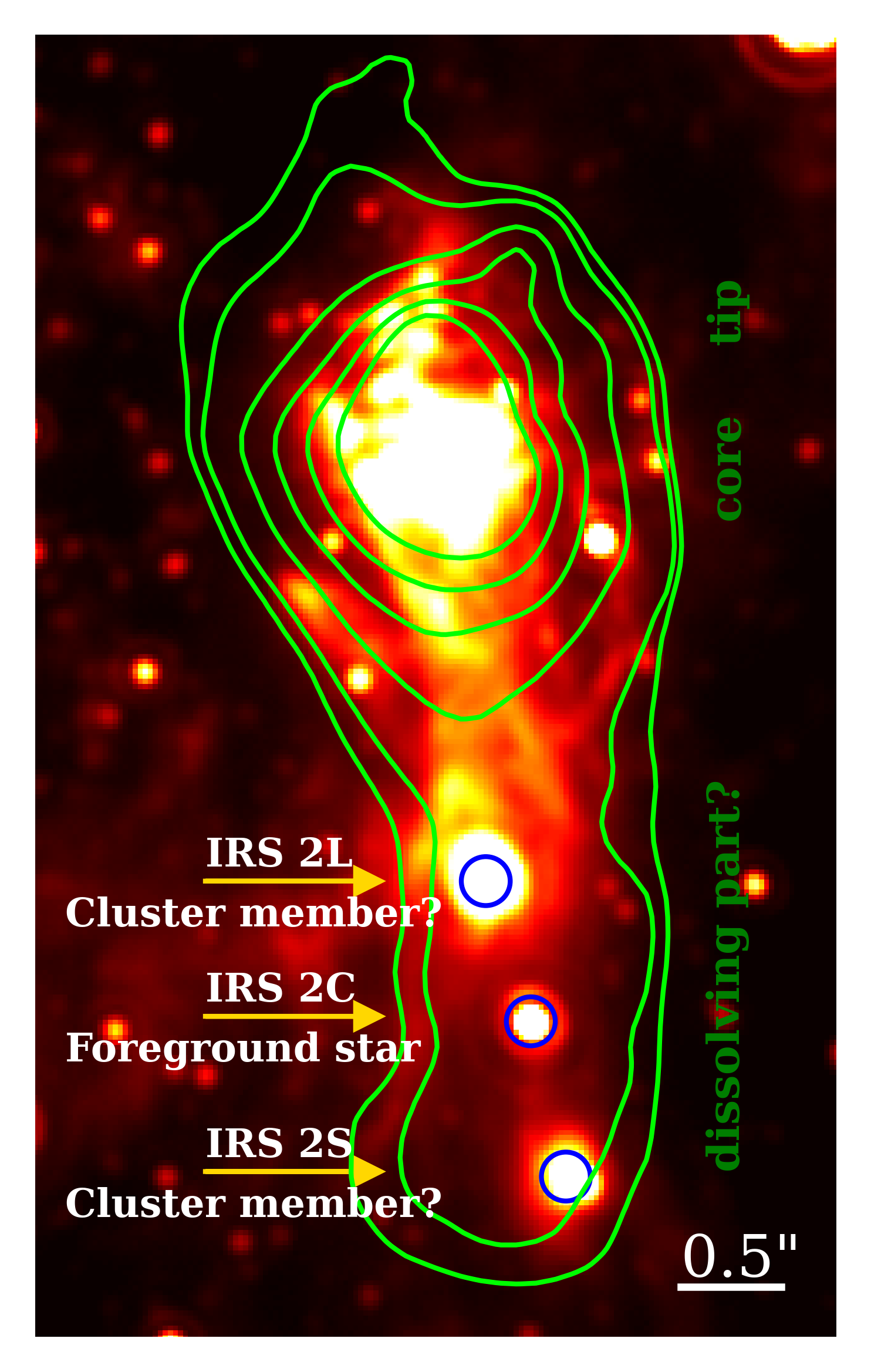}
	\caption{L-band NACO observation of IRS 13 in 2004 overlaid with [FeIII] $2.218\,\mu m$ $^3G_5\rightarrow ^3H_6$ contour lines extracted from a SINFONI 3d data cube. The contour lines represent the 26$\%$, 30$\%$, 40$\%$, 50$\%$, 60$\%$, and 70$\%$ level whereas the peak emission is at $2.5\times 10^{-10} erg s^{-1} cm^{-2} \mu m^{-1}$. The 70$\%$ contour line centered on the core region of IRS 13 resembles the projected size of the Hill radius ($\approx 22 mpc$) estimated with Eq. \ref{eq:hill} in Sec. \ref{ref:proper_motion}. In addition, the 30$\%$ contour lines enclose the southern region with respect to the core and tip components of the IRS 13 cluster. As indicated, we mark the two embedded sources IRS2L and IRS2S that might be former members of the core region of IRS 13. The compact source IRS2C is also known as AF/AHH \citep{Allen1990, Blum1996} and could be a foreground star based on the polarimetric analysis of \cite{Buchholz2013}.}
\label{fig:irs13_iron_line}
\end{figure}

Due to the complexity of possible formation scenarios for the IRS 13 cluster, we refer to Paper II. However, we briefly want to outline the basic idea to explain the findings presented in this work.
As proposed by \cite{Wang2020}\footnote{See their Fig. 10.}, the resulting trajectory of the young cluster that spirals in towards the inner parsec could have resulted in the bow shock formation caused by the supersonic motion of the cluster, cluster stellar wind, NSC winds, and the ISM. The dense region in the bow-shock shell would then be the birthplace of the second generation of IRS 13 stars (see Table \ref{tab:generation}).
\begin{table}[htbp!]
        \centering
        \begin{tabular}{|c|cc|}
        \hline\hline
              & 1. Generation  & 2. Generation   \\  
         \hline\hline     
        Approx. age [Myr] & 4 &  $<$ 1  \\
        \hline
        Birth place      & CND & Bow-shock shell \\
        \hline
        Current location  &  &  \\     
        inside IRS 13  & Core & Tip \\        
        \hline 
        Mean $K$-$L$ index & 2.49 & 5.72 \\
        \hline
        \end{tabular}   
\caption{Generation of stars inside the IRS 13 cluster with their related origin and current location. In addition, we list the mean $K$-$L$ color of the sources listed in Table \ref{tab:mag_e_sources}. We exclude E5.0 and E5.1 from the mean colors due to their missing H-band counterpart, which may be related to confusion or their nature. The majority of the sources considered in this list can be categorized as high-mass objects. See Fig. \ref{fig:irs13_iron_line} for the location of the sources.}
\label{tab:generation}
\end{table}
We note that the S-cluster \citep{Eckart1996Natur} exhibits a similar composition of stellar objects as listed in Table \ref{tab:generation} \citep[][]{Habibi2017, Peissker2020b, Ciurlo2020, peissker2021c, peissker2023a}. However, one would naturally expect a certain degree of elongation for an extended structure that gravitationally interacts with an SMBH such as Sgr~A* \citep[see simulations of][]{Hobbs2009, Jalali2014}. While the dimensions and nature of IRS 13 will be the focus of Paper II, we want to note that the [FeIII] emission of the cluster implies a larger structure as it is known from the literature.
Considering the forbidden [FeIII] line distribution presented in \cite{Lutz1993}, we find that the peak emission of the iron line clearly envelopes the IRS 13 and the IRS 2 region suggesting a combined setup of the northern and southern cluster region (see Fig. \ref{fig:irs13_iron_line}). It is already known that IRS 2C is a foreground star while IRS 2L and IRS 2S are embedded in the dust feature which is associated with IRS 13 \citep{Buchholz2013}.
However, our proposed interpretation of the dimensions of the cluster is in line with the polarization measurements of \cite{Buchholz2013} and \cite{Roche2018}. The polarimetric and magnetic field line analysis of Roche et al. reveals that the dust feature, which envelopes IRS 13 and IRS 2L/2C, is a coherent structure. The line distribution of [FeIII] but also the MIR dust emission matches the size of the polarization region of IRS 13 in \cite{Buchholz2013} and \cite{Roche2018} underlying the proposed dimensions of the cluster in this work.
In Paper II, we will present N-body simulations of an inspiralling cluster towards the inner parsec and investigate the possibility of such an event.   
\subsection{IRS 13 and the (counter-)clockwise disk}

\begin{figure}[htbp!]
	\centering
	\includegraphics[width=.5\textwidth]{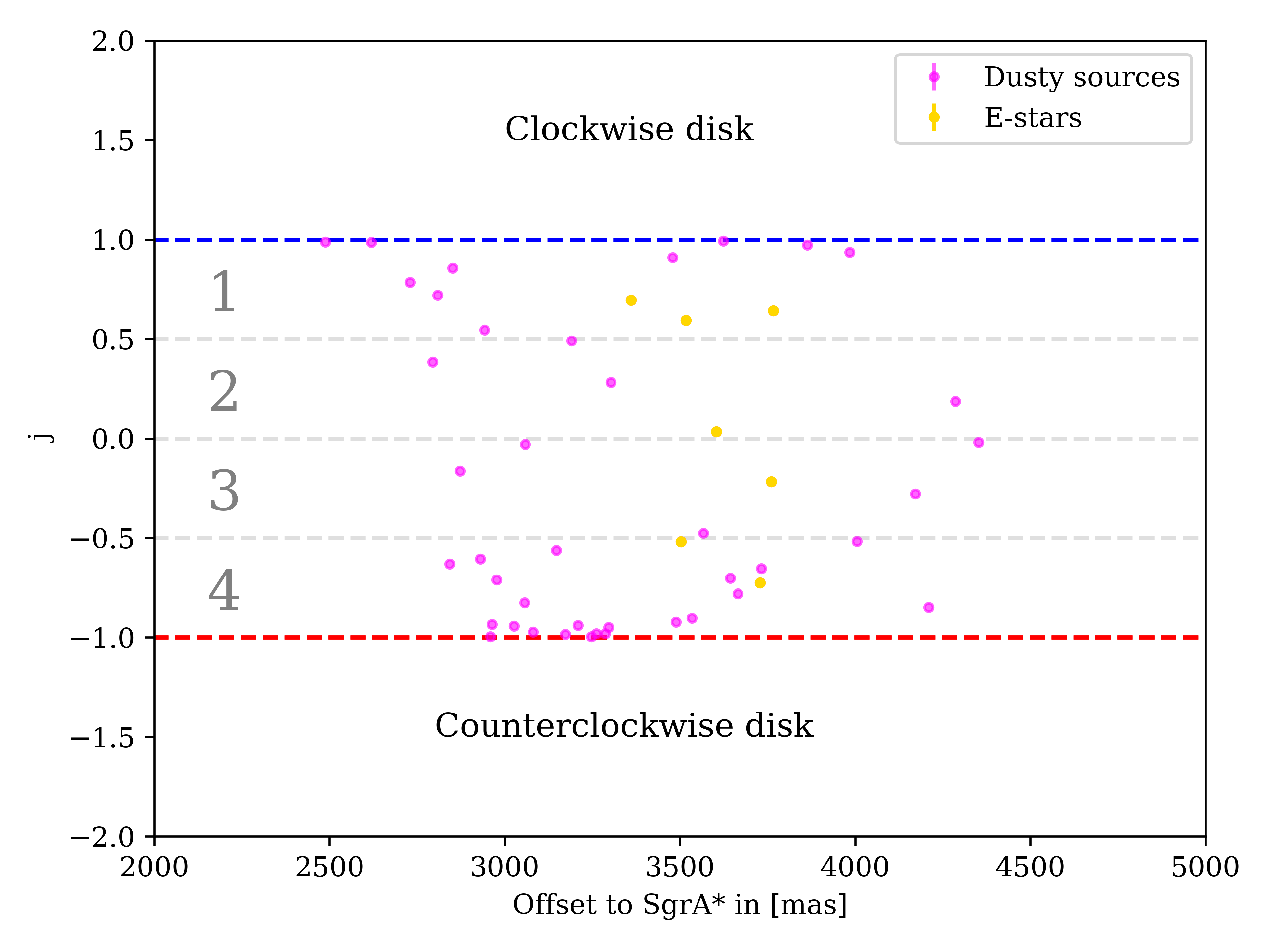}
	\caption{Normalized angular momentum $j$ as a function of distance from Sgr~A* for all the here investigated IRS 13 cluster members. We find a stellar distribution similar to the one derived by \cite{Paumard2006} for the majority of the NSC stars \citep[see also][]{Fellenberg2022}. Here, most of the investigated objects peak at j=-1, suggesting a CCW disk membership. Magenta circles represent the dusty sources analyzed in this work, the gold-filled ones indicate the E-stars (Fig. \ref{fig:finding_chart}). The size of the individual data points covers up the uncertainties calculated with error propagation.}
\label{fig:ccws_irs13}
\end{figure}

As we have shown in Fig. \ref{fig:anisotropy_parameter}, the investigated cluster member sample shows a non-uniform distribution. In addition, most of the stars in the NSC follow this nonisotropic kinematic pattern, which historically resulted in the finding of a counter- and clockwise disk, abbreviated as CCWS and CWS, respectively \citep[][]{Genzel1996, Paumard2006}. The mentioned velocity pattern is characterized by the normalized angular momentum $j$, which is defined by \cite{Genzel2003} as
\begin{equation}
    j\,=\,(xv_y\,-\,yv_x)/pv_p
\end{equation}
where $x,\,v_x,\,y,\,v_y$ refer to RA and DEC {coordinates} and their related components of proper motion, respectively, while the total distance and proper motion are given by $p$ and $v_p$. With the above equation and the numerical values given in Table \ref{tab:prop}, we find the same distribution (Fig. \ref{fig:ccws_irs13}) for IRS 13 sources as shown in \citet{Paumard2006}.
In analogy to the anisotropy parameter shown in Fig. \ref{fig:anisotropy_parameter}, we {divide} the data presented in Fig.~\ref{fig:ccws_irs13} {into} four bins. We estimate that the bin with $j\in\,\{0.5,1.0\}$ contains 22\% of the sources, for $j \in\,\{0.0,0.5\}$, we get 8\%, for $j\in\,\{0.0,-0.5\}$, there are 10\% of the sources, and finally for $j\in\,\{-0.5,-1.0\}$ we obtain 60\%. Therefore, we find an overdensity in bin 4, which strengthen our result presented in Fig. \ref{fig:anisotropy_parameter}. Since a non-uniform cluster is supposed to peak at $\gamma_{TR}=\pm 1$ as shown in Fig. \ref{fig:monte_carlo}\footnote{In addition, \cite{Genzel2000} shows Monte Carlo simulations that demonstrate in agreement with our results that non-uniform cluster exhibits an overdensity of stars at $\gamma_{TR}=\pm 1$.}, it is expected to find anisotropic structures in the angular momentum plot displayed in Fig. \ref{fig:ccws_irs13}. For our sample, we clearly identify an overdensity at $j=-1$ (Fig. \ref{fig:ccws_irs13}) suggesting a CCW disk membership \citep{Paumard2006, Ali2020, Fellenberg2022}.
Based on this finding, we propose three different scenarios:
\begin{itemize}
    \item[[a]] The (C)CWS characterization of stars in the {\it inner parsec} is valid for all (gravitationally bound) sub-regions,
    \item[[b]] The stellar overdensity of the IRS 13 cluster is the result of the intercepting disks of the CCW and CW systems,
    \item[[c]] The IRS 13 cluster shows the imprint of the CCW and CW systems.
\end{itemize}
Concerning [a], we want to highlight the theoretical work of \cite{Hobbs2009}, who predict two warped stellar discs/distributions for infalling molecular clouds. In addition, \cite{Ali2020} find a similar distribution for the S-stars, which suggests that a two-disk or even a multi-disk structure is present for other sub-regions as well, presumably those that are gravitationally bound to the SMBH or an IMBH. It needs to be verified by $N$-body numerical simulations how long the original disk-like stellar structure that bears imprints of the formation mechanism can survive within the NSC. As mentioned before, the results presented in Fig. \ref{fig:ccws_irs13} do reveal that the majority of investigated cluster members ($>50\%$) are part of at least one disk, presumably the CCW disk.\newline
In addition, \cite{Paumard2006} suggested that the IRS 13 cluster may result from the interaction of the CCW and CW disks, i.e. scenario [b]. Although this scenario cannot be excluded, it implies an underlining rotation pattern that may have been created by infalling clouds in the first place (see [a] and \citeauthor{Hobbs2009}, \citeyear{Hobbs2009}). Since the E-stars seem to be members of both distributions (Fig. \ref{fig:anisotropy_parameter}, upper left plot), the scenario seems plausible. However, we will investigate this particular point in more detail in Paper II because it would exceed the scope of this work.
For the last scenario [c], the IRS 13 cluster serves as a tracer for the underlining disk pattern. The infalling cluster IRS 13 may have intercepted the CCW and CW disks, which led to compressed gas densities that triggered star formation. Likewise, for [b], we will focus on this point in Paper II.
Independent of the exact relation between IRS 13 and the (C)CW disks, we want to stress that \cite{Hansen2003} demanded a second black hole of $\approx 10^4 M_{\odot}$ in order to explain the unusually young age of the S-cluster stars \citep{Morris1993, Ghez2003, Habibi2017}. The mass estimate is in the same order as the estimated enclosed mass for IRS 13 of $3.9 \times 10^4 M_{\odot}$. Due to the age of the S-cluster members, IRS 13 is most certainly not a suitable candidate for process explaining the presence of young stars close to Sgr~A*. But it is an interesting scientific question to explore and maybe even link possible large-scale imprints on molecular clouds in the CND from the enclosed mass of the IRS 13 cluster.

\subsection{Multiplicity fraction of the IRS13 cluster}

Considering the young age of the IRS13 cluster members, we should have detected an increased multiplicity and companion fraction \citep{Zwart2010}. Surprisingly, only one binary system close to IRS13 is known \citep{Pfuhl2014} {which might not be related to the cluster in the first place}. Taking into account the important role of binaries, especially for the evolution of massive stars \citep{Sana2012}, we expect frequent updates on the detection of binary systems in the IRS13 cluster, particularly with the upcoming Extremely Large Telescope (ELT).\newline
For example, \cite{Gautam2019} identified more than a dozen possible periodic systems. Due to resolution limitations, observation of visual binaries remains unlikely, damping the number of methods to detect such systems. Therefore, high-cadence observations with the scientific goal of identifying magnitude or LOS variations will remain the sufficient approach for multiplicity analysis.\newline
However, given the number of sources investigated in this work, we speculatively expect at least one binary system among the sample. From the analysis, we observed minor position fluctuations of $\gamma$ and $\zeta$. These uncertainties may result from a confusion problem due to the high source density \citep{Paumard2006}.\newline
Assuming that the above-mentioned uncertainties cannot be explained by source confusion, we will use the definition of \citet{Reipurth1993} and \citet{Duchene2001} for the multiplicity fraction (MF) and the related companion fraction (CF). Consequently, MF is defined as 
\begin{equation}
    \rm MF\,=\,\frac{B+T+Q+\cdot\cdot\cdot}{S+B+T+Q+\cdot\cdot\cdot}
    \label{eq:mf}
\end{equation}
where S defines the number of single stars and B is the number of binary star systems. Triple and quadruple star systems are defined by T and Q letters in the above equation. In addition, CF can be written as 
\begin{equation}
    \rm CF\,=\,\frac{2B+3T+4Q+\cdot\cdot\cdot}{S+2B+3T+4Q+\cdot\cdot\cdot}
    \label{eq:cf}
\end{equation}
and defines the ratio of stars with a companion. For simplicity, we assume that $\gamma$ and $\zeta$ are two binary systems. In total, we derived seven orbital solutions for the brightest sources of the sample. These boundary conditions are translated to MF$\sim \,28\,\%$ and CF$\sim \,44\,\%$ using Eq. \ref{eq:mf} and Eq. \ref{eq:cf}, {respectively}. If we assume a rough cluster age of $4\,\pm\,1$ Myr based on the most evolved E-stars \citep{Maillard2004, Paumard2006, Zhu2020}, we find similar MF and CF values in other clusters with a comparable age, such as RCW 108 \citep{Comeron2005, Comeron2007} or the SMC cluster NGC 330 \citep{Bodensteiner2021}. If these numbers hold, it would demonstrate comparable star formation channels between different (galactic) clusters.
We want to note that we assumed for the above discussion the presence of two unconfirmed binaries. However, we anticipate multiple opportunities for upcoming instruments and observation campaigns covering the IRS 13 cluster.

\subsection{Observations with the James Webb Space Telescope}
\label{sec:jwst}
Despite strong indications for the nature of the dusty objects, the data lack detailed spectroscopic analysis. For {YSOs of type I}, we would typically use tracers such as H$_2$ \citep{Glassgold2004}, H$_2$O \citep{Gibb2000}, and HCN \citep[][]{Lahuis2006} to confirm the classification. Considering the recent start of the scientific operations of the James Webb Space Telescope, the upcoming GTO observations\footnote{Prog. Id: GTO 1266} of the GC will minimize the uncertainties of the YSO classification for the dusty sources. 
\begin{figure}[htbp!]
	\centering
	\includegraphics[width=.5\textwidth]{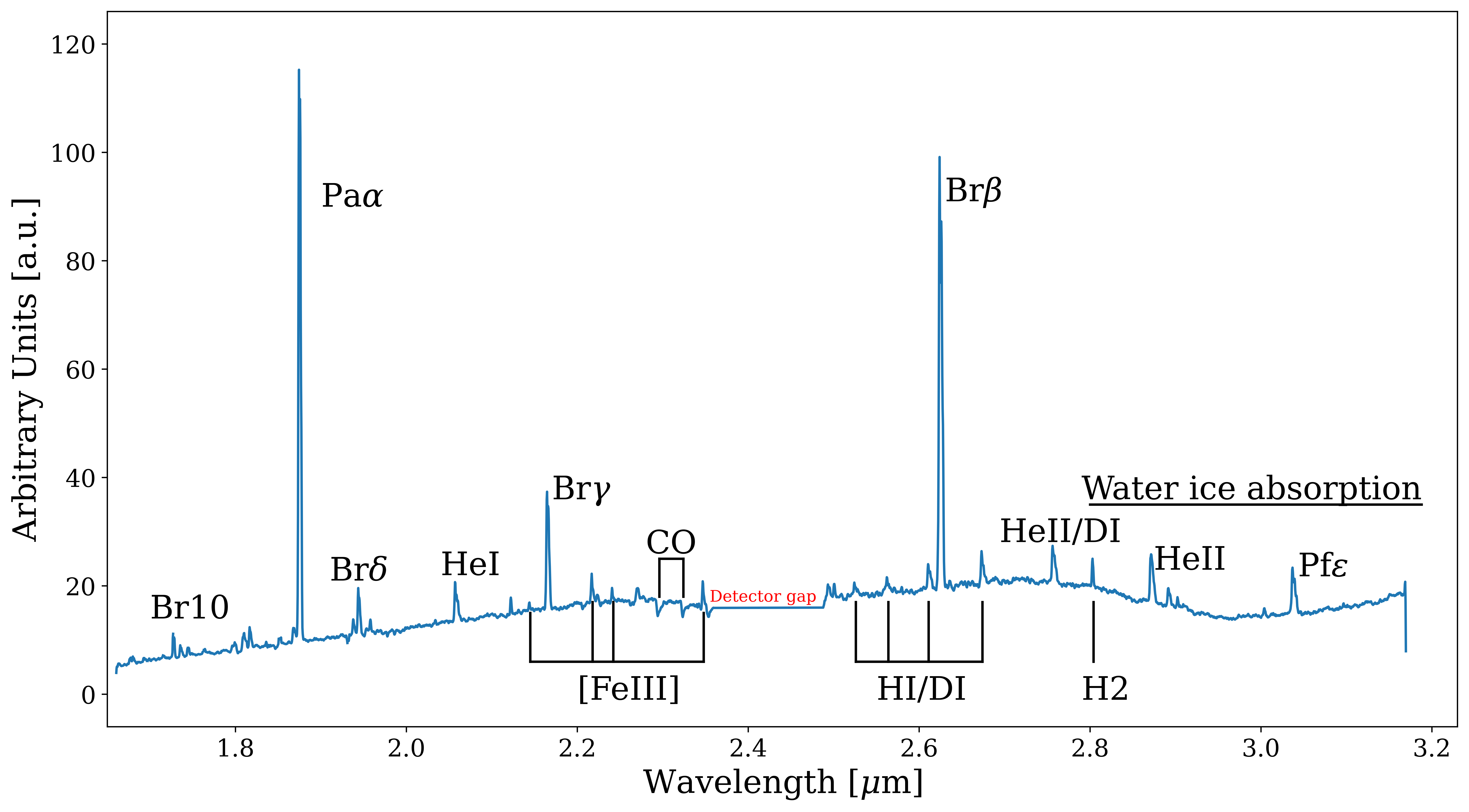}
	\caption{NIRSPEC spectrum of the inner parsec observed with the James Webb Space Telescope in 2022 (PI:Jessica Lu, Proposal ID: 1939). The NIRSPEC observation agrees with the water ice identification by, e.g., \cite{moultaka2015}. Around 2.4$\mu m$, the detector gap of NIRSPEC is marked.}
\label{fig:nirpsec_jwst}
\end{figure}
We note that there \normalfont{is publicly available} archive\footnote{Downloaded from the Barbara A. Mikulski Archive for Space Telescopes (MAST).} data observed with NIRSPEC in 2023. Consequently, we used this dataset to inspect the spectral NIR emission and parts of the $L$-band (MIR) for the presence of individual line tracers. Due to the absence of telluric emission/absorption lines, we find an unaffected Pa$\alpha$ line but also strong water absorption features \citep{moultaka2015}. With a nominal resolution power of $\sim 2700$ with the G235H/F170LP setting, we identify several single emission and absorption lines (e.g., Br$\delta$, HeI, Br$\gamma$, and CO band heads) in the $H+K$-band that are well known from SINFONI observations of the same region \citep{Peissker2020c, peissker2021, peissker2023b}. Considering the recent identification of snowlines in the spectrum of HH 48 NE, we expect similar findings in the GC with NIRSPEC in addition to YSO tracers that could potentially be identified with MIRI.
Although the spatial $L$-band resolution of NACO and MIRI (JWST) is {sufficiently} comparable, a stable PSF and longer on-source integration times could result into new insights into the {IRS13} cluster. {With the demonstrated capabilities of the JWST (Fig. \ref{fig:nirpsec_jwst}), we will search for tracers associated with YSOs \citep{peissker2023b}. Since} some dusty sources are close to the E-stars, we aim to increase the number of cluster samples to verify our findings on the warped disk structure of IRS 13. In addition to the continuum detections presented in this work, we expect an increase in the number of line-emitting sources such as G2/DSO \citep[][]{peissker2021c} observed with the MIRI and {NIRSPEC} IFU data.

\section{Conclusions} 
\label{sec:conclusion}

We analyzed the IRS 13 cluster that resides at a {projected} distance of $\sim 0.15$ pc from Sgr~A*. Based on {the here presented work}, we found a significantly higher number of cluster members {of IRS 13} compared to previous studies.
Using multi-wavelength observations resulting in a comprehensive color-color diagram classification, we applied a ray-tracing radiative transfer model to investigate the nature of the brightest dusty sources and found compelling evidence that points towards a YSO characterization. The DS sources share a comparable footprint with the bright dusty sources suggesting a YSO classification. In the following, we list our key findings:
\begin{itemize}
    \item The nature of the dusty objects can be described as MYSOs and HMYSOs in agreement with the classification of X3a,
    \item Despite the low-mass YSO $\eta$, we classify all other investigated bright dusty sources as massive class I YSOs,
    \item The majority of investigated sources in this work are arranged in a significant disk structure, presumably the CCW disk,
    \item This non-uniform arrangement of the IRS 13 cluster is in remarkable agreement with previous normalized angular momentum studies of the NSC,
    \item From the kinematics of the cluster members, we estimate a minimum mass of $4\times 10^4 M_{\odot}$ that is required for a tidally stable system,
    \item The derived {tidal (Hill) radius} shows a strong correlation with the dimensions of the peak emission distribution (dust/[FeIII]) of IRS 13 \normalfont{(Fig. \ref{fig:irs13_iron_line})},
    \item The tidally stable core of IRS 13 harbors massive O/WR stars but also HMYSOs,
    \item In total, we find two generations of stellar objects that can be distinguished by their age.
\end{itemize}
In the future, we expect {to identify} more objects as the analyzed DS objects or the bow-shock source X3 which might be {associated with the IRS 13 cluster}. The large-scale MIRI (JWST) and ERIS (VLT) observations providing IFU data will enhance the characterization and the related precise stellar age determination of individual sources in the IRS 13 cluster. 

\begin{acknowledgments}
We thank an anonymous referee for a constructive and encouraging report that helped to improve the manuscript.
This work was supported in part by the Deutsche Forschungsgemeinschaft (DFG) via the Cologne
Bonn Graduate School (BCGS), the Max Planck Society through the International Max Planck Research School
(IMPRS) for Astronomy and Astrophysics as well as special funds through the University of Cologne. Conditions and Impact of Star Formation is carried out within the Collaborative Research Centre 956, sub-project [A02], funded by the Deutsche Forschungsgemeinschaft (DFG) – project ID 184018867. B.Sh. acknowledges financial support from the State Agency for Research of the Spanish MCIU through the “Center of Excellence Severo Ochoa” award for the Instituto de Astrofisica de Andalucia (SEV-2017- 0709). MZ acknowledges the GA\v{C}R-LA grant No. GF23-04053L for financial support. Part of this work was supported by fruitful discussions with members of the European Union funded COST Action MP0905: Black Holes in a Violent Universe and the Czech Science Foundation (No.\ 21-06825X). VK has been partially supported by the Czech Ministry of Education, Youth and Sports Research Infrastructure (LM2023047). AP, JC, SE, and GB contributed useful points to the discussion. A.E. and F.P. acknowledge support through the German Space Agency DLR 50OS1501 and DLR 50OS2001  from 2015 to 2023. We also would like to thank the members of the SINFONI/NACO/VISIR and ESO's Paranal/Chile team for their support and collaboration. This paper makes use of the following ALMA data: ADS/JAO.ALMA$\#$2015.1.01080.S and ADS/JAO.ALMA$\#$2012.1.00543.S. ALMA is a partnership of ESO (representing its member states), NSF (USA) and NINS (Japan), together with NRC (Canada), MOST and ASIAA (Taiwan), and KASI (Republic of Korea), in cooperation with the Republic of Chile. The Joint ALMA Observatory is operated by ESO, AUI/NRAO and NAOJ.
\end{acknowledgments}
\facilities{VLT (SINFONI and NACO), ALMA (Band 7)}

\software{astropy \citep{2013A&A...558A..33A,2018AJ....156..123A, Astropy2022},
          SciPy \citep{SciPy2020},
          Hyperion \citep{Robitaille2011, Robitaille2017},
          DPuser \citep{Ott2013}
          }

\bibliography{bib}{}
\bibliographystyle{aasjournal}

\appendix

In this Appendix, we list the data used for the analysis. In addition, we compare the number of dusty sources of this work with the literature. We furthermore indicate the related proper motion of the sources investigated.

\section{Data}
\label{ref:data_appendix}

In Table \ref{tab:naco_data1}, we list the K-band data used in this work. Although the source confusion in the IRS 13 cluster is increased due to its high density, we identify K-band positions of the dusty objects in most epochs of the listed data. In addition, the prominent MIR emission of these dusty sources enables us to incorporate all investigated epochs of the listed L-band data observed with NACO (see Table \ref{tab:naco_data2}). The ID of the GC observation in the H+K-band with SINFONI in 2014 is listed in Table \ref{tab:data_sinfo}. 

\begin{table*}[h!]
\centering
\begin{tabular}{ccc}
\hline
\hline
\multicolumn{3}{c}{NACO K-band}\\
\hline
Date  & Observation ID & \multicolumn{1}{p{1.5cm}}{\centering number \\ of exposures } \\
\hline
2002.07.31 & 60.A-9026(A)  & 61   \\
2003.06.13 & 713-0078(A)   & 253  \\
2004.07.06 & 073.B-0775(A) & 344  \\
2004.07.08 & 073.B-0775(A) & 285  \\
2005.07.25 & 271.B-5019(A) & 330  \\
2005.07.27 & 075.B-0093(C) & 158  \\
2005.07.29 & 075.B-0093(C) & 101  \\
2005.07.30 & 075.B-0093(C) & 187  \\
2005.07.30 & 075.B-0093(C) & 266  \\
2005.08.02 & 075.B-0093(C) & 80   \\
2006.08.02 & 077.B-0014(D) & 48   \\
2006.09.23 & 077.B-0014(F) & 48   \\
2006.09.24 & 077.B-0014(F) & 53   \\
2006.10.03 & 077.B-0014(F) & 48   \\
2006.10.20 & 078.B-0136(A) & 47   \\
2007.03.04 & 078.B-0136(B) & 48   \\
2007.03.20 & 078.B-0136(B) & 96   \\
2007.04.04 & 179.B-0261(A) & 63   \\ 
2007.05.15 & 079.B-0018(A) & 116  \\ 
2008.02.23 & 179.B-0261(L) & 72   \\ 
2008.03.13 & 179.B-0261(L) & 96   \\ 
2008.04.08 & 179.B-0261(M) & 96   \\ 
2009.04.21 & 178.B-0261(W) & 96   \\ 
2009.05.03 & 183.B-0100(G) & 144  \\ 
2009.05.16 & 183.B-0100(G) & 78   \\ 
2009.07.03 & 183.B-0100(D) & 80   \\ 
2009.07.04 & 183.B-0100(D) & 80   \\ 
2009.07.05 & 183.B-0100(D) & 139  \\ 
2009.07.05 & 183.B-0100(D) & 224  \\ 
2009.07.06 & 183.B-0100(D) & 56   \\ 
2009.07.06 & 183.B-0100(D) & 104  \\ 
2009.08.10 & 183.B-0100(I) & 62   \\ 
2009.08.12 & 183.B-0100(I) & 101  \\
2010.03.29 & 183.B-0100(L) & 96   \\ 
2010.05.09 & 183.B-0100(T) & 12   \\ 
2010.05.09 & 183.B-0100(T) & 24   \\ 
2010.06.12 & 183.B-0100(T) & 24   \\ 
2010.06.16 & 183.B-0100(U) & 48   \\
2011.05.27 & 087.B-0017(A) & 305  \\
2012.05.17 & 089.B-0145(A) & 169  \\
2013.06.28 & 091.B-0183(A) & 112  \\
2017.06.16 & 598.B-0043(L) & 36   \\
2018.04.24 & 101.B-0052(B) & 120  \\
\hline  
\end{tabular}
\caption{K-band data observed with NACO between 2002 and 2018.}
\label{tab:naco_data1}
\end{table*}

\begin{table*}[h!]
\centering
\begin{tabular}{ccc}
\hline
\hline
\multicolumn{3}{c}{NACO L-band}\\
\hline
Date  & Observation ID & \multicolumn{1}{p{1.5cm}}{\centering number \\ of exposures }   \\
\hline
2002.08.30 &  060.A-9026(A) & 80 \\  
2003.05.10 &  071.B-0077(A) & 56 \\  
2004.07.06 &  073.B-0775(A) & 217\\  
2005.05.13 &  073.B-0085(E) & 108\\  
2005.06.20 &  073.B-0085(F) & 100\\  
2006.05.28 &  077.B-0552(A) &  46\\  
2006.06.01 &  077.B-0552(A) & 244\\  
2007.03.17 &  078.B-0136(B) &  78\\  
2007.04.01 &  179.B-0261(A) &  96\\  
2007.04.02 &  179.B-0261(A) & 150\\  
2007.04.02 &  179.B-0261(A) &  72\\  
2007.04.06 &  179.B-0261(A) & 175\\  
2007.06.09 &  179.B-0261(H) &  40\\  
2008.05.28 &  081.B-0648(A) &  58\\  
2008.08.05 &  179.B-0261(N) &  64\\  
2008.09.14 &  179.B-0261(U) &  49\\  
2009.03.29 &  179.B-0261(X) &  32\\  
2009.03.31 &  179.B-0261(X) &  32\\  
2009.04.03 &  082.B-0952(A) &  42\\  
2009.04.05 &  082.B-0952(A) &  12\\  
2009.09.19 &  183.B-0100(J) & 132\\  
2009.09.20 &  183.B-0100(J) &  80\\  
2010.07.02 &  183.B-0100(Q) & 485\\  
2011.05.25 &  087.B-0017(A) & 29 \\  
2012.05.16 &  089.B-0145(A) & 30 \\  
2013.05.09 &  091.C-0159(A) & 30 \\  
2015.09.21 &  594.B-0498(G) & 420   \\
2016.03.23 &  096.B-0174(A) & 60 \\  
2017.03.23 &  098.B-0214(B) & 30 \\  
2018.04.22 & 0101.B-0065(A) & 68 \\  
2018.04.24 & 0101.B-0065(A) & 50 \\  
\hline  
\end{tabular}
\caption{L-band data observed with NACO between 2002 and 2018.}
\label{tab:naco_data2}
\end{table*}

\begin{table*}[htbp!]
        \centering
        \begin{tabular}{|ccccc|}
        \hline\hline
             Date & Observation ID  & Exp. Time & Band & Instrument/Telescope  \\  
        (YYYY:MM:DD) &  & (s) &  &   \\ \hline\hline 
        
        2014.08.30 & 093.B-0218(B) &  2700  &  H+K   &  SINFONI/VLT    \\
        \hline 
        \end{tabular}   
\caption{SINFONI data used in this work for the identification of the iron line (see Fig. \ref{fig:irs13_iron_line}). We applied the standard reduction steps provided by the ESO pipeline to create the final mosaic.}
\label{tab:data_sinfo}
\end{table*}

\section{Dusty objects}
\label{ref:dusty_objects_ident}

Here, we provide an overview of the sources investigated in this work compared to previous investigations in the literature. Table \ref{tab:ID} exhibits all unknown and known dusty objects of the IRS 13 region with the corresponding id of the related publication. \normalfont{To cross-identify all sources in the literature with this work, we visually compare finding charts of the related publication listed in Table \ref{tab:ID}.}
We expect that future higher-resolution observations will most certainly establish a new nomenclature as it is commonly done for GC sources (please compare the analysis of \cite{Eckart2013} with \cite{Ciurlo2020}).

\begin{table*}[]
\centering
\begin{tabular}{@{}lccccc@{}}
\hline\hline
ID         & \cite{Maillard2004} & \cite{Schoedel2005} & \cite{muzic2008} & \cite{Fritz2010} & \cite{Eckart2013} \\ 
\hline
$\alpha$   &                &              & $\times$   &            & 8           \\
$\beta$    &                &              & $\times$   &            & 9           \\
$\gamma$   &                &              & $\times$   &            & 22          \\
$\delta$   &                &              & $\times$   &            & 17          \\
$\epsilon$ &                &              & $\times$   &            & 13          \\
$\zeta$    &                &              & $\times$   &            & 34          \\
$\eta$     & 15       & $\times$     & $\times$   &            & 19          \\
$\vartheta$   &                &              &            &            &          \\
$\iota$    &                & $\times$     &            &            & 10          \\
1          &                & $\times$     & $\times$   &            &             \\
2          &                &              &            &            &             \\
3          &                &              &            &            &             \\
4          &                & $\times$     &            &            &             \\
5          &                &              &            &            &             \\
6          &                &              &            &            &             \\
7          &                &              &            &            &             \\
8          &                &              &            &            &             \\
9          &                & $\times$     &            &            &             \\
10         &                & $\times$     &            &            &             \\
11         & 14             & $\times$     & $\times$   &            & 15          \\
12         & 13             & $\times$     & $\times$   &            & 16          \\
13         &                &              &            &            &             \\
14         &                &              &            &            &             \\
15         &                &              &            &            &             \\
16         &                &              &            &            &             \\
17         & 20             & $\times$     & $\times$   &            &             \\
18         &                &              &            &            &             \\
19         &                &              &            &            &             \\
20         &                & $\times$     &            &            &             \\
21         &                &              &            &            &             \\
22         &                &              &            &            & 23          \\
23         &                &              &            &            &             \\
24         & 18             & $\times$     &            &            &             \\
25         &                & $\times$     &            &            &             \\
26         &                &              & $\times$   &            & 7           \\
27         & 16             &              & $\times$   &            & 20          \\
28         &                &              &            &            &             \\
29         &                &              &            &            &             \\
30         &                &              &            &            &             \\
31         &                &              &            &            &             \\
32         &                &              &            &            &             \\
33         &                &              &            &            &             \\
E1         & $\times$       & $\times$     & $\times$   & $\times$   & 1           \\
E2         & $\times$       & $\times$     & $\times$   & $\times$   & 2           \\
E3         & $\times$       & $\times$     & $\times$   & $\times$   & 3           \\
E4         & $\times$       & $\times$     &            & $\times$   & 4           \\
E5.0       & $\times$       &              & $\times$   & $\times$   & 5           \\
E5.1       &                &              &            & $\times$   &             \\
E7         & 5              & $\times$     & $\times$   & $\times$   & 25          \\ 
\hline
\end{tabular}
\caption{Identification of all the sources investigated in this work. As listed, the nomenclature is not consistent throughout the literature. A $\times$ symbol indicates the detection in the related publication. If the sources were labeled with an alternative ID, we additionally list the corresponding number.}
\label{tab:ID}
\end{table*}

\section{Individual detections of the dusty sources}
\label{sec:individual_pos_app}
Here, we list all individual relative locations of the dusty sources investigated in this work. The related position of S2 is indicated. With the orbital elements of \cite{Do2019S2} and \cite{gravity2018}, the position of Sgr~A* can be estimated, which can be transferred to the absolute location of the dusty sources. The positions of the DS sources listed in Table \ref{tab:pos1}-\ref{tab:pos14} are determined using a Gaussian fit. In addition, this fit provides an uncertainty which is indicated after the related numerical R.A. and DEC. value. From these tables, the bulk motion of IRS13 can be analyzed. Since this exceeds the scope of this work, we refer to Paper II for an extended analysis. Due to the consistency of the data set which transforms into a lowered confusion, we list the L-band positions in Table \ref{tab:pos1}-\ref{tab:pos14}.
\begin{table*}[htbp!]
\centering
\setlength{\tabcolsep}{0.9pt}
\begin{tabular}{|c|cccccccccccc|}
\hline\hline
Epoch    & S2 R.A. & S2 err R.A. & S2 Dec. & S2 err Dec. & $\alpha$ R.A. & $\alpha$ err R.A. & $\alpha$ Dec. & $\alpha$ err Dec. & $\beta$ R.A. & $\beta$ err R.A. & $\beta$ Dec. & $\beta$ err Dec.\\ 
\hline
2002-08-29 & 685.59 &  0.02	& 764.57 &0.02	&785.34	&0.16	&709.63	&0.2	&793.34	&0.06	&718.19	&0.03\\
2002-08-30 & 1121.1 &  0.03	& 787.88 &0.03	&1221   &0.12	&733.01	&0.13	&1229.2	&0.08	&741.59	&0.04\\
2003-05-10 & 873.79 &  0.04	& 743.43 &0.04	&974.42	&0.1    &686.39	&0.1    &982.09	&0.04	&695.06	&0.03\\
2004-04-25 & 678.01 &  0.07	& 640.11 &0.05	&778.16	&0.1    &580.8	&0.14	&785.78	&0.04	&589.59	&0.03\\
2004-04-26 & 678.09 &  0.04	& 640.29 &0.04	&778.08	&0.1    &580.86	&0.15	&785.91	&0.06	&589.73	&0.05\\
2005-05-13 & 788.84 &  0.06	& 776.4	 &0.05	&888.24	&0.1    &717.19	&0.1    &895.85	&0.04	&725.98	&0.03\\
2006-05-29 & 744.23 &  0.04	& 747.74 &0.05	&843.14	&0.2    &687.49	&0.3    &850.76	&0.03	&696.36	&0.03\\
2007-04-01 & 1043.3 &  0.02	& 1050.8 &0.02	&1141.9	&0.1    &990.24	&0.11	&1149.7	&0.05	&999.13	&0.04\\
2007-05-15 & 1040.2 &  0.06	& 1052.2 &0.05	&1138.7	&0.1    &991.3	&0.1    &1146.7	&0.04	&1000.1	&0.04\\
2007-05-16 & 1011.6 &  0.04	& 1074.3 &0.04	&1110.1	&0.1    &1013.9	&0.2    &1118.2	&0.04	&1022.7	&0.04\\
2007-05-17 & 1038.9 &  0.08	& 1061.1 &0.08	&1137.5	&0.1    &1000.6	&0.1	&1145.4	&0.04	&1009.3	&0.04\\
2007-05-18 & 1050.3 &  0.02	& 1051	 &0.02	&1148.5	&0.11	&990.64	&0.13	&1156.8	&0.06	&999.32	&0.04\\
2007-05-19 & 972.05 &  0.05	& 1042.4 &0.04	&1070.7	&0.1  	&981.88	&0.11	&1078.7	&0.04	&991.04	&0.03\\
2007-05-22 & 808.75 &  0.02	& 812.64 &0.02	&907.15	&0.2	&752.18	&0.23	&915.17	&0.04	&760.98	&0.03\\
2007-05-23 & 1057	&  0.03	& 1082.6 &0.03	&1155.5	&0.09	&1022.2	&0.13	&1163.4	&0.04	&1031	&0.03\\
2008-05-26 & 1020	&  0.1	& 1110.9 &0.1   &1117.9	&0.1 	&1049.4	&0.1 	&1125.8	&0.1  	&1058.3	&0.1\\
2008-05-30 & 988.24 &  0.1  & 1100.9 &0.1  	&1086.1	&0.1 	&1039.5	&0.1 	&1094	&0.1  	&1048.4	&0.1\\
2008-05-31 & 1049.2 &  0.1	& 1070.8 &0.1  	&1147.2	&0.1 	&1009.5	&0.1 	&1155.2	&0.1  	&1018.4	&0.1\\
2008-06-02 & 1056.8 &  0.1  & 1071.1 &0.1   &1154.6	&0.1 	&1009.9	&0.2 	&1162.8	&0.1  	&1018.7	&0.1\\
2008-06-03 & 1060.4 &  0.1	& 1072.2 &0.1   &1158.4	&0.1 	&1010.6	&0.1 	&1166.3	&0.1  	&1019.6	&0.1\\
2011-05-25 & 866.21 &  0.02	& 850.86 &0.02	&962.21	&0.1 	&790.05	&0.11	&970.45	&0.04	&799.06	&0.03\\
2012-05-16 & 867.6	&  0.04	& 787.45 &0.04	&963.27	&0.15	&727.04	&0.17	&971.41	&0.03	&736.26	&0.03\\
2013-05-09 & 870.72 &  0.08	& 786.15 &0.07	&966.26	&0.1	&725.86	&0.13	&974.18	&0.03	&735.16	&0.03\\
2016-03-23 & 847.81 &  0.05	& 826.22 &0.04	&943.31	&0.13	&769.88	&0.13	&951.02	&0.04	&779.4	&0.03\\
2018-04-22 & 850.83 &  0.03	& 761.76 &0.06	&947.89	&0.0846	&709.41	&0.13	&956.11	&0.04	&718.83	&0.06\\
2018-04-24 & 852.93 &  0.03	& 758.05 &0.04	&950.43	&0.11	&704.96	&0.13	&958.33	&0.05	&714.66	&0.05\\
\hline
\end{tabular}
\caption{Relative pixel positions of S2 and the dusty sources. From the position of S2, the location of Sgr~A* can be estimated. With this, the absolute positions of the dusty sources can be calculated. The uncertainties represent the Gaussian fit error.}
\label{tab:pos1}
\end{table*}

\begin{table*}[htbp!]
\centering
\setlength{\tabcolsep}{0.9pt}
\begin{tabular}{|c|cccccccccccc|}
\hline\hline
Epoch    & $\gamma$ R.A. & $\gamma$ err R.A. & $\gamma$ Dec. & $\gamma$ err Dec. & $\delta$ R.A. & $\delta$ err R.A. & $\delta$ Dec. & $\delta$ err Dec. & $\epsilon$ R.A. & $\epsilon$ err R.A. & $\epsilon$ Dec.  & $\epsilon$ err Dec.\\ 
\hline
2002-08-29	&800.82	&0.06	&726.94	&0.04	&793.93	&0.05	&732.58	&0.04	&793.15	&0.12	&726.72	&0.07\\
2002-08-30	&1236.5	&0.05	&750.27	&0.03	&1229.6	&0.05	&755.81	&0.03	&1228.5	&0.08	&749.47	&0.04\\
2003-05-10	&989.62	&0.07	&704.04	&0.05	&982.75	&0.04	&709.72	&0.03	&981.74	&0.1	&703.63	&0.07\\
2004-04-25	&793.5	&0.09	&598.65	&0.07	&786.52	&0.05	&604.21	&0.04	&785.46	&0.09	&598	&0.06\\
2004-04-26	&793.4	&0.1	&598.63	&0.09	&786.37	&0.09	&604.21	&0.07	&785.41	&0.09	&598.06	&0.06\\
2005-05-13	&903.59	&0.09	&735.11	&0.08	&896.47	&0.1	&740.59	&0.08	&895.55	&0.16	&734.65	&0.1\\
2006-05-29	&858.64	&0.07	&705.52	&0.06	&851.56	&0.05	&710.96	&0.04	&850.61	&0.1	&705.13	&0.07\\
2007-04-01	&1157.6	&0.1	&1008.4	&0.07	&1150.5	&0.05	&1013.8	&0.04	&1149.6	&0.1	&1007.9	&0.06\\
2007-05-15	&1154.7	&0.08	&1009.5	&0.06	&1147.4	&0.07	&1014.8	&0.06	&1146.6	&0.15	&1008.8	&0.09\\
2007-05-16	&1126	&0.08	&1032	&0.06	&1118.7	&0.0875	&1037	&0.08	&1118.1	&0.09	&1031.3	&0.05\\
2007-05-17	&1153.5	&0.07	&1018.7	&0.05	&1146.3	&0.046	&1024	&0.03	&1145.4	&0.1	&1018.3	&0.07\\
2007-05-18	&1164.7	&0.07	&1008.6	&0.06	&1157.6	&0.05	&1013.9	&0.04	&1157	&0.15	&1008.1	&0.09\\
2007-05-19	&1086.1	&0.1	&1000.3	&0.09	&1079.2	&0.077	&1005.3	&0.07	&1078.7	&0.09	&999.5	&0.05\\
2007-05-22	&923.04	&0.1	&770.31	&0.08	&916.05	&0.069	&775.61	&0.04	&915.35	&0.1	&770.08	&0.08\\
2007-05-23	&1171.6	&0.07	&1040.3	&0.06	&1164.3	&0.06	&1045.7	&0.06	&1163.6	&0.1	&1040.2	&0.08\\
2008-05-26	&1134	&0.1	&1067.7	&0.1	&1126.8	&0.1	&1073	&0.1	&1125.9	&0.1	&1067.1	&0.1\\
2008-05-30	&1102.2	&0.1	&1057.8	&0.1	&1095	&0.1	&1063.2	&0.1	&1093.9	&0.4	&1057.3	&0.3\\
2008-05-31	&1162.6	&0.1	&1027.9	&0.1	&1155.9	&0.1	&1032.7	&0.1	&1155	&1	    &1027	&1\\
2008-06-02	&1170.7	&0.1	&1027.8	&0.1	&1163.3	&0.1	&1033.2	&0.1	&1162.8	&0.2	&1028	&0.1\\
2008-06-03	&1174.4	&0.1	&1028.9	&0.1	&1167.1	&0.1	&1034.2	&0.1	&1166.2	&0.1	&1028.4	&0.1\\
2011-05-25	&978.62	&0.1	&808.46	&0.07	&971.39	&0.08	&813.73	&0.07	&970.59	&0.1	&808.14	&0.08\\
2012-05-16	&979.97	&0.07	&745.73	&0.04	&972.58	&0.05	&750.91	&0.03	&971.4	&0.1	&745.04	&0.09\\
2013-05-09	&983.08	&0.08	&744.71	&0.05	&975.51	&0.05	&750.03	&0.04	&974.07	&0.06	&743.83	&0.04\\
2016-03-23	&960.02	&0.11	&788.97	&0.06	&952.31	&0.06	&794.27	&0.05	&951.13	&0.16	&788.2	&0.11\\
2018-04-22	&965.78	&0.07	&728.52	&0.05	&957.58	&0.07	&733.43	&0.09	&956.17	&0.05	&728.13	&0.05\\
2018-04-24	&968.05	&0.07	&724.28	&0.04	&959.85	&0.04	&729.56	&0.04	&958.42	&0.08	&723.63	&0.06\\
\hline
\end{tabular}
\caption{Relative pixel positions of the dusty sources extracted from the NACO observations between 2002 and 2018.}
\label{tab:pos2}
\end{table*}

\begin{table*}[htbp!]
\centering
\setlength{\tabcolsep}{0.9pt}
\begin{tabular}{|c|cccccccccccc|}
\hline\hline
Epoch    & $\zeta$ R.A. & $\zeta$ err R.A. & $\zeta$ Dec. & $\zeta$ err Dec. & $\eta$ R.A. & $\eta$ err R.A. & $\eta$ Dec. & $\eta$ err Dec. & $\vartheta$ R.A. &  $\vartheta$ err R.A. & $\vartheta$ Dec. &  $\vartheta$ err Dec.\\ 
\hline
2002-08-29	&803.01	&0.09	&733.65	&0.06	&801.27	&0.03	&739.99	&0.03	&786.13	&0.09	&731.19	&0.06\\
2002-08-30	&1238.8	&0.1	&756.91	&0.08	&1237	&0.04	&763.2	&0.03	&1222.4	&0.1	&754.51	&0.05\\
2003-05-10	&991.94	&0.09	&710.72	&0.06	&990.16	&0.04	&716.98	&0.05	&974.93	&0.1	&708.16	&0.07\\
2004-04-25	&795.87	&0.12	&605.23	&0.09	&793.97	&0.04	&611.38	&0.04	&778.81	&0.1	&602.59	&0.07\\
2004-04-26	&795.66	&0.1	&605.28	&0.05	&793.92	&0.04	&611.47	&0.04	&778.52	&0.09	&602.73	&0.06\\
2005-05-13	&906.12	&0.09	&741.69	&0.08	&903.98	&0.05	&747.74	&0.06	&888.56	&0.1	&738.99	&0.07\\
2006-05-29	&861.22	&0.12	&712.07	&0.1	&859.04	&0.07	&718.03	&0.09	&843.51	&0.08	&709.32	&0.07\\
2007-04-01	&1160.3	&0.06	&1014.9	&0.04	&1158.1	&0.05	&1020.9	&0.06	&1143.1	&0.15	&1012.4	&0.09\\
2007-05-15	&1157.2	&0.12	&1016	&0.07	&1155	&0.05	&1021.8	&0.06	&1139.5	&0.14	&1013.5	&0.08\\
2007-05-16	&1128.7	&0.05	&1038.5	&0.04	&1126.4	&0.03	&1044.2	&0.04	&1111.1	&0.09	&1035.9	&0.06\\
2007-05-17	&1156	&0.07	&1025.2	&0.04	&1153.7	&0.048	&1030.9	&0.1	&1138.6	&0.13	&1022.6	&0.09\\
2007-05-18	&1167.3	&0.12	&1015.2	&0.1	&1165.1	&0.08	&1021	&0.1	&1149.7	&0.11	&1012.6	&0.08\\
2007-05-19	&1088.9	&0.12	&1006.8	&0.12	&1086.8	&0.06	&1012.5	&0.08	&1070.4	&0.11	&1004.3	&0.1\\
2007-05-22	&925.42	&0.17	&776.96	&0.1	&923.49	&0.06	&782.63	&0.06	&907.89	&0.13	&774.07	&0.11\\
2007-05-23	&1174	&0.24	&1047.1	&0.23	&1171.8	&0.08	&1052.6	&0.1	&1157.1	&0.11	&1044.3	&0.06\\
2008-05-26	&1136.4	&0.2	&1074.4	&0.1	&1134.4	&0.1	&1079.9	&0.1	&1119	&0.16	&1071.5	&0.1\\
2008-05-30	&1104.6	&0.2	&1064.5	&0.1	&1102.5	&0.1	&1070	&0.1	&1087	&0.11	&1061.7	&0.09\\
2008-05-31	&1165.5	&0.1	&1034.8	&0.1	&1163.4	&0.1	&1039.8	&0.1	&1148.6	&0.12	&1031.8	&0.04\\
2008-06-02	&1173.2	&0.2	&1034.7	&0.1	&1171.1	&0.1	&1040.3	&0.1	&1155.9	&0.12	&1032	&0.07\\
2008-06-03	&1176.8	&0.3	&1035.6	&0.1	&1174.7	&0.1	&1041.1	&0.1	&1159.4	&0.12	&1032.8	&0.08\\
2011-05-25	&981.74	&0.07	&815.22	&0.05	&979.19	&0.1	&820.33	&0.2	&964.22	&0.15	&812.44	&0.06\\
2012-05-16	&983.01	&0.14	&752.27	&0.11	&980.26	&0.05	&757.52	&0.07	&964.18	&0.12	&749.43	&0.07\\
2013-05-09	&986.26	&0.13	&751.34	&0.08	&983.42	&0.05	&756.57	&0.06	&966.74	&0.09	&748.38	&0.09\\
2016-03-23	&963.58	&0.08	&795.97	&0.05	&960.2	&0.06	&800.61	&0.07	&943.94	&0.13	&792.8	&0.12\\
2018-04-22	&968.33	&0.07	&735.81	&0.06	&965.51	&0.1	&740.08	&0.36	&948.5	&0.08	&732.21	&0.09\\
2018-04-24	&971	&0.03	&731.33	&0.03	&967.68	&0.05	&735.7	&0.07	&950.88	&0.12	&728.01	&0.08\\
\hline
\end{tabular}
\caption{Relative pixel positions of the bright dusty sources observed with NACO.}
\label{tab:pos3}
\end{table*}

\begin{table*}
\centering
\setlength{\tabcolsep}{0.5pt}
\begin{tabular}{|c|cccccccccccc|}
\hline\hline
Epoch    & $\iota$ R.A. & $\iota$ err R.A. & $\iota$ Dec. & $\iota$ err Dec. & DS1 R.A. & DS1 err R.A. & DS1 Dec. & DS1 err Dec. & DS2 R.A. & DS2 err R.A. & DS2 Dec.  &  DS2 err Dec.\\ 
\hline
2002-08-29	&799	&2	    &718	&2	    &789.83	&0.0238	&753.91	&0.0204	&795	&0.5	&757	&0.5\\
2002-08-30	&1234	&2	    &742	&2   	&1225.6	&0.0356	&777.13	&0.0205	&1230.9	&0.148	&780.14	&0.538\\
2003-05-10	&987.93	&0.21	&695.36	&0.14	&978.63	&0.0343	&730.78	&0.0319	&983.96	&0.196	&735.33	&0.392\\
2004-04-25	&791.77	&0.25	&589.55	&0.16	&782.56	&0.0382	&625.02	&0.0357	&787.93	&0.123	&629.95	&0.332\\
2004-04-26	&791.93	&0.47	&589.48	&0.21	&782.55	&0.0239	&625.1	&0.0238	&788.13	&0.102	&629.32	&0.304\\
2005-05-13	&902.22	&0.3	&725.79	&0.24	&892.69	&0.0329	&761.31	&0.0318	&897.99	&0.882	&765.92	&0.628\\
2006-05-29	&856.87	&0.65	&696.53	&0.22	&847.72	&0.0245	&731.44	&0.0247	&853.87	&0.481	&735.49	&0.536\\
2007-04-01	&1156.6	&0.81	&998.79	&0.24	&1146.9	&0.0285	&1034.1	&0.0262	&1153	&1	    &1039	&1\\
2007-05-15	&1152.7	&0.21	&1000.4	&0.12	&1143.7	&0.0278	&1035.1	&0.0231	&1149.5	&0.675	&1040.6	&0.388\\
2007-05-16	&1123.6	&0.25	&1023.4	&0.15	&1115.2	&0.0305	&1057.6	&0.0294	&1121.2	&0.203	&1062.3	&0.255\\
2007-05-17	&1151.3	&0.52	&1010.5	&0.44	&1142.6	&0.0365	&1044.3	&0.0348	&1148.6	&0.231	&1049.2	&0.248\\
2007-05-18	&1162.8	&0.36	&1000.1	&0.26	&1153.9	&0.0396	&1034.3	&0.0488	&1160	&0.141	&1038.9	&0.233\\
2007-05-19	&1084.5	&0.53	&991.75	&0.22	&1075.6	&0.0702	&1025.9	&0.0615	&1084	&1	    &1030	&1\\
2007-05-22	&921.52	&0.48	&760.89	&0.2	&912.37	&0.0359	&795.98	&0.0404	&918.13	&0.127	&801.47	&0.237\\
2007-05-23	&1169.8	&0.3	&1031.4	&0.14	&1160.7	&0.0383	&1066	&0.0346	&1167	&0.0871	&1071.4	&0.182\\
2008-05-26	&1132.4	&0.465	&1058	&0.22	&1123.2	&0.0221	&1093	&0.0222	&1130.2	&0.2	&1098.5	&0.2\\
2008-05-30	&1100.3	&0.35	&1048.5	&0.146	&1091.4	&0.0197	&1083.2	&0.0179	&1096.7	&0.254	&1089.7	&0.375\\
2008-05-31	&1161	&1	    &1018	&1	    &1152.5	&0.0378	&1053.2	&0.0333	&1158.2	&0.472	&1058.9	&0.266\\
2008-06-02	&1169	&0.25	&1018.7	&0.15	&1160.2	&0.029	&1053.5	&0.0257	&1165.5	&0.481	&1058.7	&0.262\\
2008-06-03	&1173	&0.13	&1019.2	&0.09	&1163.6	&0.0191	&1054.4	&0.0187	&1169.4	&0.231	&1160.7	&0.348\\
2011-05-25	&977.02	&0.34	&798.65	&0.72	&968.29	&0.0306	&833.15	&0.027	&973.42	&0.148	&839.71	&0.237\\
2012-05-16	&978.49	&0.3	&735.83	&0.15	&969.27	&0.0197	&770.03	&0.012	&974.07	&0.235	&776.99	&0.225\\
2013-05-09	&981.34	&0.15	&734.86	&0.1	&972.36	&0.0188	&768.86	&0.018	&976.74	&0.119	&777.78	&0.0601\\
2016-03-23	&958.67	&0.3	&778.97	&0.2	&949.45	&0.0165	&812.33	&0.0128	&953.76	&0.181	&820.78	&0.275\\
2018-04-22	&963.45	&0.08	&718.82	&0.07	&954.77	&0.0235	&751.35	&0.0287	&959.57	&0.295	&760.04	&0.247\\
2018-04-24	&965.84	&0.1	&714.21	&0.057	&956.77	&0.0292	&747.19	&0.0254	&961.3	&0.155	&755.96	&0.331\\
\hline
\end{tabular}
\caption{Dusty sources DS1-DS2 observed with NACO.}
\label{tab:pos4}
\end{table*}

\begin{table*}
\centering
\setlength{\tabcolsep}{1pt}
\begin{tabular}{|c|cccccccccccc|}
\hline\hline
Epoch    & \multicolumn{4}{c|}{DS3} & \multicolumn{4}{c|}{DS4} & \multicolumn{4}{c|}{DS5}\\ 
\hline
    & R.A. & err R.A. &  Dec. &  err Dec. &  R.A. &  err R.A. &  Dec. &  err Dec. &  R.A. &  err R.A. &  Dec. &  err Dec.\\ 
\hline
2002-08-29	&799	&0.5	&755	&0.5	&786	&0.5	&746	&0.5	&795	&0.5	&749	&0.5\\
2002-08-30	&1234.6	&0.0727	&778.65	&0.0376	&1222.2	&0.39	&769.76	&0.24	&1231	&0.2	&772.64	&0.249\\
2003-05-10	&988.54	&0.145	&723.11	&0.0782	&975.07	&0.64	&723.28	&0.431	&984.09	&0.0576	&725.65	&0.0811\\
2004-04-25	&792.24	&0.0762	&626.57	&0.0397	&778.86	&0.196	&617.58	&0.109	&787.84	&0.165	&620.1	&0.32\\
2004-04-26	&792.84	&0.0743	&626.51	&0.0588	&778.57	&0.248	&617.72	&0.088	&787.78	&0.314	&620.6	&0.479\\
2005-05-13	&902	&0.112	&763.41	&0.084	&888.09	&0.417	&754.14	&0.334	&897.58	&0.366	&756.33	&0.45\\
2006-05-29	&858.13	&0.155	&733.36	&0.117	&844.18	&0.501	&724.17	&0.158	&853.01	&0.181	&726.97	&0.246\\
2007-04-01	&1156	&1	    &1037	&1	    &1142.1	&0.198	&1026.7	&0.0923	&1152	&0.388	&1030.8	&0.567\\
2007-05-15	&1153.9	&0.235	&1037.5	&0.074	&1140.4	&0.339	&1027.9	&0.309	&1149.2	&0.157	&1032.1	&0.39\\
2007-05-16	&1125.3	&0.214	&1059.5	&0.325	&1110.5	&0.325	&1050.2	&0.201	&1120	&0.5	&1054	&0.5\\
2007-05-17	&1152.5	&0.159	&1046.3	&0.463	&1137.5	&0.228	&1036.9	&0.0974	&1147.5	&0.239	&1040.2	&0.15\\
2007-05-18	&1164.4	&0.142	&1036.2	&0.375	&1148.4	&0.363	&1027.1	&0.208	&1159	&0.5	&1031	&0.5\\
2007-05-19	&1086	&0.5	&1028	&0.5	&1072	&1	    &1018	&1	    &1080	&1	    &1021	&1\\
2007-05-22	&922.28	&0.0658	&798.13	&0.0572	&907.78	&0.123	&788.53	&0.0738	&918	&0.5	&793	&0.5\\
2007-05-23	&1170.4	&0.188	&1068.4	&0.0806	&1156	&0.124	&1058.7	&0.0698	&1165.3	&0.363	&1062.2	&0.185\\
2008-05-26	&1133	&1	    &1097	&1	    &1118.9	&0.2	&1085.8	&0.08	&1128.5	&0.07	&1089	&0.11\\
2008-05-30	&1100.6	&0.0907	&1086.3	&0.0893	&1086.9	&0.115	&1076	&0.0979	&1096.2	&0.275	&1079.6	&0.168\\
2008-05-31	&1163.3	&0.234	&1055.1	&0.0875	&1148.3	&0.415	&1046	&0.0983	&1157.9	&0.127	&1049.4	&0.281\\
2008-06-02	&1169.7	&0.216	&1057.1	&0.931	&1154.7	&0.212	&1046.3	&0.0859	&1165.4	&0.356	&1051.1	&0.752\\
2008-06-03	&1173.7	&0.104	&1057	&0.0659	&1159.2	&0.144	&1047.1	&0.0724	&1168.4	&0.514	&1049.8	&0.352\\
2011-05-25	&977.7	&0.225	&837.07	&0.419	&962.98	&0.194	&826.46	&0.125	&972.83	&0.315	&830.35	&0.448\\
2012-05-16	&978.41	&0.0598	&774.5	&0.059	&964.72	&0.0619	&763.64	&0.0698	&973.8	&0.387	&766.15	&0.42\\
2013-05-09	&981.05	&0.207	&773.85	&0.211	&967.74	&0.22	&762.16	&0.134	&977.75	&0.406	&766.15	&0.804\\
2016-03-23	&957.71	&0.0873	&818.43	&0.0413	&944.44	&0.239	&805.76	&0.152	&-   	&-	    &-	    &-\\
2018-04-22	&962.9	&0.0404	&757.89	&0.0258	&948.69	&0.815	&745.21	&0.808	&-   	&-	    &-	    &-\\
2018-04-24	&965.33	&0.33	&753.46	&0.386	&951.7	&0.193	&740.42	&0.186	&960	&1	    &747	&1\\
\hline
\end{tabular}
\caption{Dusty sources DS3-DS5 observed with NACO.}
\label{tab:pos5}
\end{table*}

\begin{table*}
\centering
\setlength{\tabcolsep}{1pt}
\begin{tabular}{|c|cccccccccccc|}
\hline\hline
Epoch    & \multicolumn{4}{c|}{DS6} & \multicolumn{4}{c|}{DS7} & \multicolumn{4}{c|}{DS8}\\ 
\hline
    & R.A. & err R.A. &  Dec. &  err Dec. &  R.A. &  err R.A. &  Dec. &  err Dec. &  R.A. &  err R.A. &  Dec. &  err Dec.\\ 
\hline
2002-08-29	&804.21	&0.122	&748.7	&0.332	&805	&0.5	&745	&0.5	&812.61	&0.08	&738.53	&0.136\\
2002-08-30	&1240.2	&0.0891	&771.37	&0.221	&1241	&0.5	&768	&0.5	&1247.7	&0.535	&761.85	&0.621\\
2003-05-10	&992.89	&0.128	&725.53	&0.138	&994.67	&0.168	&721.76	&0.135	&994.98	&0.0835	&721.57	&0.115\\
2004-04-25	&796.74	&0.323	&619.99	&0.328	&798.56	&0.333	&616.42	&0.335	&806.32	&0.193	&613.13	&0.271\\
2004-04-26	&797	&0.5	&620	&0.5	&798	&0.5	&617	&0.5	&806.51	&0.129	&612.94	&0.156\\
2005-05-13	&907.2	&0.0847	&756.52	&0.0934	&909.03	&0.116	&752.49	&0.136	&916.92	&0.107	&749.95	&0.134\\
2006-05-29	&861.78	&0.213	&728.96	&0.239	&863.53	&0.765	&724.09	&0.339	&871.96	&0.166	&720.41	&0.21\\
2007-04-01	&1161.6	&0.2	&1030.4	&0.984	&1163	&0.148	&1026.1	&0.101	&1171.1	&0.0548	&1023.2	&0.0627\\
2007-05-15	&1158.7	&0.34	&1031.7	&17899	&1159.4	&0.157	&1027.1	&0.0619	&1168.1	&0.0398	&1024.5	&0.0427\\
2007-05-16	&1130	&0.5	&1054	&0.5	&1131	&0.5	&1051	&0.5	&1140	&0.123	&1046.9	&0.131\\
2007-05-17	&1157.4	&0.125	&1040.1	&0.515	&1158.4	&0.158	&1036.4	&0.119	&1167.3	&0.0408	&1033.7	&0.0578\\
2007-05-18	&1168.1	&0.144	&1032.6	&0.168	&1169.6	&0.699	&1027.9	&0.298	&1179	&0.118	&1023.8	&0.12\\
2007-05-19	&1090	&0.764	&1023.2	&0.69	&1090.4	&0.0898	&1019.3	&0.124	&	-	&-    &   -   &-\\
2007-05-22	&927	&0.5	&792	&0.5	&928	&1	    &790	&1	    &936.99	&0.0386	&785.18	&0.0479\\
2007-05-23	&1175.1	&0.25	&1062.1	&0.25	&1176.1	&0.251	&1058.5	&0.288	&1185.3	&0.0327	&1055.2	&0.0447\\
2008-05-26	&1134.4	&0.0414	&1080.2	&0.04	&1136.2	&0.2	&1074.5	&0.07	&1147.6	&0.05	&1082.5	&0.07\\
2008-05-30	&1106.1	&0.215	&1079.8	&0.571	&1107.3	&0.0848	&1076	&0.0642	&1116	&0.0446	&1072.4	&0.166\\
2008-05-31	&1167.2	&0.201	&1049.4	&0.29	&1168.4	&0.0693	&1045.9	&0.0588	&1176.5	&0.176	&1043.2	&0.107\\
2008-06-02	&1175	&1	    &1050	&1   	&1176	&1	    &1047	&1	    &1184.6	&0.0547	&1043.2	&0.0502\\
2008-06-03	&1178.3	&0.122	&1050.4	&0.47	&1179.6	&0.0842	&1046.8	&0.158	&1188.3	&0.052	&1043.9	&0.0551\\
2011-05-25	&983.31	&0.0809	&830.56	&0.238	&984.15	&0.105	&826.74	&0.289	&993.23	&0.0803	&824.08	&0.115\\
2012-05-16	&983.65	&0.864	&769	&1	    &985.23	&0.174	&764.03	&0.274	&994.96	&0.0706	&761.31	&0.0853\\
2013-05-09	&987.23	&0.376	&768	&1	    &997.67	&0.0474	&732.72	&0.0755	&994.38	&0.0732	&751.09	&0.0947\\
2016-03-23	&964.59	&0.162	&811.7	&0.22	&965.77	&0.559	&808.1	&0.266	&974.83	&0.211	&805.25	&0.193\\
2018-04-22	&970	&1	    &752	&1	    &971.13	&0.076	&748.82	&0.134	&981	&1	    &745	&1\\
2018-04-24	&972.71	&0.278	&747.36	&0.484	&973.13	&0.921	&744.03	&0.77	&980.42	&0.0492	&741.86	&0.133\\
\hline
\end{tabular}
\caption{Dusty sources DS6-DS8 observed with NACO.}
\label{tab:pos6}
\end{table*}

\begin{table*}
\centering
\setlength{\tabcolsep}{1pt}
\begin{tabular}{|c|cccccccccccc|}
\hline\hline
Epoch    & \multicolumn{4}{c|}{DS9} & \multicolumn{4}{c|}{DS10} & \multicolumn{4}{c|}{DS11}\\ 
\hline
    & R.A. & err R.A. &  Dec. &  err Dec. &  R.A. &  err R.A. &  Dec. &  err Dec. &  R.A. &  err R.A. &  Dec. &  err Dec.\\ 
\hline
2002-08-29	&819.57	&0.102	&747.94	&0.0753	&828.36	&0.036	&749.57	&0.036	&772.44	&0.0615	&731.23	&0.023\\
2002-08-30	&1255.3	&0.0857	&771.47	&0.0882	&1264.3	&0.0153	&773.08	&0.0141	&1208.2	&0.0402	&754.51	&0.0119\\
2003-05-10	&1008.3	&0.254	&725.37	&0.487	&1017.1	&0.0386	&726.26	&0.0362	&961.09	&0.0171	&708.15	&0.018\\
2004-04-25	&812.07	&0.11	&619.2	&0.139	&820.87	&0.047	&620.17	&0.0446	&764.93	&0.0364	&602.42	&0.0362\\
2004-04-26	&812.06	&0.103	&619.42	&0.115	&820.88	&0.0466	&620.23	&0.0433	&764.96	&0.0207	&602.49	&0.0194\\
2005-05-13	&921.92	&0.067	&755.87	&0.068	&930.68	&0.0483	&756.29	&0.038	&875.12	&0.0346	&738.58	&0.0219\\
2006-05-29	&877.02	&0.088	&725.63	&0.087	&885.65	&0.0353	&726.07	&0.0355	&830.01	&0.0375	&708.81	&0.0371\\
2007-04-01	&1176.1	&0.0599	&1028.1	&0.0811	&1184.7	&0.0347	&1028.6	&0.0339	&1129.3	&0.0312	&1011.6	&0.0234\\
2007-05-15	&1173.2	&0.0924	&1028.5	&0.217	&1182.2	&0.0555	&1029.6	&0.0408	&1126	&0.0619	&1012.4	&0.0231\\
2007-05-16	&1144	&0.0903	&1051.1	&0.0603	&1153	&0.0608	&1051.8	&0.0718	&1097.3	&0.0452	&1035.1	&0.0263\\
2007-05-17	&1171.6	&0.0442	&1038.3	&0.0411	&1180.3	&0.018	&1038.8	&0.02	&1124.7	&0.0186	&1021.8	&0.00854\\
2007-05-18	&1183	&0.0931	&1028.1	&0.087	&1191.7	&0.0183	&1028.7	&0.0172	&1136.2	&0.0194	&1011.8	&0.0114\\
2007-05-19	&-		&-      & -     &  -    &   -   &    -  &     - &       &  -    &   -   &    -  &     -  \\								
2007-05-22	&941.33	&0.0916	&789.78	&0.0719	&950.02	&0.0209	&790.41	&0.0215	&894.57	&0.0232	&773.37	&0.0115\\
2007-05-23	&1189.8	&0.062	&1059.7	&0.118	&1198.3	&0.0168	&1060.4	&0.0155	&1143	&0.0259	&1043.4	&0.0191\\
2008-05-26	&1152.4	&0.29	&1087	&0.297	&1160.7	&0.04	&1086.9	&0.04	&1105.2	&0.0244	&1070.6	&0.024\\
2008-05-30	&1120.6	&0.186	&1077.2	&0.184	&1128.9	&0.0204	&1077.1	&0.0202	&1073.5	&0.0518	&1060.8	&0.0644\\
2008-05-31	&1180.6	&45261	&1047.1	&0.231	&1190.1	&0.0642	&1047.3	&0.0362	&1134.5	&0.0784	&1030.8	&0.0309\\
2008-06-02	&1189	&0.488	&1047.1	&0.294	&1198.1	&0.0545	&1047.5	&0.0337	&1142.1	&0.0383	&1031	&0.0108\\
2008-06-03	&1192.7	&0.141	&1048.2	&0.147	&1201.2	&0.0271	&1048.3	&0.0263	&1145.6	&0.0373	&1031.9	&0.0173\\
2011-05-25	&997.37	&0.181	&826.41	&0.291	&1004.9	&0.124	&826.06	&0.114	&950.12	&0.0364	&810.77	&0.0112\\
2012-05-16	&-	    &-     	&-	    &-	    &1005.7	&0.0337	&762.54	&0.066	&951.12	&0.0441	&747.82	&0.0176\\
2013-05-09	&1008.4	&0.0761	&760.85	&0.106	&1008.3	&0.0415	&760.71	&0.049	&954.01	&0.0662	&746.58	&0.0246\\
2016-03-23	&977.83	&0.235	&806.76	&0.188	&984	&1	    &803	&1	    &930.94	&0.108	&790.04	&0.0606\\
2018-04-22	&982	&1	    &746	&1	    &988.41	&0.392	&743	&0.737	&936.61	&0.698	&729.79	&0.559\\
2018-04-24	&983.32	&1	    &740.96	&0.174	&984.41	&0.355	&741.18	&0.144	&938.45	&0.214	&724.95	&0.059\\
\hline
\end{tabular}
\caption{Dusty sources DS9-DS11 observed with NACO.}
\label{tab:pos7}
\end{table*}

\begin{table*}
\centering
\setlength{\tabcolsep}{1pt}
\begin{tabular}{|c|cccccccccccc|}
\hline\hline
Epoch    & \multicolumn{4}{c|}{DS12} & \multicolumn{4}{c|}{DS13} & \multicolumn{4}{c|}{DS14}\\ 
\hline
    & R.A. & err R.A. &  Dec. &  err Dec. &  R.A. &  err R.A. &  Dec. &  err Dec. &  R.A. &  err R.A. &  Dec. &  err Dec.\\ 
\hline
2002-08-29	&778.39	&0.0164	&733.27	&0.0157	&819.89	&0.0795	&737.12	&0.0646	&820.37	&0.0745	&732.08	&0.035\\
2002-08-30	&1214.2	&0.0204	&756.54	&0.0141	&1255	&0.542	&761.28	&0.352	&1256.2	&0.0897	&755.17	&0.103\\
2003-05-10	&967.2	&0.0317	&710.15	&0.0308	&1008.5	&0.0983	&714.64	&0.118	&1009.8	&0.106	&709.02	&0.0552\\
2004-04-25	&770.9	&0.0618	&604.19	&0.0603	&812.4	&0.0786	&609.41	&0.0854	&814.2	&0.0982	&603.68	&0.0748\\
2004-04-26	&770.77	&0.0226	&604.26	&0.0214	&812.83	&0.173	&609	&0.251	&813.83	&0.091	&603.75	&0.0619\\
2005-05-13	&880.82	&0.0318	&740.16	&0.0304	&923.13	&0.0847	&745.92	&0.0698	&924.83	&0.19	&740.44	&0.304\\
2006-05-29	&835.58	&0.03	&710.14	&0.0282	&878.05	&0.0644	&716.71	&0.075	&880.13	&0.0905	&710.83	&0.0688\\
2007-04-01	&1134.4	&0.0324	&1012.6	&0.0227	&1177.5	&0.0228	&1019.4	&0.032	&1179.4	&0.0527	&1013.7	&0.0454\\
2007-05-15	&1131.4	&0.0522	&1013.6	&0.0243	&1174.2	&0.109	&1020.2	&0.0821	&1176.4	&0.145	&1014.4	&0.296\\
2007-05-16	&1102.5	&0.0377	&1036	&0.0239	&1145.6	&0.113	&1043.1	&0.118	&1147.8	&0.0696	&1037.6	&0.0669\\
2007-05-17	&1130.2	&0.0826	&1022.8	&0.0333	&1173.2	&0.0451	&1029.7	&0.0422	&1175.5	&0.419	&1023.9	&0.14\\
2007-05-18	&1141.3	&0.0611	&1012.8	&0.0258	&1184.6	&0.048	&1019.6	&0.0456	&1186.4	&0.198	&1014.3	&0.161\\
2007-05-19	&	-	&	-	&	-	&-		&	-	&-	&   -   &  -   &       -    &    -  &     -   &- \\
2007-05-22	&899.67	&0.0245	&774.34	&0.0212	&943.27	&0.0582	&781.19	&0.0655	&945.31	&45047	&775.97	&0.229\\
2007-05-23	&1147.9	&0.0323	&1044.3	&0.0274	&1191.4	&0.0907	&1051.2	&0.102	&1192.8	&0.57	&1045.8	&0.36\\
2008-05-26	&1110.6	&0.025	&1071.4	&0.024	&1154.3	&0.089	&1078.6	&0.09	&1156.1	&0.06	&1072.7	&0.06\\
2008-05-30	&1078.6	&0.0395	&1061.5	&0.0237	&1122.6	&0.0715	&1068.7	&0.0719	&1124.3	&0.0484	&1062.9	&0.0411\\
2008-05-31	&1139.2	&0.0223	&1031.4	&0.014	&1183.9	&0.093	&1038.6	&0.124	&1185.9	&0.0352	&1033.7	&0.045\\
2008-06-02	&1147	&0.0764	&1031.7	&0.0236	&1191.2	&0.0631	&1039.2	&0.0753	&1192.9	&0.0489	&1033.4	&0.0477\\
2008-06-03	&1150.7	&0.0221	&1032.6	&0.0283	&1194.8	&0.158	&1039.7	&0.185	&1196.4	&0.0975	&1034	&0.0564\\
2011-05-25	&954.56	&0.0407	&810.77	&0.0395	&999.97	&0.0261	&820.07	&0.0278	&1002	&0.466	&814.64	&0.237\\
2012-05-16	&955.18	&0.0535	&747.54	&0.0862	&1001.8	&0.117	&757.12	&0.103	&1003.4	&0.0804	&751.12	&0.118\\
2013-05-09	&958.17	&0.0493	&746.09	&0.052	&1004.8	&0.0492	&756.16	&0.0371	&1007.2	&0.0803	&750.21	&0.166\\
2016-03-23	&935.18	&0.0916	&788.87	&0.0446	&981.95	&0.59	&800.84	&0.806	&985.3	&0.107	&795.28	&0.0828\\
2018-04-22	&939.86	&0.591	&726.99	&0.601	&988.94	&0.015	&741.21	&0.0189	&991.5	&0.047	&734.61	&0.082\\
2018-04-24	&942.11	&0.095	&723.05	&0.058	&990.97	&0.0481	&737.03	&0.0252	&993.38	&0.022	&730.18	&0.02\\
\hline
\end{tabular}
\caption{Dusty sources DS12-DS14 observed with NACO.}
\label{tab:pos8}
\end{table*}

\begin{table*}
\centering
\setlength{\tabcolsep}{1pt}
\begin{tabular}{|c|cccccccccccc|}
\hline\hline
Epoch    & \multicolumn{4}{c|}{DS15} & \multicolumn{4}{c|}{DS16} & \multicolumn{4}{c|}{DS17}\\ 
\hline
    & R.A. & err R.A. &  Dec. &  err Dec. &  R.A. &  err R.A. &  Dec. &  err Dec. &  R.A. &  err R.A. &  Dec. &  err Dec.\\ 
\hline
2002-08-29	&- &- &- &-	    &831.09	&0.0448	&735.79	&0.0558	&838.69	&0.0737	&729.72	&0.0619\\
2002-08-30	&- &- &- &-	    &1266.8	&0.0254	&759.13	&0.0205	&1274.1	&0.0822	&753.06	&0.257\\
2003-05-10	&- &- &- &-	    &1019.8	&0.0523	&712.91	&0.055	&1027.4	&0.0287	&706.69	&0.0225\\
2004-04-25	&810.08	&0.279	&602.35	&0.281	&823.42	&0.0815	&607.19	&0.081	&830.81	&0.0639	&601.06	&0.0476\\
2004-04-26	&- &- &- &-		    &823.81	&0.0652	&606.89	&0.0579	&830.81	&0.0947	&601.16	&0.178\\
2005-05-13	&921	&1	    &739  	&1	    &934.02	&0.073	&743	&0.133	&941.49	&0.0541	&737.59	&0.0497\\
2006-05-29	&875.92	&0.196	&708.92	&0.116	&888.65	&0.0371	&713.36	&0.0637	&896.01	&0.108	&707.86	&0.0866\\
2007-04-01	&1175	&0.0569	&1011.6	&0.077	&1187.7	&0.0345	&1016	&0.033	&1194.8	&0.277	&1010.6	&0.444\\
2007-05-15	&1172.3	&0.0565	&1012.6	&0.0747	&1184.7	&0.0283	&1016.6	&0.0432	&1191.9	&0.061	&1011.7	&0.178\\
2007-05-16	&- &- &- &-	  	&1156.4	&0.0783	&1038.8	&0.0623	&1163.5	&0.155	&1034.6	&0.274\\
2007-05-17	&1171.8	&0.0745	&1022.9	&0.059	&1183.5	&0.0171	&1025.6	&0.0282	&1190.6	&0.123	&1021.4	&0.278\\
2007-05-18	&1182.3	&0.172	&1012.2	&0.115	&1194.9	&0.0239	&1016.1	&0.0233	&1202	&0.268	&1011	&0.159\\
2007-05-19	&-		&	-	&	-	&	-	&	-	&	-&     -     &     -   &   -    &  -     &  -     &\\
2007-05-22	&941.45	&0.0965	&774.18	&0.061	&953.41	&0.0221	&777.85	&0.0309	&960.39	&0.078	&772.62	&0.175\\
2007-05-23	&1191	&1	    &1044	&1	    &1201.5	&0.0165	&1047.5	&0.0302	&1208.9	&0.0646	&1043.3	&0.251\\
2008-05-26	&1151.4	&0.13	&1070.6	&0.162	&1164.2	&0.134	&1074.6	&0.202	&1171.3	&0.0893	&1069.5	&0.092\\
2008-05-30	&1119.8	&0.206	&1060.8	&0.485	&1132.3	&0.0133	&1064.8	&0.0182	&1139.7	&0.0265	&1059.6	&0.0241\\
2008-05-31	&-	    &-	    &-	    &-	    &1193.6	&0.12	&1034.1	&0.0957	&1200.6	&0.151	&1029.6	&0.135\\
2008-06-02	&1191	&1	    &1032	&1	    &1200.3	&0.0544	&1035.6	&0.0484	&1208	&0.184	&1030.1	&0.185\\
2008-06-03	&1191.7	&0.522	&1031.7	&0.248	&1204.4	&0.0439	&1036.1	&0.0668	&1211.7	&0.116	&1031.7	&0.132\\
2011-05-25	&-    &-	    &-    &-	    &1008.9	&0.0563	&814.4	&0.0856	&1016	&0.232	&810.04	&0.13\\
2012-05-16	&999.73	&0.0372	&749.53	&0.0256	&1010	&0.0557	&751.7	&0.0511	&1017.2	&0.265	&747.16	&0.136\\
2013-05-09	&1002.9	&0.398	&748.4	&0.363	&1013.7	&0.0517	&750.74	&0.0649	&1020.5	&0.0227	&745.97	&0.0193\\
2016-03-23	&981.45	&0.21	&793.37	&0.158	&989.71	&0.228	&794.81	&0.421	&997.8	&0.0261	&790.36	&0.0236\\
2018-04-22	&-    &-    	&-	    &-	    &995.56	&1	    &734.33	&0.34	&1003.1	&0.021	&729.85	&0.03\\
2018-04-24	&988.89	&0.101	&728.67	&0.145	&-      &-   	&-	    &-	    &1005.4	&0.021	&725.45	&0.023\\
\hline
\end{tabular}
\caption{Dusty sources DS15-DS17 observed with NACO.}
\label{tab:pos9}
\end{table*}

\begin{table*}
\centering
\setlength{\tabcolsep}{1pt}
\begin{tabular}{|c|cccccccccccc|}
\hline\hline
Epoch    & \multicolumn{4}{c|}{DS18} & \multicolumn{4}{c|}{DS19} & \multicolumn{4}{c|}{DS20}\\ 
\hline
    & R.A. & err R.A. &  Dec. &  err Dec. &  R.A. &  err R.A. &  Dec. &  err Dec. &  R.A. &  err R.A. &  Dec. &  err Dec.\\ 
\hline
2002-08-29	&844.88	&42005	&733.7	&0.823	&0      &0	&0	    &0	    &835.43	&0.0321	&722.83	&0.042\\
2002-08-30	&1280.1	&0.246	&755.86	&0.612	&1275	&1	    &753	&1	    &1271.3	&0.0282	&746	&0.0303\\
2003-05-10	&1033.3	&0.0362	&709.79	&0.0408	&1029	&1	    &706	&1	    &1024.2	&0.0343	&699.47	&0.0461\\
2004-04-25	&837.46	&0.129	&603.93	&0.382	&835.02	&0.237	&599.78	&0.368	&828.12	&0.0615	&593.73	&0.0868\\
2004-04-26	&836.89	&0.247	&604.02	&0.3	&832	&1	    &600	&1	    &827.71	&0.0695	&593.68	&0.0881\\
2005-05-13	&947.73	&0.0539	&740.28	&0.122	&-	& -	&-   	&  -	&938.22	&0.125	&730.57	&0.107\\
2006-05-29	&902.6	&0.193	&710.4	&0.418	&899.33	&0.252	&706.19	&0.188	&893.45	&0.0872	&700.51	&0.102\\
2007-04-01	&1201.9	&552	&1013.1	&45231	&1198.6	&0.175	&1009.1	&0.219	&1192.2	&0.063	&1003.2	&0.0722\\
2007-05-15	&1198.3	&0.0542	&1013.4	&0.109	&1196.1	&0.44	&1009.2	&0.513	&1189.5	&0.0861	&1004.3	&0.104\\
2007-05-16	&1169.7	&0.144	&1035.6	&0.13	&1166.3	&0.662	&1031.7	&0.213	&1160.9	&0.106	&1026.9	&0.115\\
2007-05-17	&1196.7	&0.0812	&1022.4	&0.165	&1193.8	&0.159	&1019.5	&0.0768	&1187.9	&0.0834	&1013.6	&0.1\\
2007-05-18	&1208.6	&0.455	&1012.7	&0.553	&1205.6	&0.143	&1009.2	&0.106	&1199.5	&0.0517	&1003.5	&0.0506\\
2007-05-19	&	-	&	-	&	-	&	-	&	-	&-	&      -    &   -     &   -    &  -     &   -    &   -       \\
2007-05-22	&966.77	&0.366	&774.83	&0.483	&964.25	&0.0853	&771.08	&0.0695	&958.09	&0.0442	&765.18	&0.0563\\
2007-05-23	&1215.5	&0.403	&1045	&45017	&1212.8	&0.399	&1041.1	&0.18	&1206.2	&0.0299	&1035.2	&0.0497\\
2008-05-26	&1178.2	&0.2	&1071.8	&0.4	&1176	&1	    &1067	&2	    &1168.7	&0.141	&1062	&0.112\\
2008-05-30	&1146.3	&0.582	&1061.1	&0.823	& -     &	-	&-	    &-  	&1136.7	&0.0487	&1052.2	&0.0598\\
2008-05-31	&1208.4	&0.528	&1032.4	&0.81	&1203.9	&0.58	&1026.8	&45079	&1198.2	&0.079	&1022.1	&0.0489\\
2008-06-02	&1216.7	&0.052	&1040.5	&0.0531	&1213.8	&0.0614	&1030.4	&0.109	&1205.5	&0.0934	&1022.3	&0.111\\
2008-06-03	&1218	&1	    &1032.9	&1	    &1215.5	&0.209	&1028.8	&0.104	&1209.2	&0.0432	&1023.2	&0.0478\\
2011-05-25	&1023.1	&0.756	&811.3	&0.221	&1019.8	&0.207	&807.49	&0.504	&1014.2	&0.304	&802.77	&0.104\\
2012-05-16	&1024.7	&0.0532	&748.97	&0.165	&1021.7	&0.104	&745.13	&0.0894	&1014.9	&0.0545	&739.29	&0.0486\\
2013-05-09	&1028	&1	    &747	&1	    &1026	&1	    &742	&1	    &1018.3	&0.0678	&737.92	&0.0563\\
2016-03-23	&1005.4	&0.0491	&792.05	&0.0802	&1002.5	&0.0776	&787.07	&0.117	&996.28	&0.0443	&782.35	&0.0276\\
2018-04-22	&1011.7	&1	    &732.4	&1	    &1008.4	&0.249	&726.21	&0.331	&1001.8	&0.043	&721.75	&0.053\\
2018-04-24	&1013.2	&0.125	&726.97	&0.168	&1010.4	&0.125	&722.29	&0.194	&1004.2	&0.015	&717.51	&0.015\\
\hline
\end{tabular}
\caption{Dusty sources DS18-DS20 observed with NACO.}
\label{tab:pos10}
\end{table*}

\begin{table*}
\centering
\setlength{\tabcolsep}{1pt}
\begin{tabular}{|c|cccccccccccc|}
\hline\hline
Epoch    & \multicolumn{4}{c|}{DS21} & \multicolumn{4}{c|}{DS22} & \multicolumn{4}{c|}{DS23}\\ 
\hline
    & R.A. & err R.A. &  Dec. &  err Dec. &  R.A. &  err R.A. &  Dec. &  err Dec. &  R.A. &  err R.A. &  Dec. &  err Dec.\\ 
\hline
2002-08-29	&826.01	&0.116	&715.61	&0.0884	&811.52	&0.182	&726.4	&0.0594	&780.03	&0.16	&721.22	&0.286\\
2002-08-30	&1262	&0.131	&738.9	&0.13	&1247.5	&0.189	&749.76	&0.0447	&1216.5	&0.36	&744.14	&0.45\\
2003-05-10	&1015.4	&0.14	&692.3	&0.111	&1000.7	&0.155	&703.49	&0.061	&969.52	&0.0683	&698.48	&0.132\\
2004-04-25	&818.38	&0.184	&586.9	&0.106	&804.19	&0.206	&597.79	&0.207	&773.4	&0.0905	&593	&0.107\\
2004-04-26	&819.06	&0.206	&587.05	&0.0998	&804.26	&0.178	&598.14	&0.0704	&772.94	&0.147	&592.75	&0.177\\
2005-05-13	&929.3	&0.203	&723.24	&0.121	&914.48	&0.257	&734.44	&0.141	&883.21	&0.214	&729.35	&0.193\\
2006-05-29	&884.49	&0.267	&693.62	&0.109	&869.5	&0.602	&705.11	&0.645	&838.17	&0.131	&699.84	&0.203\\
2007-04-01	&1184.4	&0.0869	&996.26	&0.0471	&1168.7	&0.275	&1008	&0.0585	&1136.7	&0.133	&1001.8	&0.391\\
2007-05-15	&1181.2	&0.121	&997.32	&0.0656	&1165.7	&0.115	&1008.9	&0.117	&1132.9	&0.256	&1002	&0.58\\
2007-05-16	&1153.5	&0.125	&1019.8	&0.118	&1138	&0.151	&1031.6	&0.0804	&1103.6	&0.277	&1025.1	&0.586\\
2007-05-17	&1180.4	&0.159	&1006.5	&0.0866	&1165.3	&0.171	&1018.2	&0.112	&1132.3	&0.267	&1012.1	&0.82\\
2007-05-18	&1191.8	&0.113	&996.49	&0.0701	&1176.3	&0.156	&1008.1	&0.0456	&1143.4	&0.181	&1001.2	&0.785\\
2007-05-19	&-		&	-	&	-	&	-	&	-	&-	&  -  &        -      &    -   &    -   &  -     &    -     \\      
2007-05-22	&949.79	&0.104	&758.26	&0.117	&934.54	&0.117	&769.99	&0.047	&901.72	&0.369	&763.48	&0.946\\
2007-05-23	&1198.4	&0.0849	&1028.2	&0.0776	&1182.8	&0.101	&1039.7	&0.136	&1150	&1   	&1032	&1\\
2008-05-26	&1160.9	&0.15	&1055.1	&0.08	&1145.2	&0.3	&1066.8	&0.118	&1113.1	&0.142	&1061.6	&0.22\\
2008-05-30	&1129.5	&0.117	&1045.4	&0.0845	&1113.6	&0.124	&1057.1	&0.0951	&1080.6	&0.132	&1051	&0.375\\
2008-05-31	&1190.7	&0.212	&1015.4	&0.0718	&1174.5	&0.168	&1027.7	&0.0941	&1140.6	&0.844	&1019.8	&0.45\\
2008-06-02	&1198.5	&0.0969	&1015.5	&0.0765	&1182.5	&0.355	&1027.6	&0.0862	&1148.6	&0.276	&1020.6	&0.828\\
2008-06-03	&1201.6	&0.109	&1016.5	&0.0673	&1186	&0.101	&1028.3	&0.052	&1152.4	&0.333	&1020.1	&1\\
2011-05-25	&1006.6	&0.0764	&796.04	&0.0765	&991.08	&0.143	&807.95	&0.0719	&956.75	&0.56	&801	&1\\
2012-05-16	&1007.2	&0.222	&733.16	&0.111	&992.11	&0.153	&745.49	&0.0667	&957.69	&0.555	&739.01	&1\\
2013-05-09	&1011.1	&0.0696	&731.87	&0.0409	&994.87	&0.242	&744.18	&0.1	&960.97	&0.474	&737.12	&0.367\\
2016-03-23	&988.95	&0.0444	&776.23	&0.0318	&972.45	&0.233	&789.01	&0.0742	&937.27	&0.349	&781.06	&0.44\\
2018-04-22	&994.91	&0.095	&715.37	&0.27	&978.32	&0.161	&728.42	&0.15	&941.9	&1	    &718.62	&2\\
2018-04-24	&997.04	&0.079	&711.32	&0.165	&980.13	&0.251	&724.09	&0.107	&945.45	&0.483	&717.15	&0.47\\
\hline
\end{tabular}
\caption{Dusty sources DS21-DS23 observed with NACO.}
\label{tab:pos11}
\end{table*}

\begin{table*}
\centering
\setlength{\tabcolsep}{1pt}
\begin{tabular}{|c|cccccccccccc|}
\hline\hline
Epoch    & \multicolumn{4}{c|}{DS24} & \multicolumn{4}{c|}{DS25} & \multicolumn{4}{c|}{DS26}\\ 
\hline
    & R.A. & err R.A. &  Dec. &  err Dec. &  R.A. &  err R.A. &  Dec. &  err Dec. &  R.A. &  err R.A. &  Dec. &  err Dec.\\ 
\hline
2002-08-29	&-    &-	    &-  	&-	    &776.07	&18629	&701.27	&0.951	&790.11	&0.118	&699.64	&0.127\\
2002-08-30	&1213	&0.456	&733.99	&0.516	&1211	&1	    &725	&1	    &1225.9	&0.15	&722.78	&0.133\\
2003-05-10	&966	&0.5	&687	&0.5	&964.53	&0.0737	&678.91	&0.0643	&979.01	&0.506	&676.56	&0.0989\\
2004-04-25	&769.59	&0.203	&581.31	&0.317	&768.23	&0.168	&573.24	&0.12	&782.48	&0.0989	&570.97	&0.0644\\
2004-04-26	&770	&0.5	&582	&0.5	&767.77	&0.0881	&573.24	&0.114	&782.22	&0.245	&570.94	&0.105\\
2005-05-13	&879.27	&0.327	&716.87	&0.624	&878.57	&0.128	&709.17	&0.207	&892.41	&0.299	&707.26	&0.117\\
2006-05-29	&834.43	&0.0807	&688.04	&0.134	&833.31	&0.315	&679.98	&0.385	&847.54	&0.047	&677.17	&0.0361\\
2007-04-01	&1133.9	&0.164	&991.25	&0.155	&1131.5	&0.0739	&982.25	&0.108	&1146.3	&0.0469	&980.14	&0.0297\\
2007-05-15	&1130.6	&0.115	&992.38	&0.206	&1128.7	&0.0627	&983.15	&0.124	&1143.5	&0.242	&981.26	&0.063\\
2007-05-16	&1102.7	&0.768	&1015	&0.672	&1100	&0.5	&1006	&0.5	&1114.3	&0.843	&1004.4	&0.341\\
2007-05-17	&1129.4	&0.0922	&1001.5	&0.0933	&1127.7	&0.0793	&992.56	&0.0634	&1147.7	&0.0671	&993.41	&0.148\\
2007-05-18	&1140.3	&0.693	&990.59	&0.315	&1139.1	&0.516	&982.14	&0.189	&1153.7	&0.345	&980.13	&0.125\\
2007-05-19	&	-	&	-	&	-	&	-	&	-	&	-&   -        &   -    & -      &  -     &  -     &   -       \\ 
2007-05-22	&900	&1	    &753	&1	    &896.92	&0.14	&744.04	&0.0981	&911.9	&0.0775	&742.02	&0.0343\\
2007-05-23	&1148	&1	    &1023	&1	    &1146	&0.875	&1014.1	&0.761	&1159.4	&0.4	&1012.3	&0.292\\
2008-05-26	&1109	&0.065	&1049.4	&0.1	&1107.9	&0.12	&1041.4	&0.14	&1122.3	&0.0938	&1039.4	&0.0984\\
2008-05-30	&1077.8	&0.186	&1040.3	&0.178	&1076.1	&0.372	&1031.8	&0.183	&1090.6	&0.0572	&1029.3	&0.0314\\
2008-05-31	&1138.4	&0.468	&1010.9	&0.205	&1137.1	&0.117	&1001.7	&0.322	&1151.5	&0.248	&999.39	&0.0841\\
2008-06-02	&1145.9	&0.686	&1010.9	&0.444	&1144.2	&0.187	&1001.8	&0.327	&1159.8	&0.331	&999.35	&0.0898\\
2008-06-03	&1150.5	&0.0997	&1012.1	&0.105	&1147.6	&0.0938	&1002.7	&0.0901	&1162.8	&0.227	&1000.5	&0.068\\
2011-05-25	&954	&1	    &790	&1	    &952.16	&0.144	&781.95	&0.0778	&967.08	&0.17	&779.59	&0.0859\\
2012-05-16	&954.87	&0.0961	&727.35	&0.126	&952.72	&0.0758	&718.88	&0.205	&967.42	&1	    &716.84	&0.3\\
2013-05-09	&957.14	&0.0226	&725.74	&0.0208	&955.48	&0.0911	&718.06	&0.0858	&970.02	&0.602	&715.61	&0.17\\
2016-03-23	&933.97	&0.0344	&769.41	&0.0194	&933.07	&0.725	&761.84	&0.202	&947.52	&0.166	&759.77	&0.216\\
2018-04-22	&938.75	&0.021	&708.34	&0.021	&937.93	&1	    &701.08	&1	    &952.38	&0.277	&699.15	&0.173\\
2018-04-24	&941.04	&0.0396	&704.25	&0.0359	&940.72	&0.359	&696.36	&0.147	&954.33	&0.369	&964.82	&0.692\\
\hline
\end{tabular}
\caption{Dusty sources DS24-DS26 observed with NACO.}
\label{tab:pos12}
\end{table*}

\begin{table*}
\centering
\setlength{\tabcolsep}{1pt}
\begin{tabular}{|c|cccccccccccc|}
\hline\hline
Epoch    & \multicolumn{4}{c|}{DS27} & \multicolumn{4}{c|}{DS28} & \multicolumn{4}{c|}{DS29}\\ 
\hline
    & R.A. & err R.A. &  Dec. &  err Dec. &  R.A. &  err R.A. &  Dec. &  err Dec. &  R.A. &  err R.A. &  Dec. &  err Dec.\\ 
\hline
2002-08-29	&812.13 &0.4  &737.12 &0.66 &781.45	&0.336	&715.13	&0.411	&772.6	&687.3	&0.11	&0.23\\
2002-08-30	&1248   &0.33 &760.76 &0.26 &1215	&1	    &739	&1	    &1208.1	&710.94	&0.09	&0.06\\
2003-05-10	&1000.8 &0.08 &714.58 &0.11 &369.65	&0.249	&691.95	&0.142	&961.54	&664.01	&0.08	&0.77\\
2004-04-25	&804.67 &0.13 &608.42 &0.14 &773	&1	    &587	&1	    &765.05	&558.44	&0.06	&0.05\\
2004-04-26	&804.7  &0.13 &608.56 &0.14	&772.75	&0.603	&587.54	&0.261	&765.09	&558.46	&0.08	&0.07\\
2005-05-13	&915.13 &0.12 &744.33 &0.24 &883	&1	    &724	&    1	&875.04	&694.48	&0.08	&0.06\\
2006-05-29	&870.14 &0.11 &714.39 &0.18 &838.76	&0.224	&692.96	&0.221	&829.74	&664.34	&0.23	&0.58\\
2007-04-01	&899.22 &0.10 &746.71 &0.30	&1136.9	&0.365	&996.67	&0.25	&861.87	&705.18	&0.29	&0.12\\
2007-05-15	&-       &-   &-       &-		&1133.3	&0.406	&998.16	&0.598	&	-&	-&          -&-\\
2007-05-16	&-       &-   &-       &-		&1104	&1	    &1020	&1		&	-&   -&          -&-\\
2007-05-17	&-       &-   &-       &-		&1132.2	&0.435	&1007.6	&0.816	&	-&	-&          -&-\\
2007-05-18	&-       &-   &-       &-		&1143	&1	    &997	&1		&	-&   -&          -&-\\
2007-05-19	&-       &-   &-       &-		&		&		&		&	    &   -  &   -&          -&-\\
2007-05-22	&934.74  &0.06&778.68  &0.08	&901.6	&45658	&758.67	&0.344	&896.04	&736.98	&0.32	&0.24\\
2007-05-23	&1183.1  &0.06&1048.7  &0.07	&1151.2	&0.303	&1027.8	&0.182	&1145	&1007.5	&1	    &1\\
2008-05-26	&875.38	 &0.77&805.8   &0.21    &1112.4	&0.348	&1056	&0.153	&837.52	&764.24	&0.11	&0.1\\
2008-05-30	&-       &-   &-       &-		&1080.8	&0.0448	&1046.1	&0.454	&-	&	-&         - &-\\
2008-05-31	&-       &-   &-       &-		&1141	&1	    &1016	&1		&-	&   -&         - &-\\
2008-06-02	&-       &-   &-       &-		&1149.4	&0.193	&1016.1	&0.368	&-	&	-&         - &-\\
2008-06-03	&-       &-   &-       &-		&1153.8	&0.673	&1016.9	&0.288	&-	&	-&         - &-\\
2011-05-25	&882.54  &0.09&708.52  &0.05    &957.5	&0.248	&795.85	&0.199	&844.37	&667.22	&0.12	&0.1\\
2012-05-16	&991.46  &0.12&752.45  &0.05	&957.73	&0.523	&733.9	&0.203	&953.32	&710.75	&0.08	&0.2\\
2013-05-09	&994.12  &0.22&751.14  &0.13	&960.14	&0.731	&733.12	&0.168	&956.39	&710.13	&0.08	&0.11\\
2016-03-23	&971.33  &0.21&794.48  &0.09	&937.11	&0.514	&777.51	&0.892	&933.37	&753.72	&0.19	&0.11\\
2018-04-22	&907.13  &0.10&659.31  &0.24	&942.56	&0.448	&716.18	&0.508	&868.76	&620.78	&0.17	&0.35\\
2018-04-24	&816.18  &0.11&771.9   &0.09	&944.69	&0.178	&711.86	&0.11	&778.26	&731.67	&0.12	&0.1\\
\hline
\end{tabular}
\caption{Dusty sources DS27-DS29 observed with NACO.}
\label{tab:pos13}
\end{table*}

\begin{table*}
\centering
\setlength{\tabcolsep}{1pt}
\begin{tabular}{|c|cccccccccccccccc|}
\hline\hline
Epoch    & \multicolumn{4}{c|}{DS30} & \multicolumn{4}{c|}{DS31} & \multicolumn{4}{c|}{DS32}& \multicolumn{4}{c|}{DS33}\\ 
\hline
    & R.A. & err R.A. &  Dec. &  err Dec. &  R.A. &  err R.A. &  Dec. &  err Dec. &  R.A. &  err R.A. &  Dec. &  err Dec. &  R.A. &  err R.A. &  Dec. &  err Dec.\\ 
\hline
2002-08-29	&792.35	&690.38	&0.05	&0.15	&793.05	&711.58	&0.38	&0.15	&780.49	&721.44	&0.11	&0.16	&783.52	&713.28	&0.18	&0.51\\
2002-08-30	&1228.4	&713.99	&0.31	&0.13	&1228.5	&734.73	&0.48	&0.15	&1216.2	&744.45	&0.16	&0.24	&1218.9	&736.46	&0.11	&0.22\\
2003-05-10	&980.6	&667.73	&0.31	&0.35	&981.44	&687.99	&0.16	&0.35	&969.62	&698.43	&0.08	&0.1	&972.22	&689.99	&0.11	&0.16\\
2004-04-25	&785.01	&561.78	&0.06	&0.09	&784.94	&582.35	&0.14	&0.1	&773.32	&592.86	&0.07	&0.09	&775.55	&584.97	&0.06	&0.11\\
2004-04-26	&785.02	&561.77	&0.05	&0.11	&784.87	&582.37	&0.13	&0.1	&773.2	&592.92	&0.08	&0.1	&775.35	&585.34	&0.09	&0.16\\
2005-05-13	&875.04	&694.48	&0.08	&0.06	&894.84	&717.9	&0.35	&0.94	&882.99	&729.18	&0.11	&0.1	&885.91	&720.75	&0.12	&0.21\\
2006-05-29	&850.44	&667.83	&0.05	&0.08	&849.62	&689.18	&0.14	&0.24	&838.11	&699.61	&0.07	&0.11	&840.57	&691.4	&0.09	&0.18\\
2007-04-01	&879.19	&701.06	&0.05	&0.09	&879.18	&722.46	&0.11	&0.22	&867.06	&732.18	&0.33	&0.43	&869.77	&723.58	&0.17	&0.33\\
2007-05-15	&	-	&	-	&	-	&	-	&	-	&	-	&	-	&	-    & -      &-       &  -     &  -     &   -    &   -    &  -     & -        \\             
2007-05-16	&	-	&	-	&	-	&	-	&	-	&	-	&	-	&	-    & -      &-       &  -     &  -     &   -    &   -    &  -     & -        \\             
2007-05-17	&	-	&	-	&	-	&	-	&	-	&	-	&	-	&	-    & -      &-       &  -     &  -     &   -    &   -    &  -     & -        \\             
2007-05-18	&	-	&	-	&	-	&	-	&	-	&	-	&	-	&	-    & -      &-       &  -     &  -     &   -    &   -    &  -     & -        \\             
2007-05-19	&	-	&	-	&	-	&	-	&	-	&	-	&	-	&	-    & -      &-       &  -     &  -     &   -    &   -    &  -     & -        \\             
2007-05-22	&914.92	&732.45	&0.06	&0.15	&913.96	&754.4	&0.19	&0.42	&902.04	&763.65	&0.16	&0.45	&904.87	&756.03	&0.11	&0.18\\
2007-05-23	&1163.5	&1001.9	&0.07	&0.14	&1162	&1024	&0.11	&0.43	&1150.6	&1033.6	&0.4	&0.49	&1153.1	&1026.1	&0.11	&0.17\\
2008-05-26	&855.37	&759.97	&0.04	&0.08	&854.71	&780.73	&0.14	&0.28	&843.48	&791.47	&0.21	&0.3	&845.66	&783.38	&0.18	&0.22\\
2008-05-30	&	-	&	-	&	-	&	-	&	-	&	-	&	-	&	-    &  -     &   -    &   -    &   -    &   -    &   -    &   -    &  -       \\             
2008-05-31	&	-	&	-	&	-	&	-	&	-	&	-	&	-	&	-    &  -     &   -    &   -    &   -    &   -    &   -    &   -    &  -       \\             
2008-06-02	&	-	&	-	&	-	&	-	&	-	&	-	&	-	&	-    &  -     &   -    &   -    &   -    &   -    &   -    &   -    &  -       \\             
2008-06-03	&	-	&	-	&	-	&	-	&	-	&	-	&	-	&	-    &  -     &   -    &   -    &   -    &   -    &   -    &   -    &  -       \\             
2011-05-25	&861.91	&663.14	&0.08	&0.16	&861.28	&684.15	&0.23	&0.12	&848.99	&693.74	&0.33	&0.12	&852.28	&686.26	&0.12	&0.22\\
2012-05-16	&970.66	&707.61	&0.13	&0.1	&970.18	&729.19	&0.23	&0.44	&958.45	&738.6	&0.25	&0.52	&961.01	&730.87	&0.15	&0.28\\
2013-05-09	&973.47	&706.72	&0.13	&0.09	&972.99	&727.45	&0.07	&0.05	&961.13	&737.9	&0.12	&0.13	&964.05	&729.66	&0.08	&0.16\\
2016-03-23	&950.6	&750.81	&0.08	&0.15	&949.7	&770.57	&0.41	&0.71	&936.74	&778.47	&0.76	&0.15	&941.17	&773.2	&0.14	&0.44\\
2018-04-22	&887.54	&612.9	&0.13	&0.44	&885.67	&638.42	&0.24	&0.28	&872.07	&645.23	&0.31	&0.3	&875.86	&642.45	&0.13	&0.31\\
2018-04-24	&795.38	&728.79	&0.2	&0.07	&795.01	&749.25	&0.07	&0.05	&782	&758	&1	    &1	    &785.93	&751.05	&0.1	&0.25\\
\hline
\end{tabular}
\caption{Dusty sources DS30-DS33 observed with NACO.}
\label{tab:pos14}
\end{table*}

\section{K-band counterpart of the dusty sources}
\label{sec:kband-counter-app}

\begin{figure*}[htbp!]
	\centering
	\includegraphics[width=1.\textwidth]{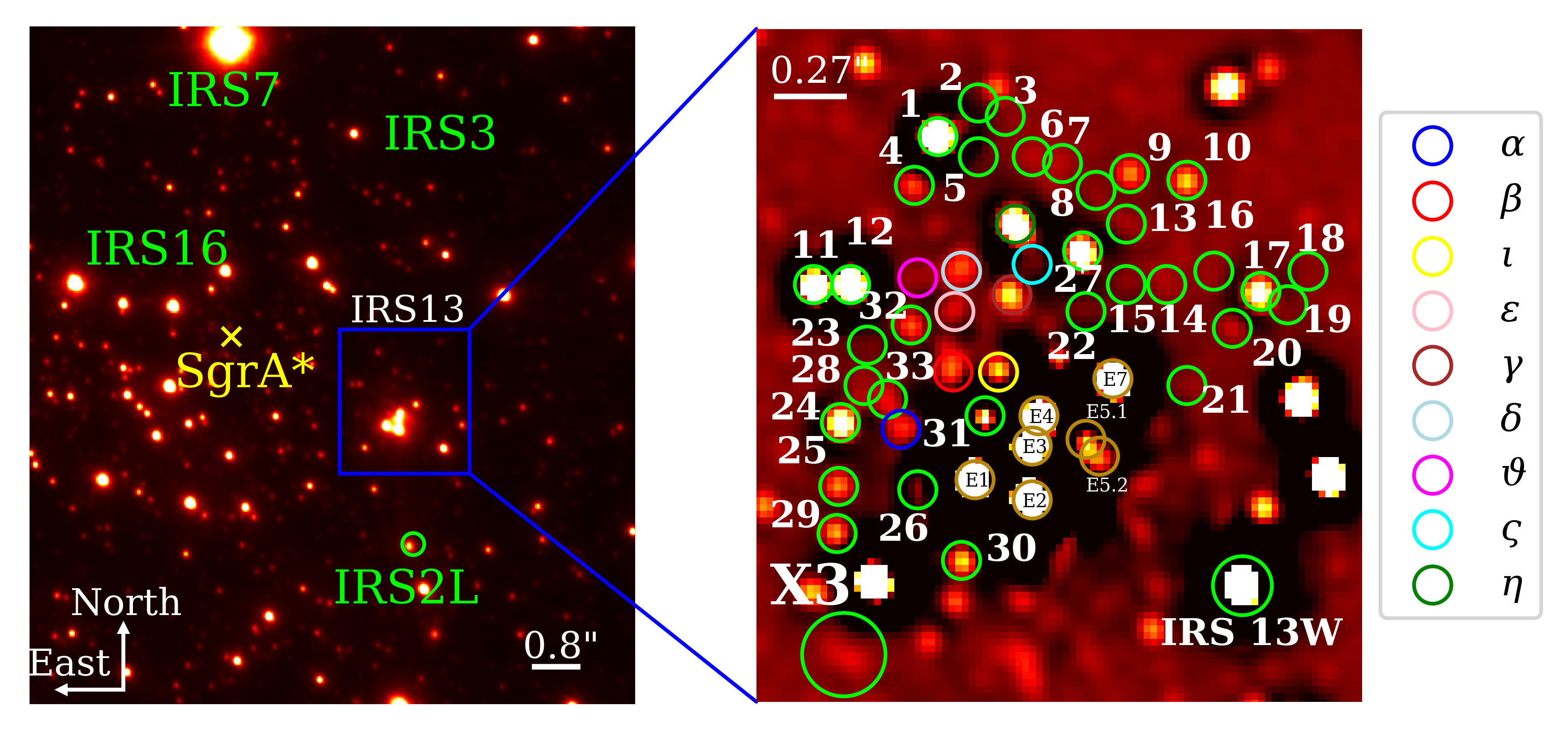}
	\caption{K-band observation of IRS 13 in 2009.}
\label{fig:finding_chart_kband}
\end{figure*}

Here, we display the K-band counterparts for dusty sources analyzed in this work. The K-band data presented in Fig. \ref{fig:finding_chart_kband} was observed in 2009 with NACO. Because of the proper motion of the DS sources, the detectability may be hindered for individual objects.
Single detections of these objects, including light curves, are presented in Appendix \ref{sec:light-curves-app}.

\section{H-band counterpart of the dusty sources}
\label{sec:hband-counter-app}

\begin{figure*}[htbp!]
	\centering
	\includegraphics[width=1.\textwidth]{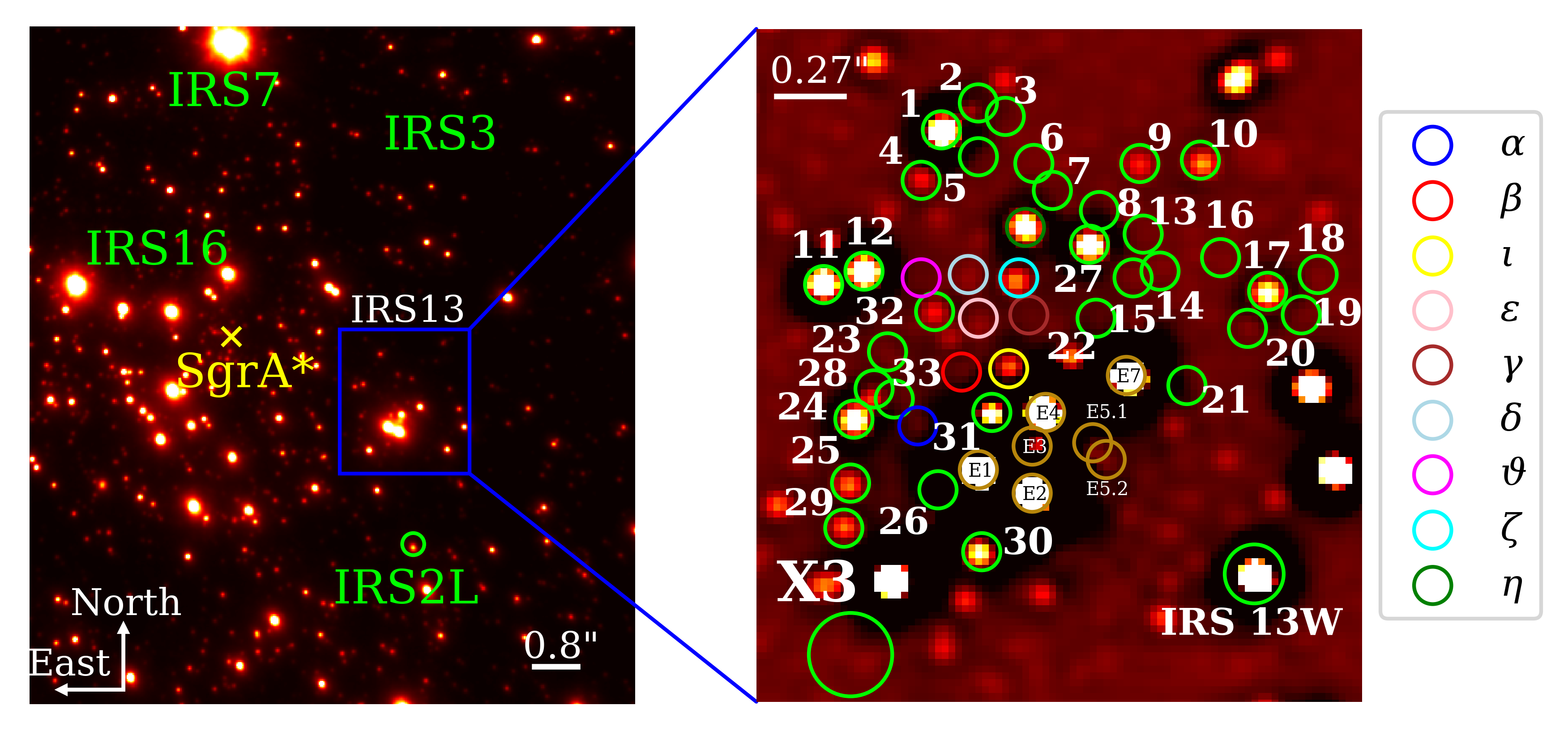}
	\caption{H-band observation of IRS 13 in 2004.}
\label{fig:finding_chart_hband}
\end{figure*}

Near-infrared NACO observations of the DS sources in the H-band. The observations were carried out in 2004 and show the same FOV as it is presented in Fig. \ref{fig:finding_chart} and Fig. \ref{fig:finding_chart_kband}. The detection of some sources in Fig, \ref{fig:finding_chart_hband} may be infected by confusion and interference, which is why we present individual detections in Appendix \ref{sec:light-curves-app}.

\section{Multi-wavelength identification of the dusty sources}
\label{sec:light-curves-app}

We use the $L$-band emission presented in Fig. \ref{fig:finding_chart} as a starting point to identify the here discussed dusty sources (abbreviated with DS). Whenever feasible, we use the continuum data without any applied high-pass filter to detect the DS sources in the $H$-, $K$-, and $L$- bands. For the photometric analysis, we use IRS 2L and S65 with an L-band magnitude of $10.59\pm0.03$ \citep{Hosseini2020, peissker2021} as a calibrator. To determine the impact of the high-pass filter on the photometric analysis, we will compare the $L$-band magnitude of one of the brightest dusty sources, DS1, with and without the applied smooth-subtract algorithm. 

As it is demonstrated in Table \ref{tab:filter_comparison}, the impact on the magnitude after applying a high-pass filter is negligible. This result is expected since \citet{Ott1999} investigated the impact of applying different high-pass filters to the data in detail. The authors found no trend regarding a specific filter for almost 30 individual sources in agreement with the comparison presented in Table \ref{tab:filter_comparison}. Although individual sources may be {affected} by confusion, which is translated to brighter/fainter magnitude values, we note that non-filtered data is exposed to crowding problems.

\subsection{DS1}

\begin{figure*}[htbp!]
	\centering
	\includegraphics[width=1.\textwidth]{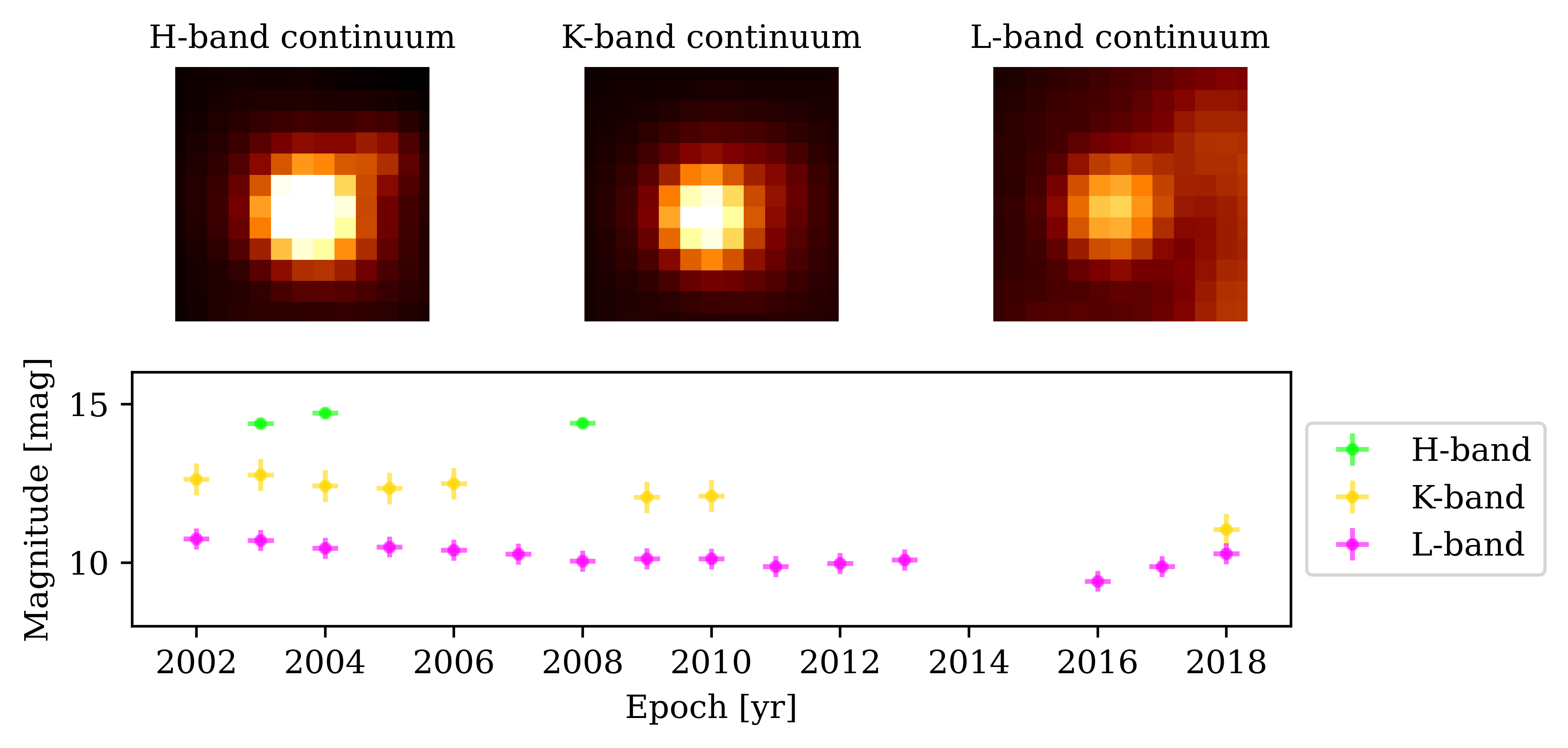}
	\caption{Multi-wavelength identification of DS1 with NACO (see Fig. \ref{fig:finding_chart}, Fig. \ref{fig:finding_chart_hband}, and Fig. \ref{fig:finding_chart_kband}). The lower subplot displays the light curve with a fairly constant magnitude distribution between 2002 and 2018.}
\label{fig:ident_ds1}
\end{figure*}

The source DS1 is one of the brightest H- and K-band object of the here presented sample. In Fig. \ref{fig:ident_ds1}, we show a confusion-free detection of DS1 in 2004 including the light curve based on the H-, K-, and L-band NACO data observed between 2002 and 2018.
In Table \ref{tab:ds1_mag}, we list the mean and median magnitude with the related standard deviation based on the here analyzed data set.
\begin{table*}
\centering
\begin{tabular}{|cccc|}
\hline
\hline
       & mean & median & std  \\
H-band & 14.40 & 14.49 & 0.15 \\
K-band & 12.37 & 12.23 & 0.50 \\
L-band & 10.13 & 10.19 & 0.33 \\
\hline
\hline
\end{tabular}
\caption{Estimated magnitudes for DS1 using multi-wavelength observations carried out with NACO between 2002 and 2018.}
\label{tab:ds1_mag}
\end{table*}

\subsection{DS2 $\&$ DS3}

\begin{figure*}[htbp!]
	\centering
	\includegraphics[width=1.\textwidth]{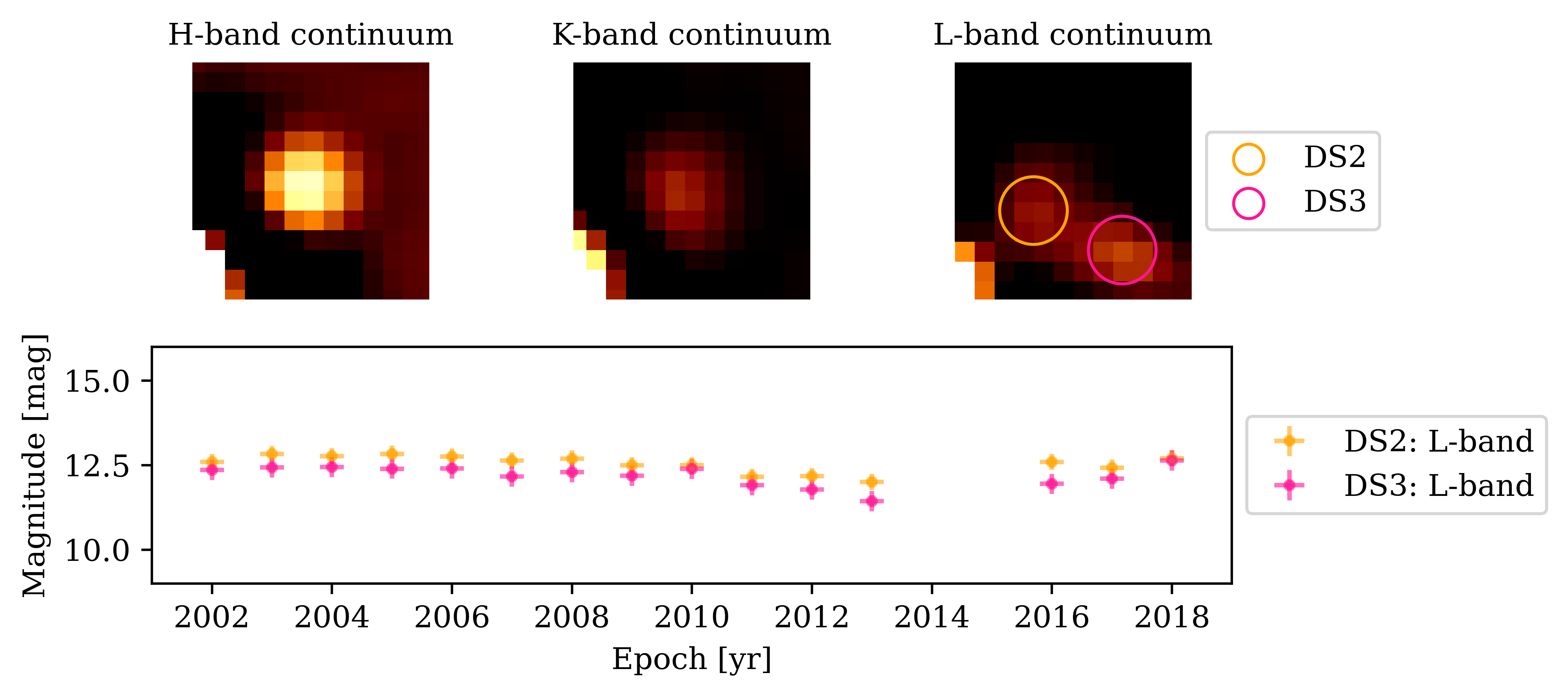}
	\caption{Detection of DS2 and DS3 observed with NACO. Both sources are located between DS1 and the star shown in the H- and K-band continuum subplots. As implied, no NIR counterpart of DS2 and DS3 can be detected above the noise. If there are NIR counterparts, the detection is hampered due to the PSF wings of the two stars shown in the H- and K-band continuum subplots.}
\label{fig:ident_ds2_ds3}
\end{figure*}

In Figure \ref{fig:ident_ds2_ds3}, we show the detection of DS2 and DS3. We do not detect a K- and H-band counterpart above the noise level for both sources, which is due to the high level of crowding. However, NACO K-band data of 2018 reveals some promising candidates that might be associated with DS2 and DS3. 
However, both faint sources demonstrate the challenges of the analysis and the need for high-resolution IFU data. We list the related L-band magnitude in Table \ref{tab:ds2_3_mag}.
\begin{table*}
\centering
\begin{tabular}{|cccc|}
\hline
\hline
       & mean & median & std  \\
L-band (DS2) & 12.60 & 12.54 & 0.24 \\
L-band (DS3) & 12.30 & 12.19 & 0.30 \\
\hline
\hline
\end{tabular}
\caption{Estimated magnitudes for DS2 and DS3 using L-band observations carried out with NACO between 2002 and 2018.}
\label{tab:ds2_3_mag}
\end{table*}

\subsection{DS4}

\begin{figure*}[htbp!]
	\centering
	\includegraphics[width=1.\textwidth]{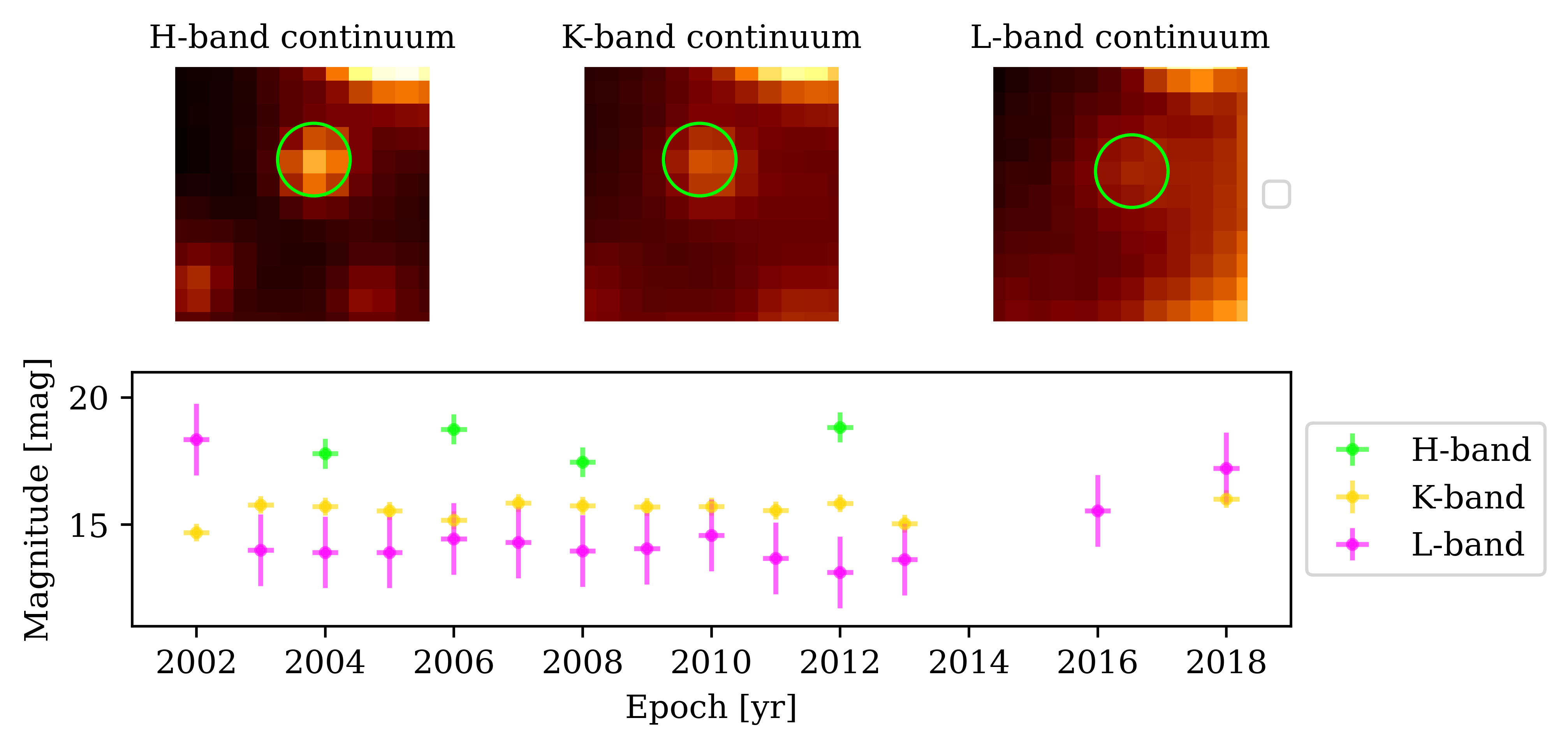}
	\caption{Detection of DS4 in the H-, K-, and L-band close to DS1 (upper right boundary of the images).}
\label{fig:ident_ds4}
\end{figure*}

In Fig. \ref{fig:ident_ds4}, we present the multi-wavelength identification of DS4. Despite the close distance to DS1, the object DS4 can be observed without confusion. This object is a prime example of magnitude-confusion susceptibility; if the investigated object is bright enough, the close distance to a brighter object does not impact the observability. An example of a toxic magnitude-confusion susceptibility is displayed in Fig. \ref{fig:ident_ds2_ds3}.
Table \ref{tab:ds4_mag} lists the magnitude of DS4 in several IR bands.
\begin{table*}
\centering
\begin{tabular}{|cccc|}
\hline
\hline
       & mean & median & std  \\
H-band & 18.27 & 18.20 & 0.59 \\
K-band & 15.71 & 15.56 & 0.35 \\
L-band & 14.03 & 14.62 & 1.41 \\
\hline
\hline
\end{tabular}
\caption{Estimated magnitudes for DS4 using multi-wavelength observations carried out with NACO between 2002 and 2018.}
\label{tab:ds4_mag}
\end{table*}

\subsection{DS5}

The L-band detection of DS5 close to DS1. As for DS2 and DS3, the magnitude-confusion susceptibility is high, which might be the reason for the non-detection of DS5 in the H- and K-band. The classification of a core-less dust blob is rather unlikely because of evaporation time scales in a radiative-dominated environment. For example, \cite{Stewart2016} and \cite{Hoefner2019} report variations of dust clouds close to stars on time scales of a few years.
\begin{figure*}[htbp!]
	\centering
	\includegraphics[width=1.\textwidth]{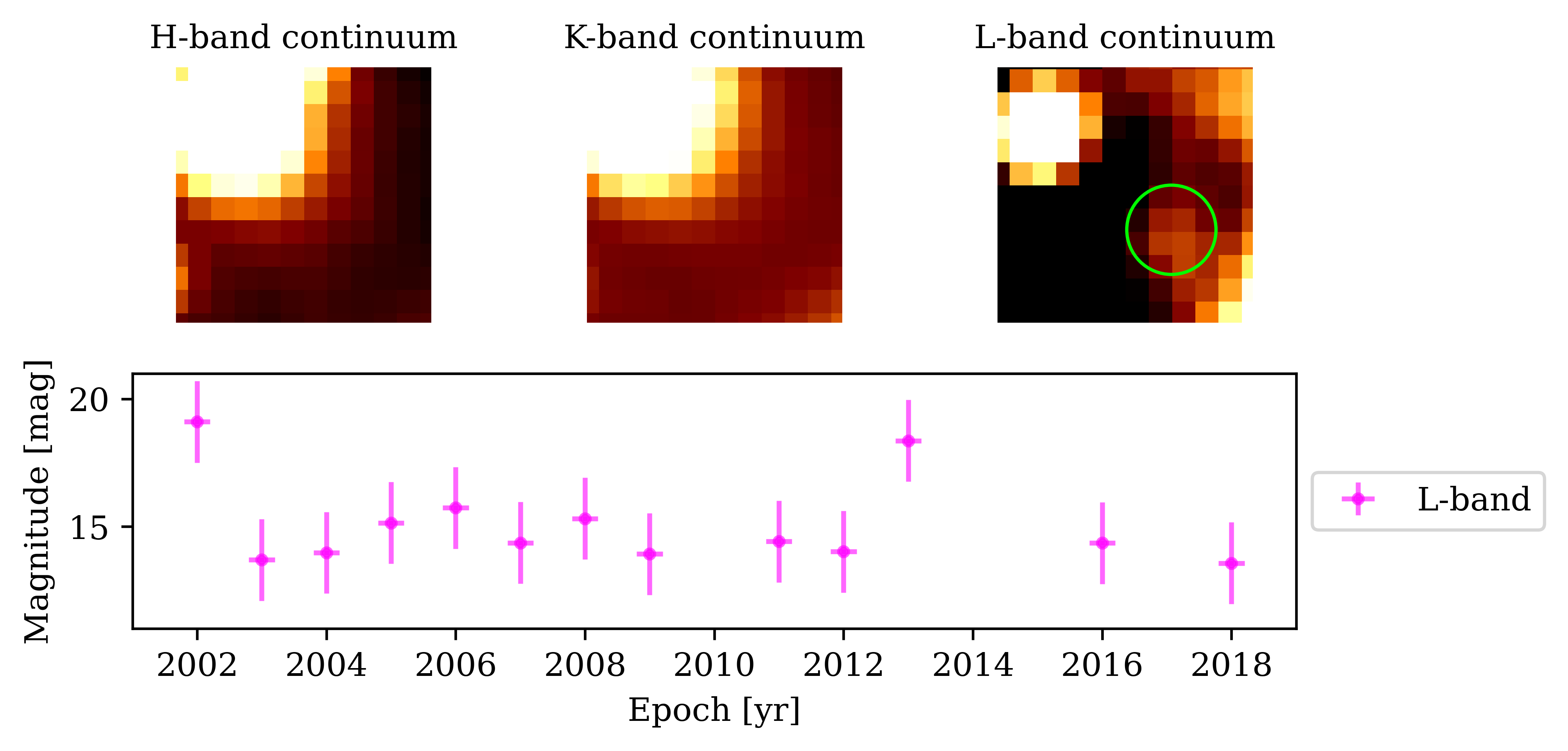}
	\caption{Dusty source DS5 close to DS1 that is located in the upper left corner of the subplots. We mark DS5 with a lime-colored circle. Due to the dominant PSF of DS1, the detection of DS5 is challenging. We estimate a median L-band magnitude of $14.36\pm 1.68$. Despite the two outliers in 2002 and 2013, the L-band magnitude is constant inside the uncertainty range.}
\label{fig:ident_ds5}
\end{figure*}
As before, we list the corresponding L-band magnitude for DS5 in Table \ref{tab:ds5_mag}.
\begin{table*}
\centering
\begin{tabular}{|cccc|}
\hline
\hline
       & mean & median & std  \\
L-band & 14.37 & 15.08 & 1.68 \\
\hline
\hline
\end{tabular}
\caption{Estimated magnitudes for DS5 using multi-wavelength observations carried out with NACO between 2002 and 2018.}
\label{tab:ds5_mag}
\end{table*}

\subsection{DS6 $\&$ DS7}

\begin{figure*}[htbp!]
	\centering
	\includegraphics[width=1.\textwidth]{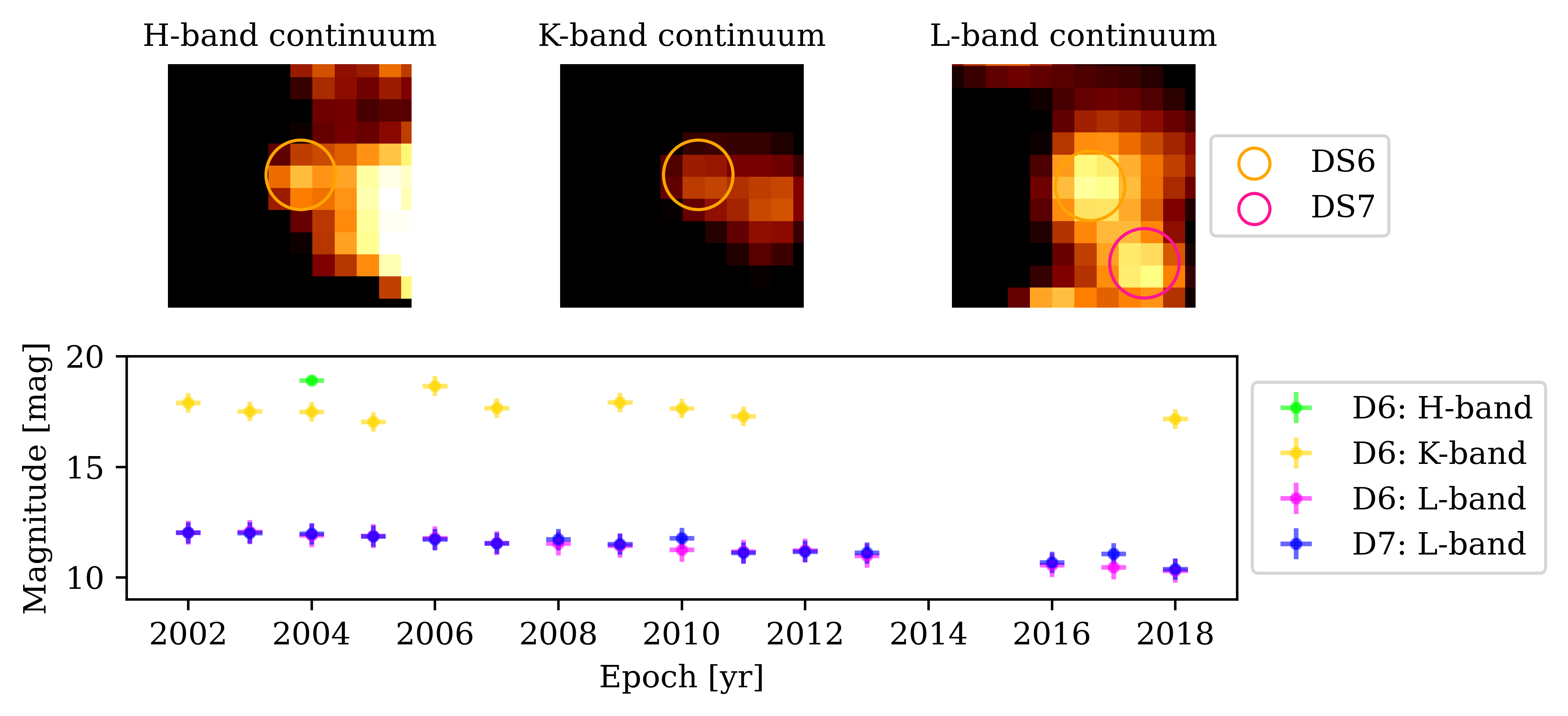}
	\caption{Multi-wavelength detection of DS6 and DS7 as observed with NACO.}
\label{fig:ident_ds6_ds7}
\end{figure*}

In Fig. \ref{fig:ident_ds6_ds7}, we present the identification of DS6 in the H-, K-, and L-band. Close to DS6, a bright star is observable in the H- and K-band with no L-band counterpart, demonstrating the challenges of the multi-wavelength analysis. In addition, Fig. \ref{fig:ident_ds6_ds7} indicates the location of the L-band source DS7.  
Table \ref{tab:ds6_7_mag} lists the corresponding DS6 and DS7 magnitudes. 
\begin{table*}
\centering
\begin{tabular}{|cccc|}
\hline
\hline
       & mean & median & std  \\
H-band (DS6) & 18.91 & 18.91 & 0.00 \\
K-band (DS6) & 17.62 & 11.62 & 0.44 \\
L-band (DS6) & 11.43 & 11.33 & 0.54 \\
L-band (DS7) & 11.54 & 11.44 & 0.48 \\
\hline
\hline
\end{tabular}
\caption{Estimated magnitudes for DS6 and DS7 using multi-wavelength observations carried out with NACO between 2002 and 2018.}
\label{tab:ds6_7_mag}
\end{table*}

\subsection{DS8}

\begin{figure*}[htbp!]
	\centering
	\includegraphics[width=1.\textwidth]{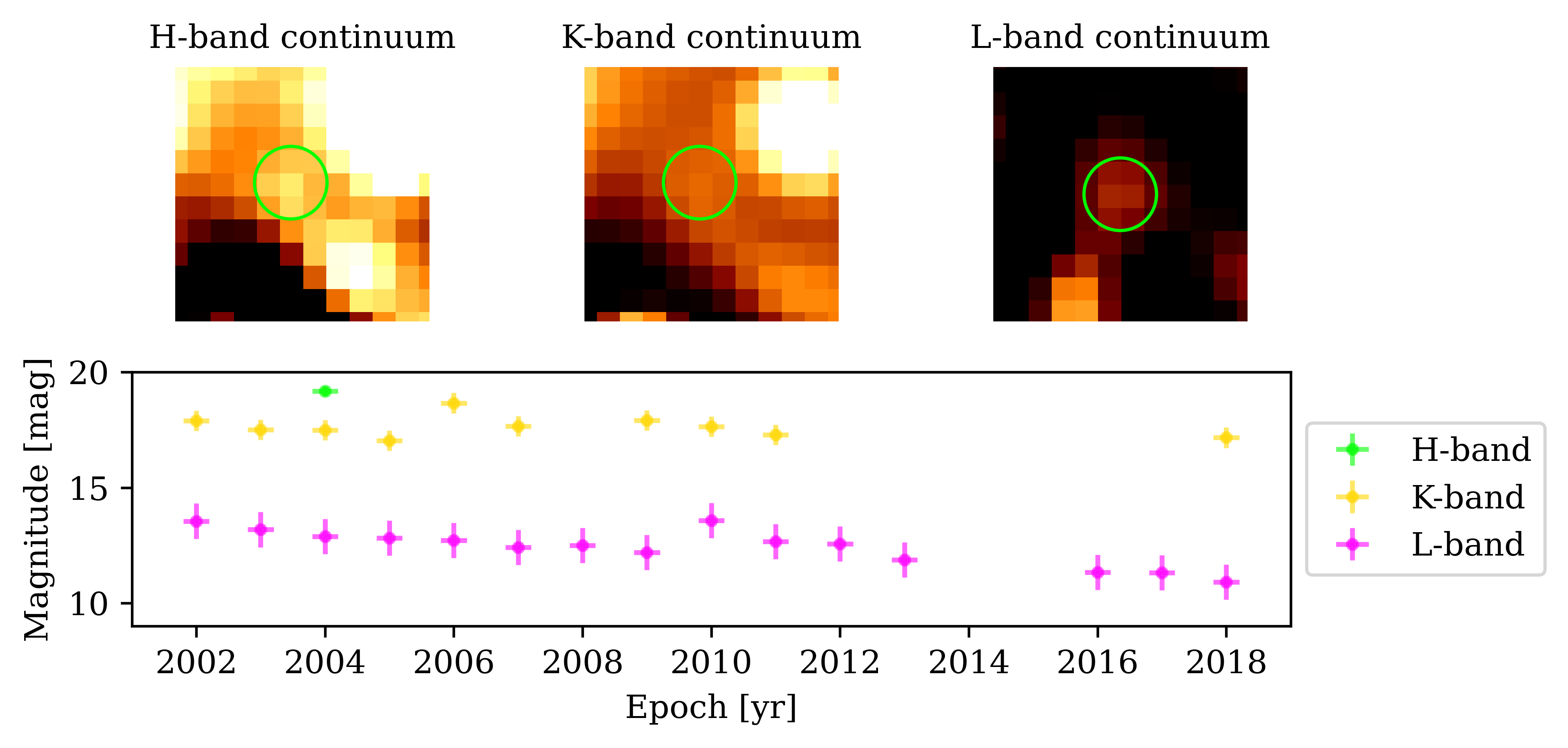}
	\caption{Multi-wavelength observation of DS8 with NACO.}
\label{fig:ident_ds8}
\end{figure*}

Like other fainter sources, the observation of DS8 is challenging because of the surrounding stars as it is shown in Fig. \ref{fig:ident_ds8}. While the L-band emission exhibits a low level of confusion, the detection of DS8 in the NIR bands is hindered due to the presence of close-by stars.
The related L-, K-, and H-band magnitude is displayed in Fig. \ref{fig:ident_ds8} and listed in Table \ref{tab:ds8_mag}.
\begin{table*}
\centering
\begin{tabular}{|cccc|}
\hline
\hline
       & mean & median & std  \\
H-band & 19.17 & 19.17 & 0.00 \\
K-band & 17.57 & 17.62 & 0.44 \\
L-band & 12.56 & 12.43 & 0.76 \\
\hline
\hline
\end{tabular}
\caption{Estimated magnitudes for DS8 using multi-wavelength observations carried out with NACO between 2002 and 2018.}
\label{tab:ds8_mag}
\end{table*}

\subsection{DS9 $\&$ DS10}
\begin{figure*}[htbp!]
	\centering
	\includegraphics[width=1.\textwidth]{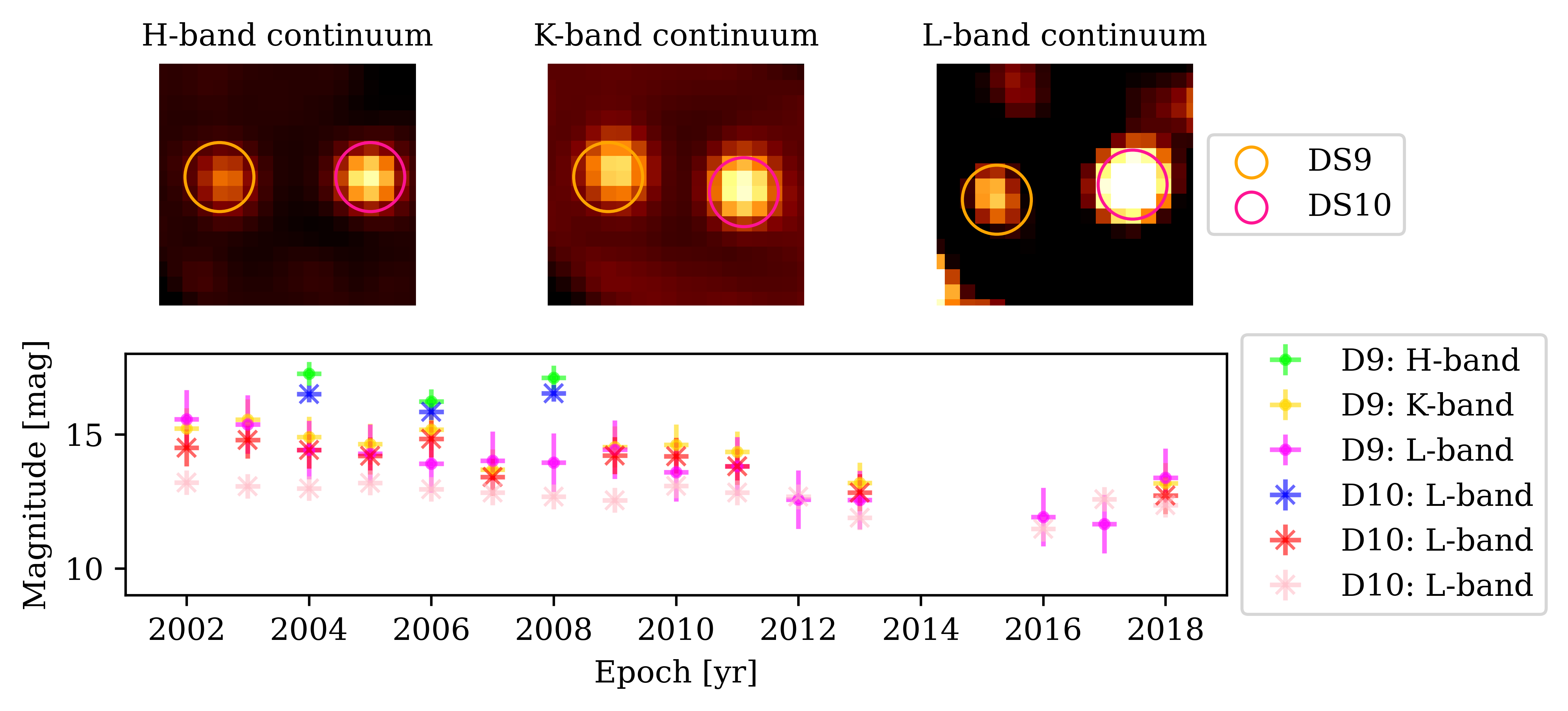}
	\caption{Detection of DS9 and DS10 with NACO.}
\label{fig:ident_ds9_ds10}
\end{figure*}

Due to their colors presented in Fig. \ref{fig:color_color_diagram}, DS9 and DS10 are most likely embedded stars such as IRS3. In Fig. \ref{fig:ident_ds9_ds10}, we show the multi-wavelength detection of the stars. Because of cosmetic reasons, we use the K-band observation of 2009 which translates in a minor offset compared to the H- and L-band data from 2004. This apparent misalignment is not in conflict with the overall identification of the two stars. 
Both stars show comparable magnitude values in different bands as shown in Fig. \ref{fig:ident_ds9_ds10} and listed in Table \ref{tab:ds9_10_mag}
\begin{table*}
\centering
\begin{tabular}{|cccc|}
\hline
\hline
       & mean & median & std  \\
H-band (DS9)  & 17.11 & 16.86 & 0.45 \\
K-band (DS9)  & 14.61 & 14.45 & 0.76 \\
L-band (DS9)  & 13.91 & 13.69 & 1.09 \\
H-band (DS10) & 16.50 & 16.29 & 0.31 \\
K-band (DS10) & 14.19 & 13.98 & 0.69 \\
L-band (DS10) & 12.82 & 12.68 & 0.46 \\
\hline
\hline
\end{tabular}
\caption{Estimated magnitudes for DS9 and DS10 using multi-wavelength observations carried out with NACO between 2002 and 2018.}
\label{tab:ds9_10_mag}
\end{table*}

\subsection{DS11 $\&$ DS12}
\begin{figure*}[htbp!]
	\centering
	\includegraphics[width=1.\textwidth]{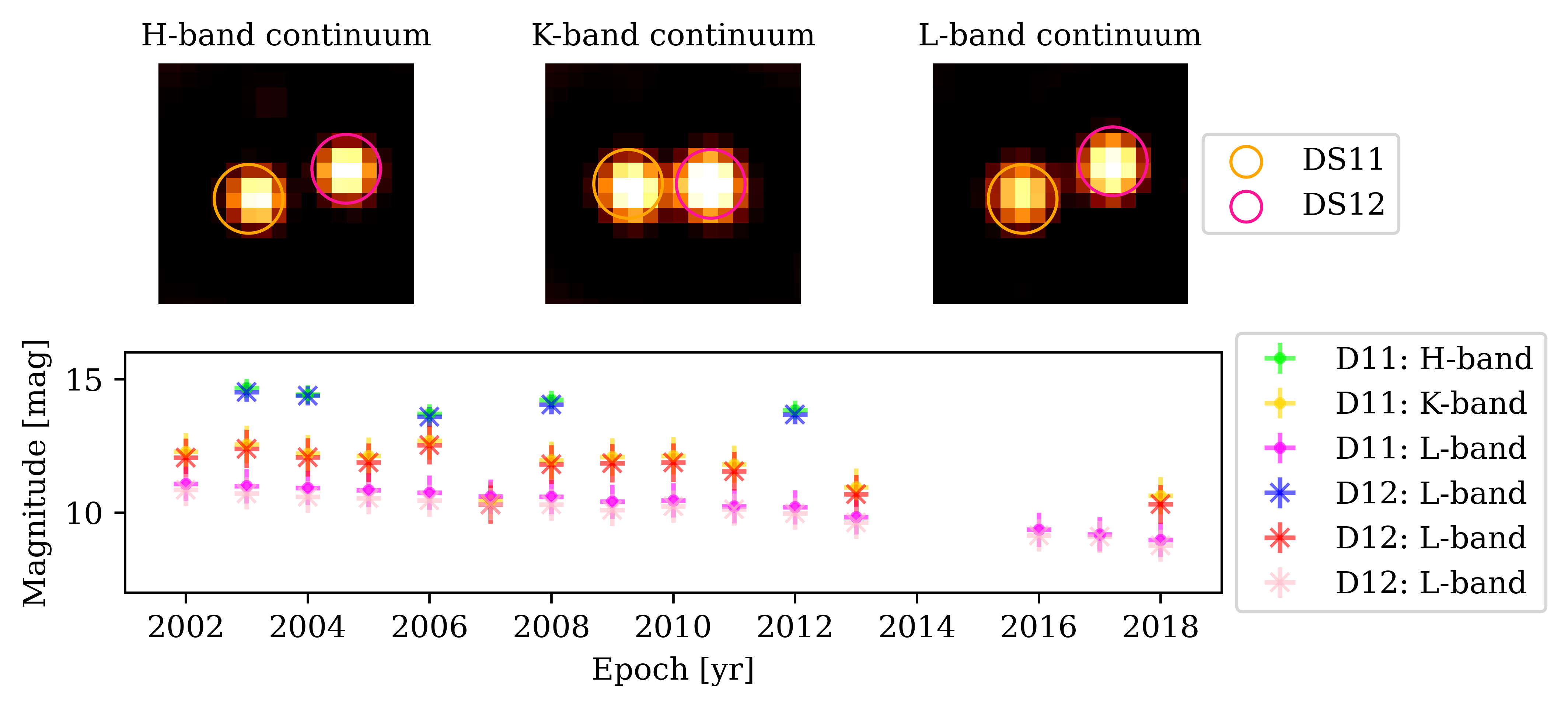}
	\caption{Identification of DS11 and DS12 in various bands.}
\label{fig:ident_ds11_ds12}
\end{figure*}

As for DS9 and DS10, we use K-band data of 2009 to complement the observations in the H- and L-band of 2004. Both sources, DS11 and DS12, can be most likely classified as stellar sources encircled by a dusty envelope. We present the related identification in Fig. \ref{fig:ident_ds11_ds12} and list the magnitudes in Table \ref{tab:ds11_12_mag}.

\begin{table*}
\centering
\begin{tabular}{|cccc|}
\hline
\hline
       & mean & median & std  \\
H-band (DS11)  & 14.22 & 14.16 & 0.35 \\
K-band (DS11)  & 12.10 & 11.82 & 0.70 \\
L-band (DS11)  & 10.46 & 10.29 & 0.64 \\
H-band (DS12) & 14.04 & 14.03 & 0.36 \\
K-band (DS12) & 11.86 & 11.60 & 0.72 \\
L-band (DS12) & 10.23 & 10.05 & 0.60 \\
\hline
\hline
\end{tabular}
\caption{Estimated magnitudes for DS11 and DS12 using multi-wavelength observations carried out with NACO between 2002 and 2018.}
\label{tab:ds11_12_mag}
\end{table*}

\subsection{DS13}

\begin{figure*}[htbp!]
	\centering
	\includegraphics[width=1.\textwidth]{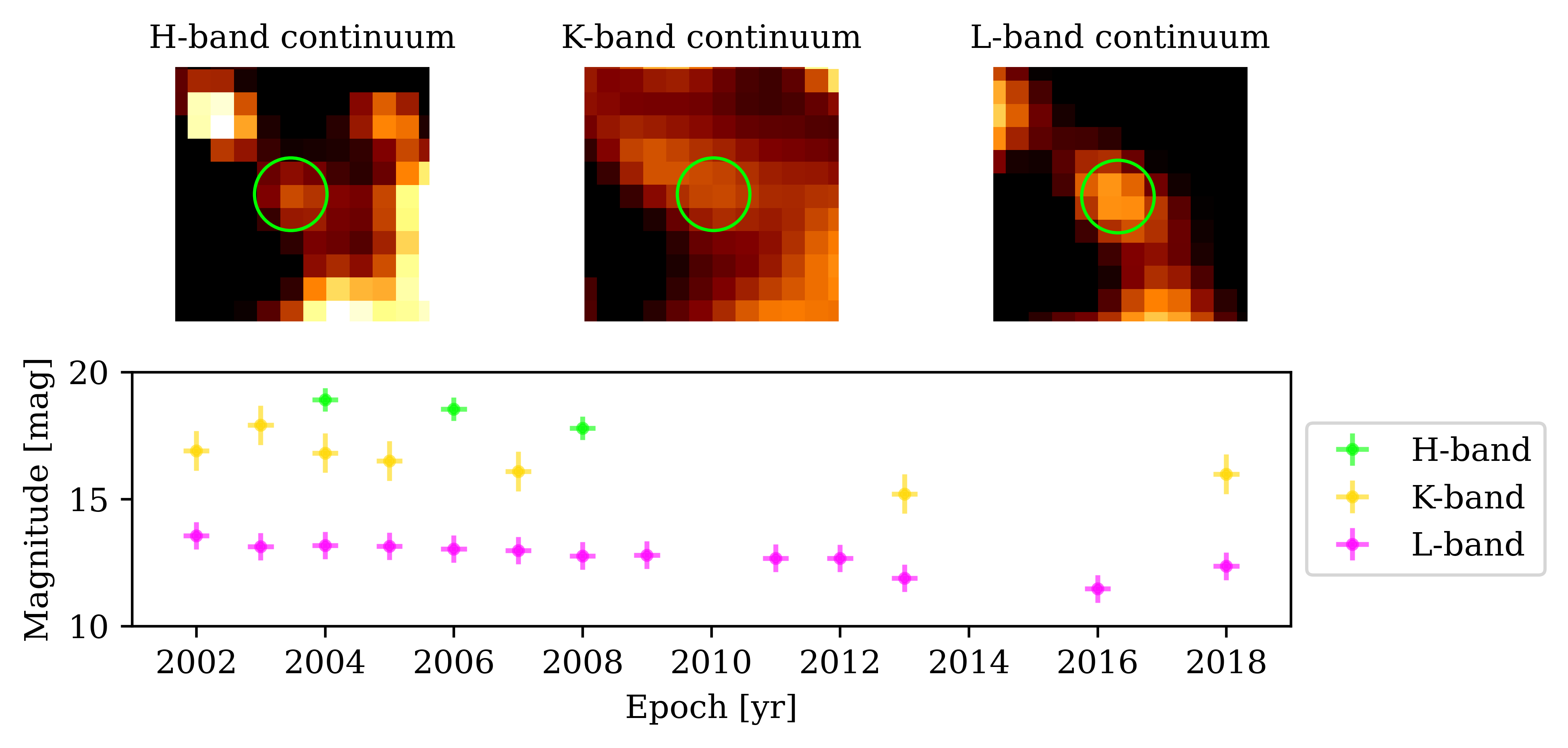}
	\caption{Marked DS13 observed in the H-, K-, and L-band, demonstrating the high confusion reflecting the exceptionally high stellar core density.}
\label{fig:ident_ds13}
\end{figure*}

For DS13, we note an increased magnitude-confusion susceptibility throughout the accessible IR bands. The observation of fainter DS sources is even more challenging due to the high stellar density. In Fig. \ref{fig:ident_ds13}, we exhibit the detection of DS8 in the NIR and MIR. Table \ref{tab:ds13_mag} lists the related magnitude values.

\begin{table*}
\centering
\begin{tabular}{|cccc|}
\hline
\hline
       & mean & median & std  \\
H-band & 18.55 & 18.42 & 0.46 \\
K-band & 16.51 & 16.49 & 0.78 \\
L-band & 12.80 & 12.74 & 0.54 \\
\hline
\hline
\end{tabular}
\caption{Estimated magnitudes for DS13 using multi-wavelength observations carried out with NACO between 2002 and 2018.}
\label{tab:ds13_mag}
\end{table*}

\subsection{DS14 $\&$ DS15}

\begin{figure*}[htbp!]
	\centering
	\includegraphics[width=1.\textwidth]{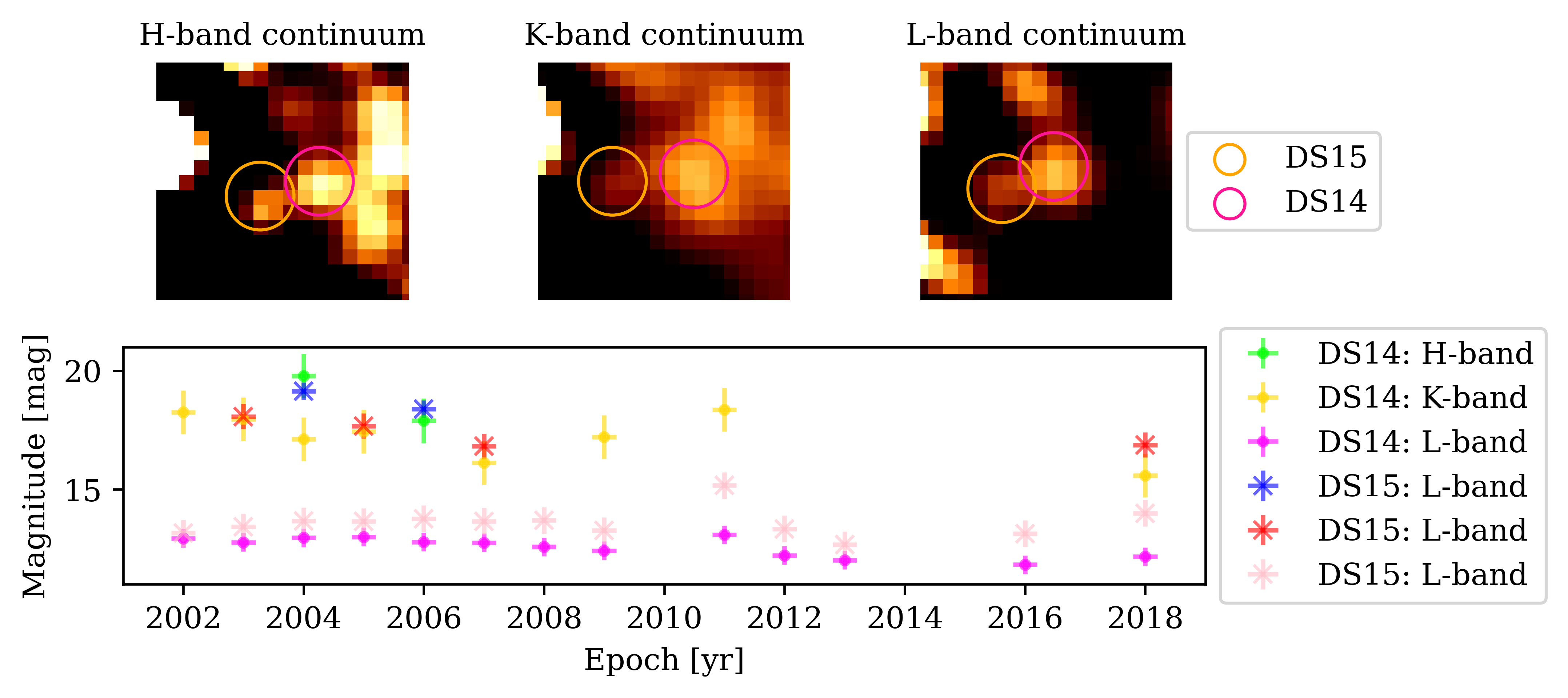}
	\caption{Observation of DS14 and DS15 in the H-, K-, and L-band.}
\label{fig:ident_ds14_ds15}
\end{figure*}

Both sources, DS14 and DS15, are detected in the H-, K-, and L-band but suffer from increased confusion. In Fig. \ref{fig:ident_ds14_ds15}, we show data observed in 2004 (H- and L-band) and 2009 (K-band). Especially the K-band detection of DS15 is exposed to the dominant emission of close-by stars, including their respective PSF wings. Despite these challenges, we managed to estimate the K-band magnitude of DS15 for several epochs as indicated in the light curve presented in Fig. \ref{fig:ident_ds14_ds15}. The related magnitudes are listed in Table \ref{tab:ds14_15_mag}.

\begin{table*}
\centering
\begin{tabular}{|cccc|}
\hline
\hline
       & mean & median & std  \\
H-band (DS14)  & 18.84 & 18.84 & 0.94 \\
K-band (DS14)  & 17.33 & 17.25 & 0.92 \\
L-band (DS14)  & 12.76 & 12.58 & 0.39 \\
H-band (DS15)  & 18.77 & 18.77 & 0.37 \\
K-band (DS15)  & 17.28 & 17.36 & 0.53 \\
L-band (DS15)  & 13.65 & 13.58 & 0.56 \\
\hline
\hline
\end{tabular}
\caption{Estimated magnitudes for DS14 and DS15 using multi-wavelength observations carried out with NACO between 2002 and 2018.}
\label{tab:ds14_15_mag}
\end{table*}

\subsection{DS16 $\&$ DS17}

\begin{figure*}[htbp!]
	\centering
	\includegraphics[width=1.\textwidth]{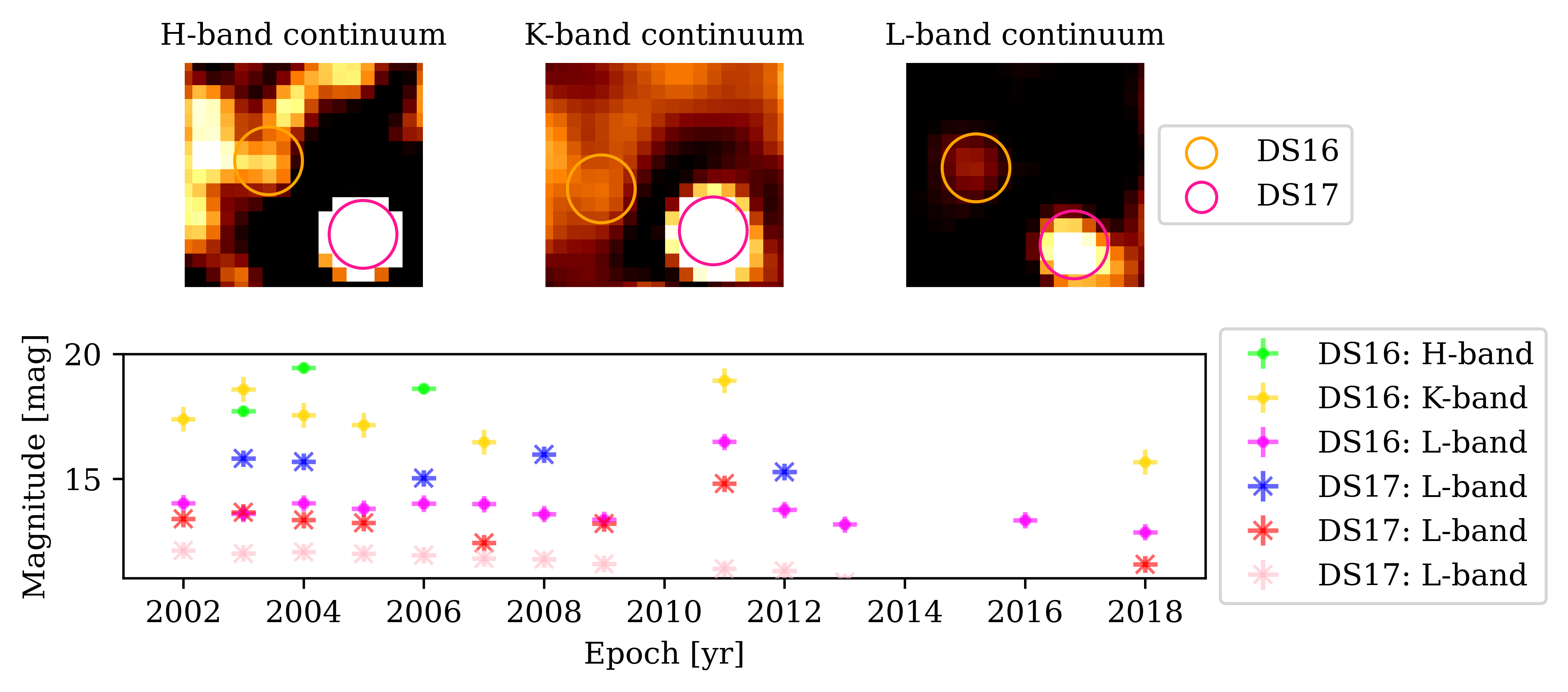}
	\caption{Detection of DS16 and DS17 in the NIR and MIR with NACO. We use data from 2004 (H- and L-band) and 2009 (K-band) which is reflected in minor offsets between the two presented sources.}
\label{fig:ident_ds16_ds17}
\end{figure*}

While Fig. \ref{fig:ident_ds16_ds17} indicates a confusion-free detection of DS17, the close-by and fainter object DS16 experiences the dominant imprint surrounding stars. As for DS14 and DS15, we use data from 2004 (H- and L-band) and 2009 (K-band) for cosmetic reasons. Especially the detection of DS16 demonstrates the challenges of this analysis. While the observation of DS16 is confusion-free in the L-band, the detection in the NIR bands requires the astrometric identification in the MIR band to avoid a false association. Please see Table \ref{tab:ds16_17_mag}.

\begin{table*}
\centering
\begin{tabular}{|cccc|}
\hline
\hline
       & mean & median & std  \\
H-band (DS16)  & 18.61 & 18.59 & 0.71 \\
K-band (DS16)  & 17.39 & 17.39 & 1.05 \\
L-band (DS16)  & 13.75 & 13.83 & 0.83 \\
H-band (DS17)  & 15.68 & 15.55 & 0.35 \\
K-band (DS17)  & 13.28 & 13.19 & 0.87 \\
L-band (DS17)  & 11.76 & 11.47 & 0.64 \\
\hline
\hline
\end{tabular}
\caption{Estimated magnitudes for DS16 and DS17 using multi-wavelength observations carried out with NACO between 2002 and 2018.}
\label{tab:ds16_17_mag}
\end{table*}

\subsection{DS18 $\&$ DS19}

\begin{figure*}[htbp!]
	\centering
	\includegraphics[width=1.\textwidth]{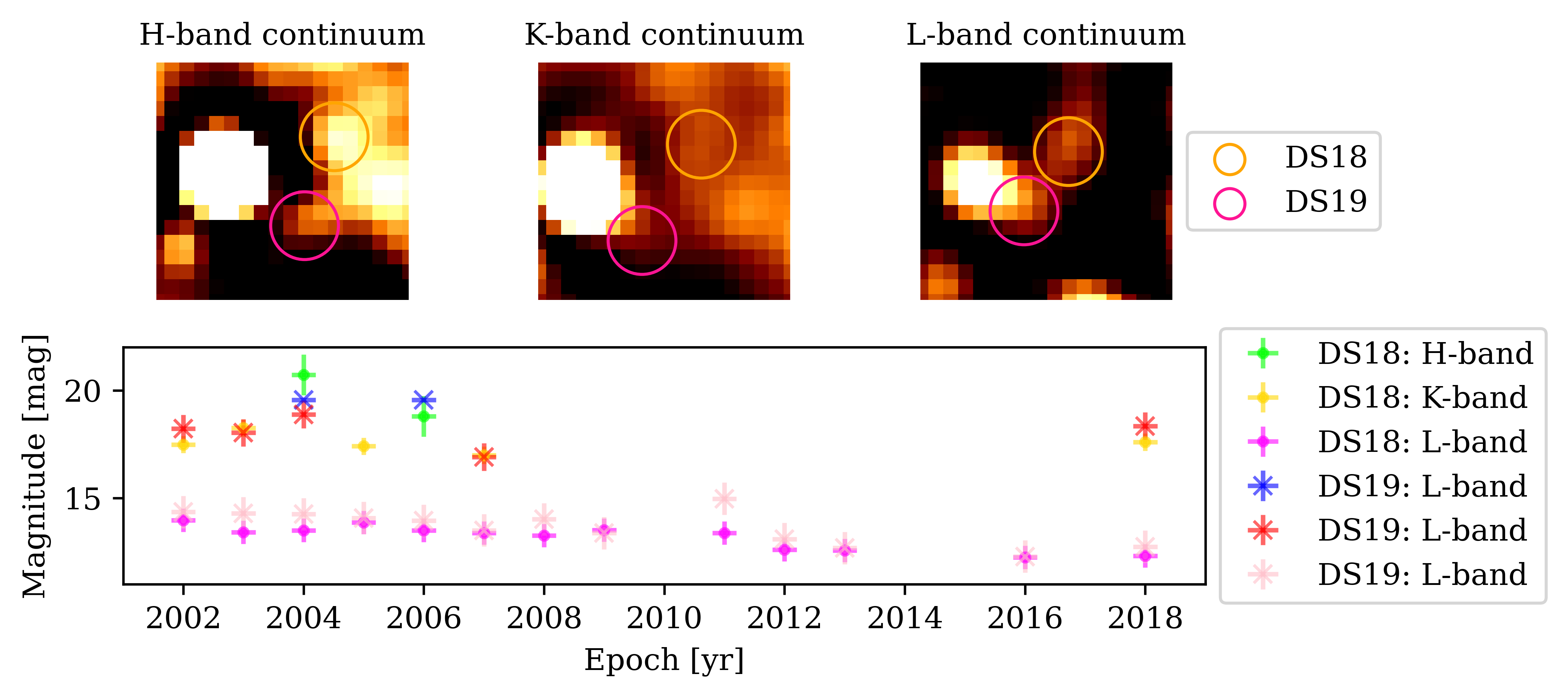}
	\caption{Multi-wavelength detection of DS18 and DS19 in the H-,K-, and L-band.}
\label{fig:ident_ds18_ds19}
\end{figure*}

The observation of DS18 and DS19 is presented in Fig. \ref{fig:ident_ds18_ds19}. Due to the close distance to the bright star DS17 (see Fig. \ref{fig:ident_ds16_ds17}), the detection of DS18 and DS19 is hindered. Hence, the magnitude-confusion susceptibility is increased. However, we list the related magnitudes for DS18 and DS19 based on the NACO data set observed between 2002 and 2018 in Table \ref{tab:ds18_19_mag}.

\begin{table*}
\centering
\begin{tabular}{|cccc|}
\hline
\hline
       & mean & median & std  \\
H-band (DS18)  & 19.75 & 19.75 & 0.95 \\
K-band (DS18)  & 17.49 & 17.54 & 0.40 \\
L-band (DS18)  & 13.39 & 13.19 & 0.54 \\
H-band (DS19)  & 19.55 & 19.55 & 0.01 \\
K-band (DS19)  & 18.23 & 18.08 & 0.64 \\
L-band (DS19)  & 13.95 & 13.66 & 0.75 \\
\hline
\hline
\end{tabular}
\caption{Estimated magnitudes for DS18 and DS19 using multi-wavelength observations carried out with NACO between 2002 and 2018.}
\label{tab:ds18_19_mag}
\end{table*}

\subsection{DS20}

\begin{figure*}[htbp!]
	\centering
	\includegraphics[width=1.\textwidth]{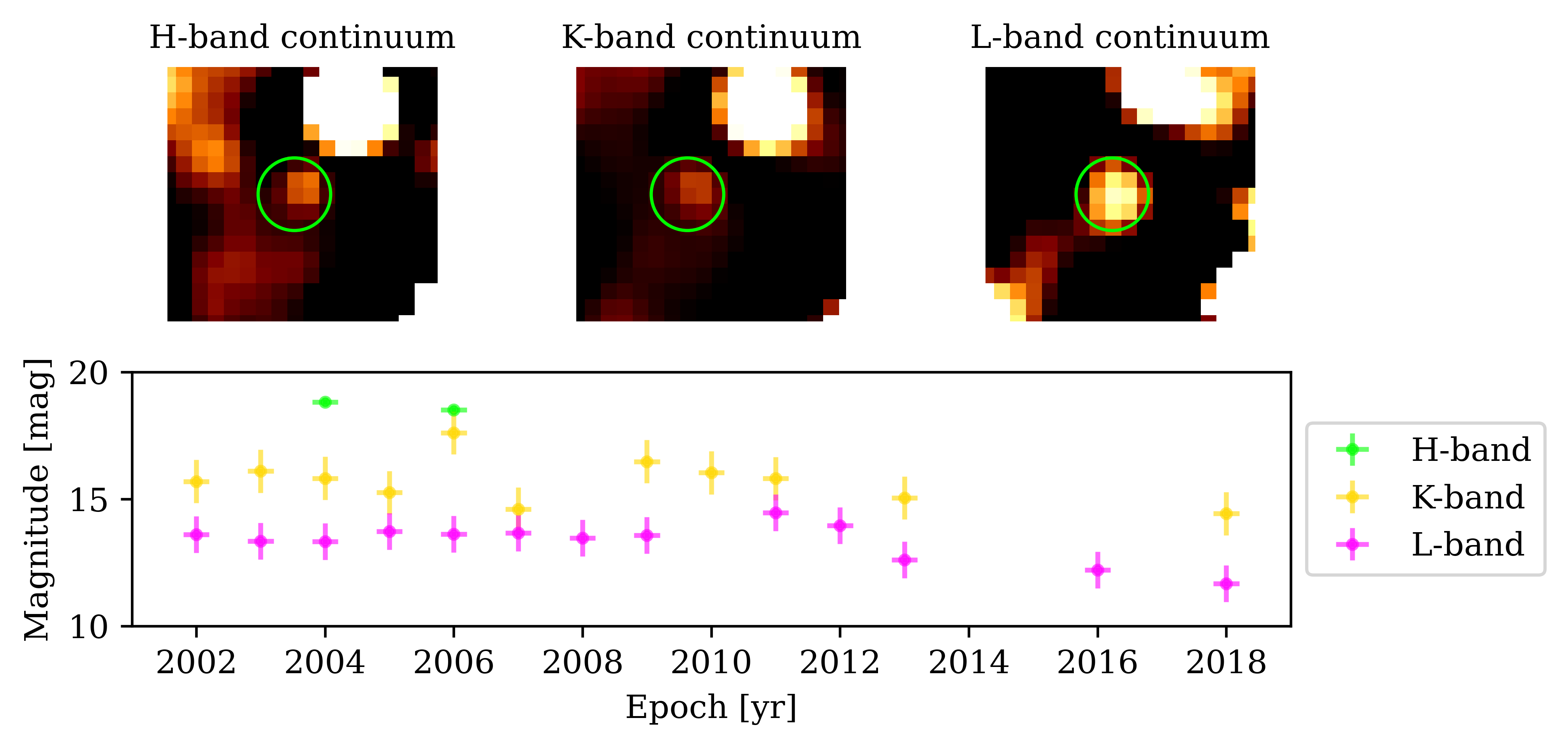}
	\caption{Observation of DS20 close to the bright star DS17 (Fig. \ref{fig:ident_ds16_ds17}).}
\label{fig:ident_ds20}
\end{figure*}

For DS20, we do not find increased confusion due to crowding (Fig. \ref{fig:ident_ds20}). The magnitude-confusion susceptibility is lowered. We list the related multi-wavelength magnitudes in Table \ref{tab:ds20_mag}.

\begin{table*}
\centering
\begin{tabular}{|cccc|}
\hline
\hline
       & mean & median & std  \\
H-band & 18.67 & 18.67 & 0.15 \\
K-band & 15.81 & 15.71 & 0.85 \\
L-band & 13.58 & 13.32 & 0.72 \\
\hline
\hline
\end{tabular}
\caption{Estimated magnitudes for DS20 using multi-wavelength observations carried out with NACO between 2002 and 2018.}
\label{tab:ds20_mag}
\end{table*}

\subsection{DS21}

\begin{figure*}[htbp!]
	\centering
	\includegraphics[width=1.\textwidth]{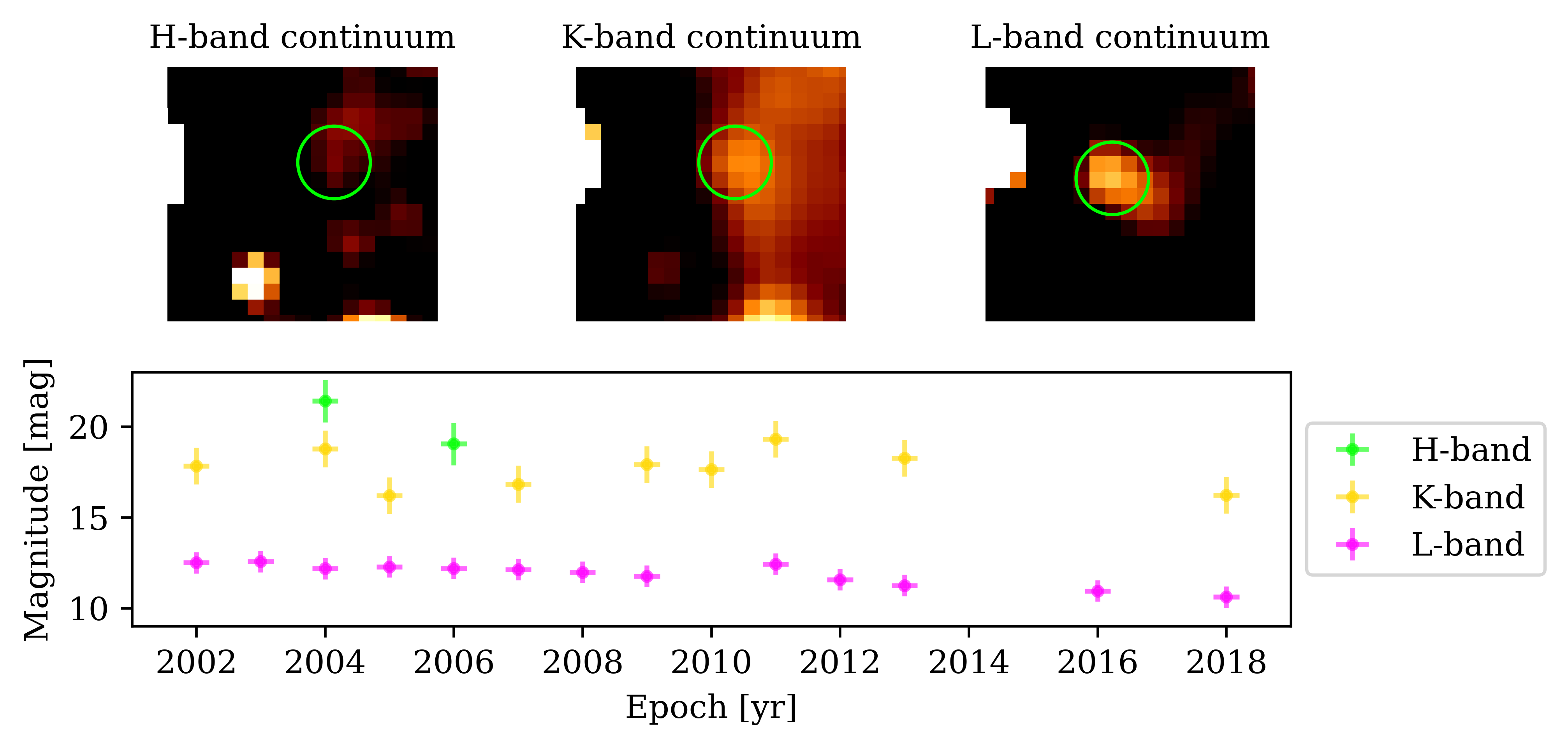}
	\caption{Detection of the source DS21 close to E7.}
\label{fig:ident_ds21}
\end{figure*}

The source DS21 is located close to the E-star E7. Because of the bright emission of the E-stars, the source suffers from blending and dominant PSF wings, although DS21 seems isolated (Fig. \ref{fig:ident_ds21}). Despite the challenges of the identification related to DS21, we identify the source in several epochs and list the magnitude in Table \ref{tab:ds21_mag}.

\begin{table*}
\centering
\begin{tabular}{|cccc|}
\hline
\hline
       & mean & median & std  \\
H-band & 20.23 & 20.23 & 1.17 \\
K-band & 17.83 & 17.66 & 1.01 \\
L-band & 12.12 & 11.87 & 0.59 \\
\hline
\hline
\end{tabular}
\caption{Estimated magnitudes for DS21 using multi-wavelength observations carried out with NACO between 2002 and 2018.}
\label{tab:ds21_mag}
\end{table*}

\subsection{DS22}

\begin{figure*}[htbp!]
	\centering
	\includegraphics[width=1.\textwidth]{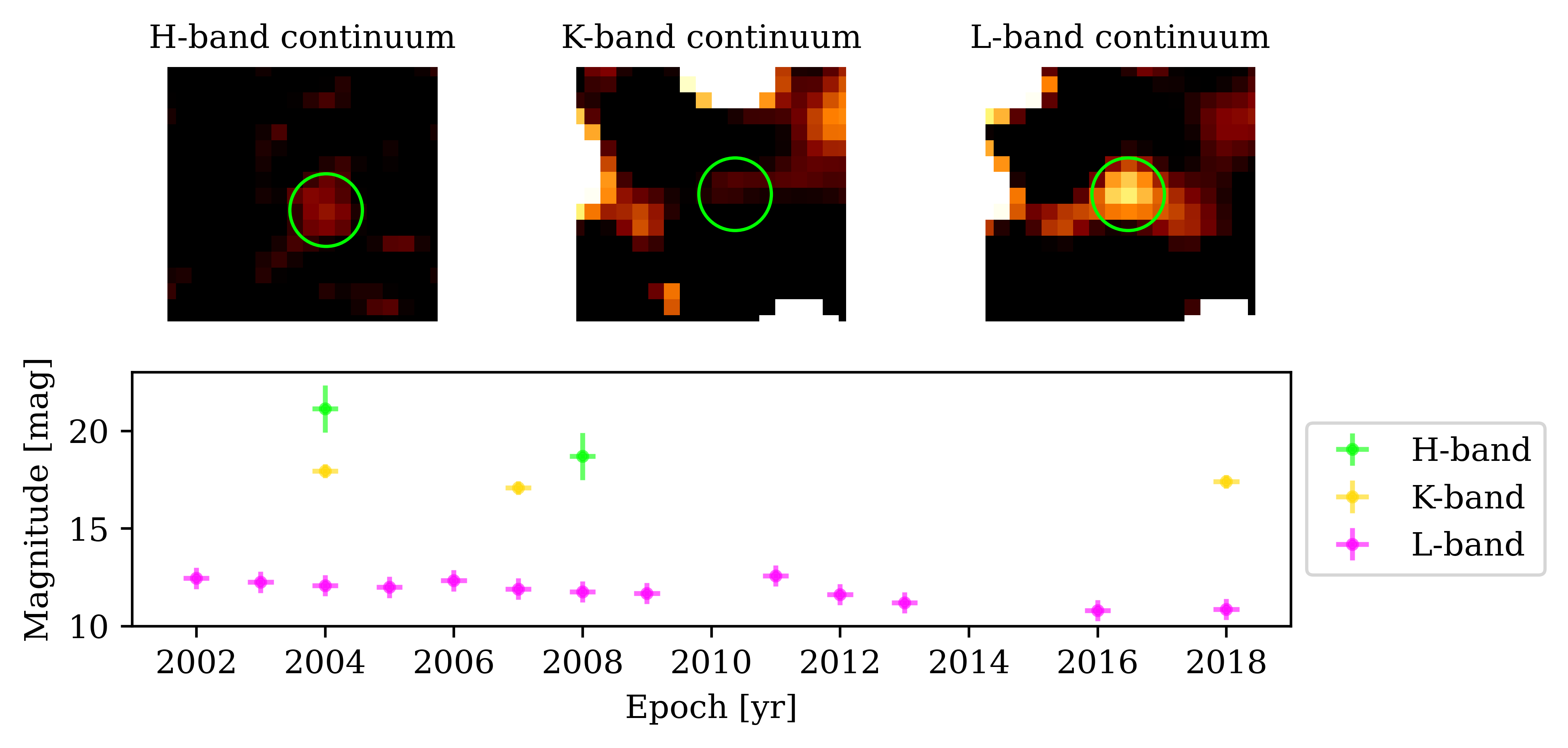}
	\caption{Observation of DS22 in the H-, K-, and L-band.}
\label{fig:ident_ds22}
\end{figure*}

The detection of DS22 in the H- and K-band is challenging due to the close distance of several surrounding stars (Fig. \ref{fig:ident_ds22}). As demonstrated for DS20 (Fig. \ref{fig:ident_ds20}), the close distance of bright stars does not have to result in lowered detectability. However, due to the low K- and H-band magnitude of DS22, the detection of the source is limited to a few epochs (Fig. \ref{fig:ident_ds22}). The related magnitude is indicated in Table \ref{tab:ds22_mag}.

\begin{table*}
\centering
\begin{tabular}{|cccc|}
\hline
\hline
       & mean & median & std  \\
H-band & 19.90 & 19.90 & 1.21 \\
K-band & 17.40 & 17.47 & 0.35 \\
L-band & 11.91 & 11.80 & 0.54 \\
\hline
\hline
\end{tabular}
\caption{Estimated magnitudes for DS22 using multi-wavelength observations carried out with NACO between 2002 and 2018.}
\label{tab:ds22_mag}
\end{table*}

\subsection{DS23}

Like DS22 (Fig. \ref{fig:ident_ds22}), the source DS23 suffers from close-by stars and increased confusion. The magnitude of the source is listed in Table \ref{tab:ds23_mag}.
\begin{figure*}[htbp!]
	\centering
	\includegraphics[width=1.\textwidth]{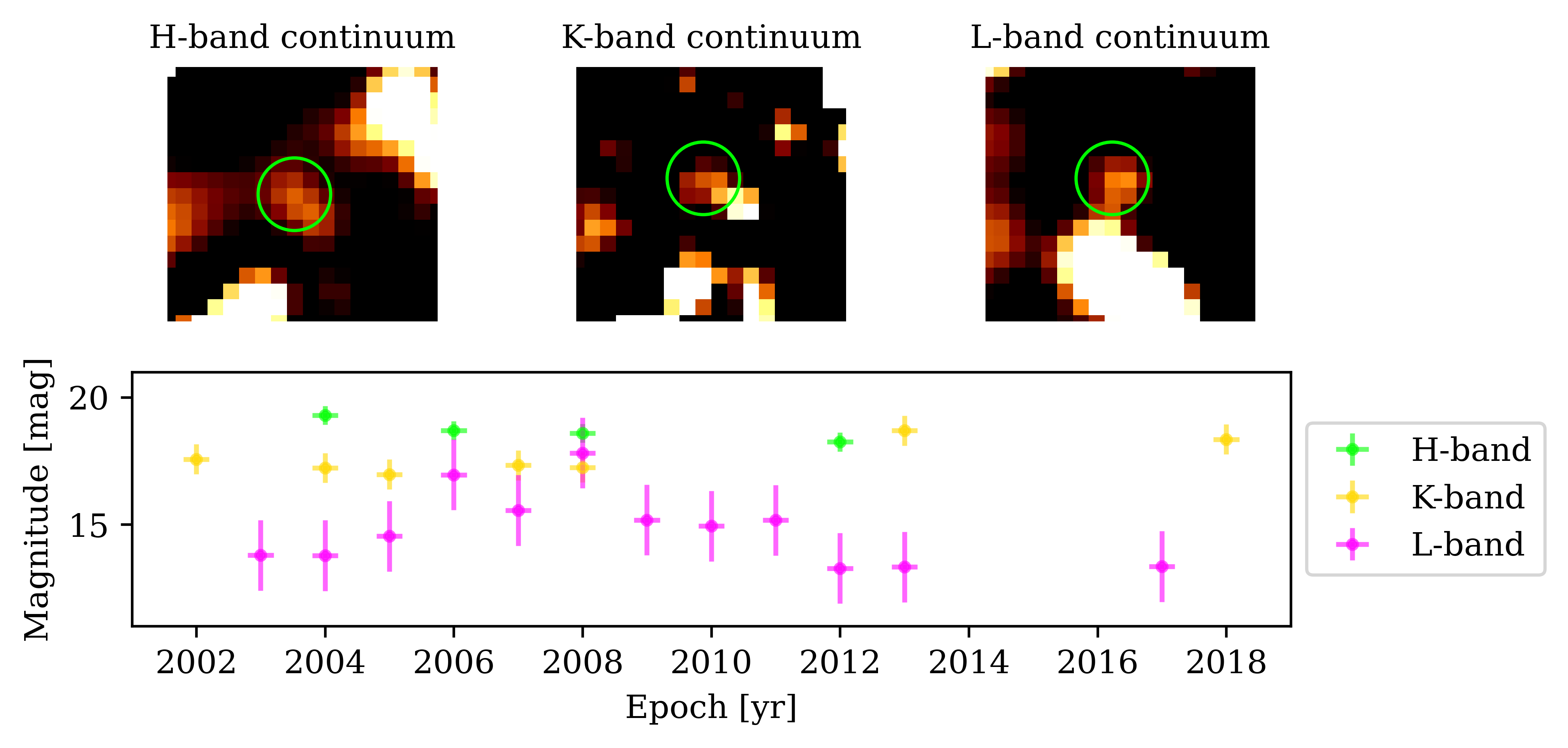}
	\caption{Detection of DS23 in various bands observed with NACO.}
\label{fig:ident_ds23}
\end{figure*}

\begin{table*}
\centering
\begin{tabular}{|cccc|}
\hline
\hline
       & mean & median & std  \\
H-band & 18.65 & 18.71 & 0.37 \\
K-band & 17.33 & 17.62 & 0.59 \\
L-band & 14.73 & 14.80 & 1.39 \\
\hline
\hline
\end{tabular}
\caption{Estimated magnitudes for DS23 using multi-wavelength observations carried out with NACO between 2002 and 2018.}
\label{tab:ds23_mag}
\end{table*}

\subsection{DS24}

\begin{figure*}[htbp!]
	\centering
	\includegraphics[width=1.\textwidth]{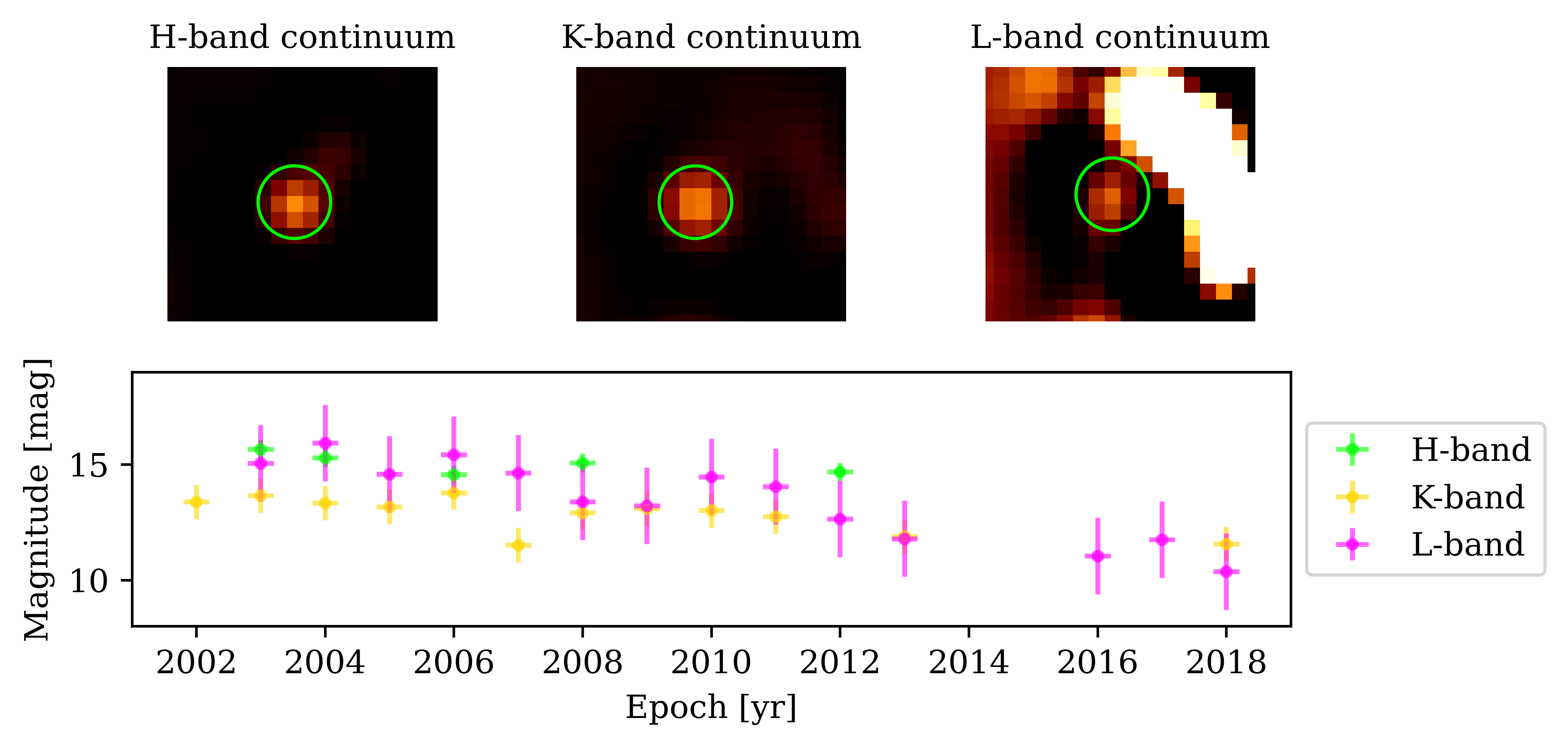}
	\caption{Dusty source 24 in several bands observed with NACO between 2002 and 2018.}
\label{fig:ident_ds24}
\end{figure*}

The bright source is located at the eastern edge of the IRS 13 cluster and, therefore, is not affected by the dominant crowding of the inner core region. We detect DS24 without confusion in various epochs (Fig. \ref{fig:ident_ds24}) and list the related magnitude in Table \ref{tab:ds24_mag}.

\begin{table*}
\centering
\begin{tabular}{|cccc|}
\hline
\hline
       & mean & median & std  \\
H-band & 15.07 & 15.05 & 0.40 \\
K-band & 13.05 & 12.83 & 0.74 \\
L-band & 13.71 & 13.45 & 1.65 \\
\hline
\hline
\end{tabular}
\caption{Estimated magnitudes for DS24 using multi-wavelength observations carried out with NACO between 2002 and 2018.}
\label{tab:ds24_mag}
\end{table*}

\subsection{DS25}

\begin{figure*}[htbp!]
	\centering
	\includegraphics[width=1.\textwidth]{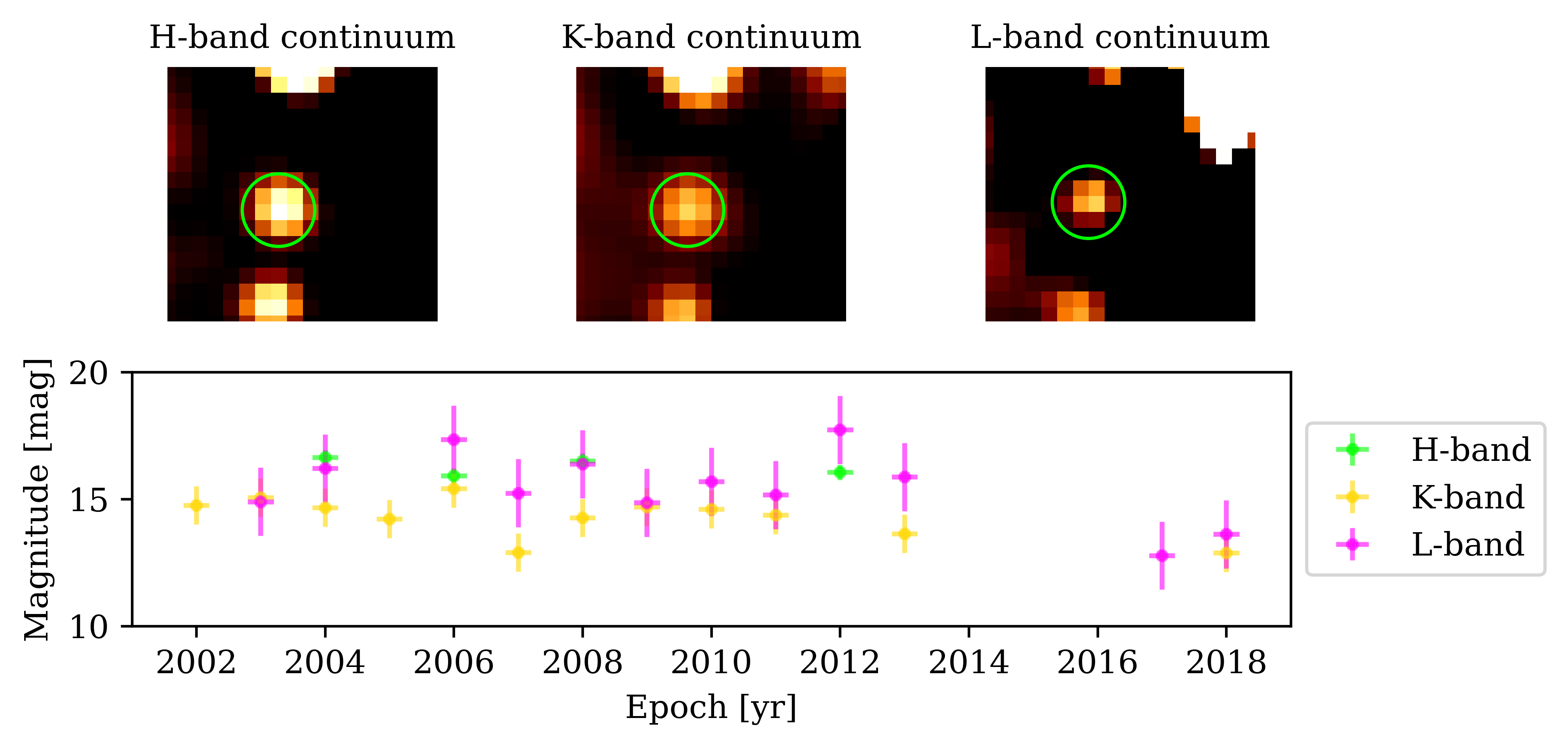}
	\caption{The source DS25 located south to DS24. We detect the source in all investigated bands.}
\label{fig:ident_ds25}
\end{figure*}

Like DS24 (Fig. \ref{fig:ident_ds24}), the source DS25 is located at the edge of the cluster. The chance for confusion is lowered, which is why we detect the source in all analyzed bands (Fig. \ref{fig:ident_ds25}). The related magnitude is listed in Table \ref{tab:ds25_mag}.

\begin{table*}
\centering
\begin{tabular}{|cccc|}
\hline
\hline
       & mean & median & std  \\
H-band & 16.28 & 16.28 & 0.30 \\
K-band & 14.49 & 14.29 & 0.75 \\
L-band & 15.46 & 15.48 & 1.34 \\
\hline
\hline
\end{tabular}
\caption{Estimated magnitudes for DS25 using multi-wavelength observations carried out with NACO between 2002 and 2018.}
\label{tab:ds25_mag}
\end{table*}

\subsection{DS26}

\begin{figure*}[htbp!]
	\centering
	\includegraphics[width=1.\textwidth]{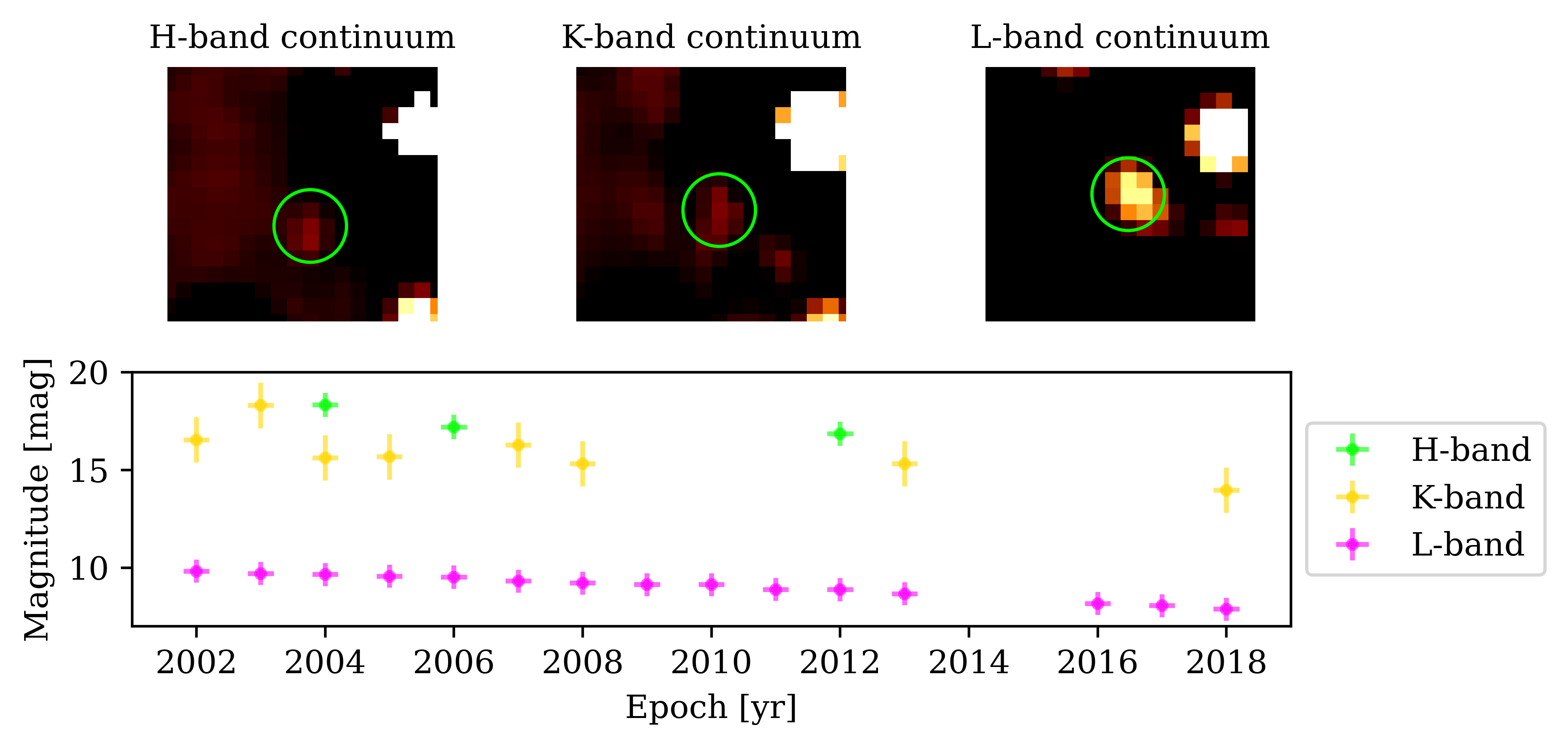}
	\caption{Detection of DS26 in the H-, K-, and L-band. The imprint of the dominant PSF wings of the O supergiant E1 in the NIR bands is recognizable.}
\label{fig:ident_ds26}
\end{figure*}

The source DS26 is affected by the dominant PSF wings of E1, the O supergiant in the core of IRS 13. Therefore, the detection of DS26 is limited to a few epochs (Fig. \ref{fig:ident_ds26}). Please consult Table \ref{tab:ds26_mag} for the related magnitude values of DS26.

\begin{table*}
\centering
\begin{tabular}{|cccc|}
\hline
\hline
       & mean & median & std  \\
H-band & 17.20 & 17.46 & 0.62 \\
K-band & 15.64 & 15.87 & 1.16 \\
L-band & 9.13 & 9.03 & 0.59 \\
\hline
\hline
\end{tabular}
\caption{Estimated magnitudes for DS26 using multi-wavelength observations carried out with NACO between 2002 and 2018.}
\label{tab:ds26_mag}
\end{table*}

\subsection{DS27}

The bright source DS27 can be observed without confusion in the available data set covering the epochs between 2002 and 2018. The related magnitude is listed in Table \ref{tab:ds27_mag}.

\begin{figure*}[htbp!]
	\centering
	\includegraphics[width=1.\textwidth]{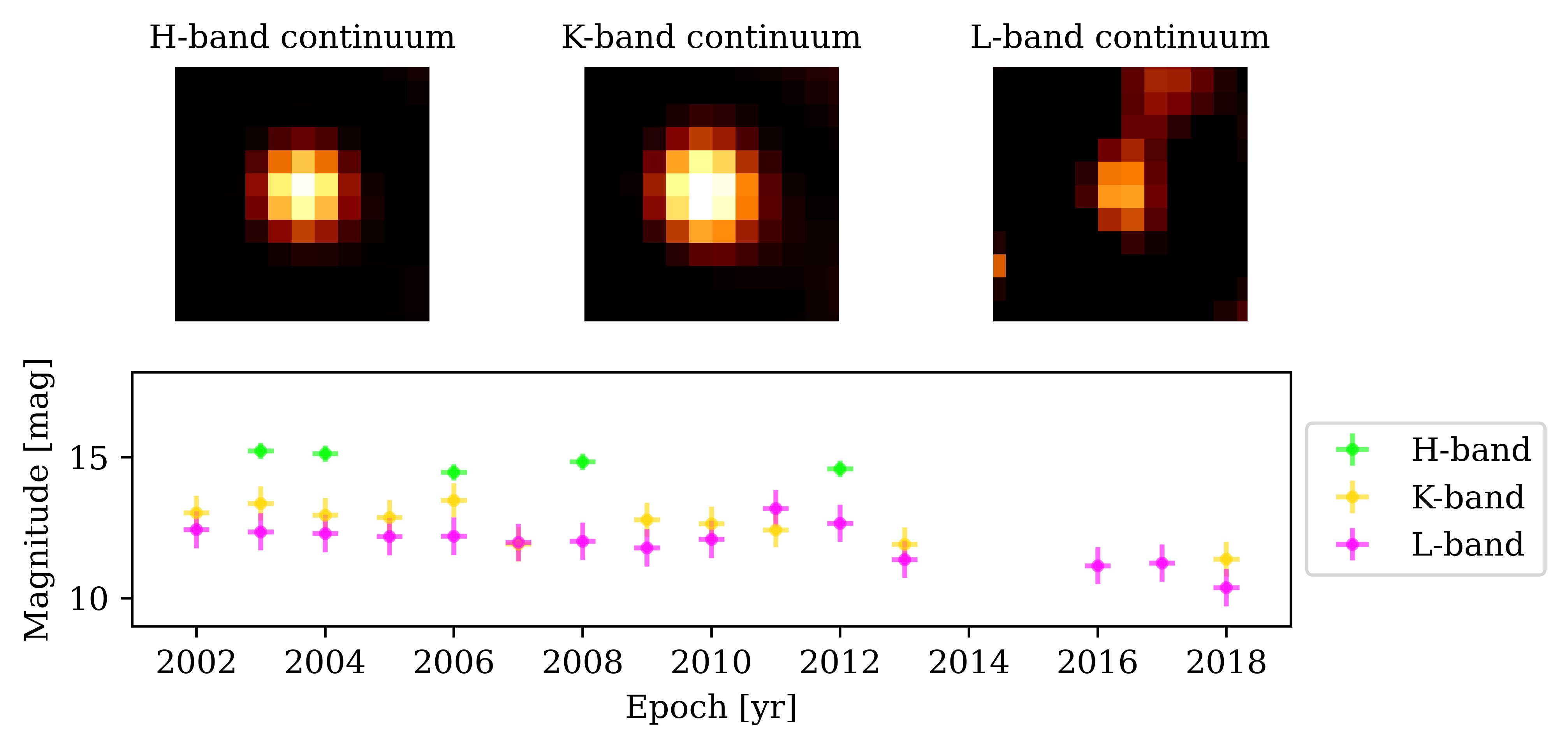}
	\caption{Dusty source 27 observed in the H-, K-, and L-band. The source can be observed without confusion close to DS8 (Fig. \ref{fig:ident_ds8}).}
\label{fig:ident_ds27}
\end{figure*}

\begin{table*}
\centering
\begin{tabular}{|cccc|}
\hline
\hline
       & mean & median & std  \\
H-band & 14.83 & 14.83 & 0.29 \\
K-band & 12.77 & 12.60 & 0.61 \\
L-band & 12.08 & 11.94 & 0.66 \\
\hline
\hline
\end{tabular}
\caption{Estimated magnitudes for DS27 using multi-wavelength observations carried out with NACO between 2002 and 2018.}
\label{tab:ds27_mag}
\end{table*}

\subsection{DS28}

\begin{figure*}[htbp!]
	\centering
	\includegraphics[width=1.\textwidth]{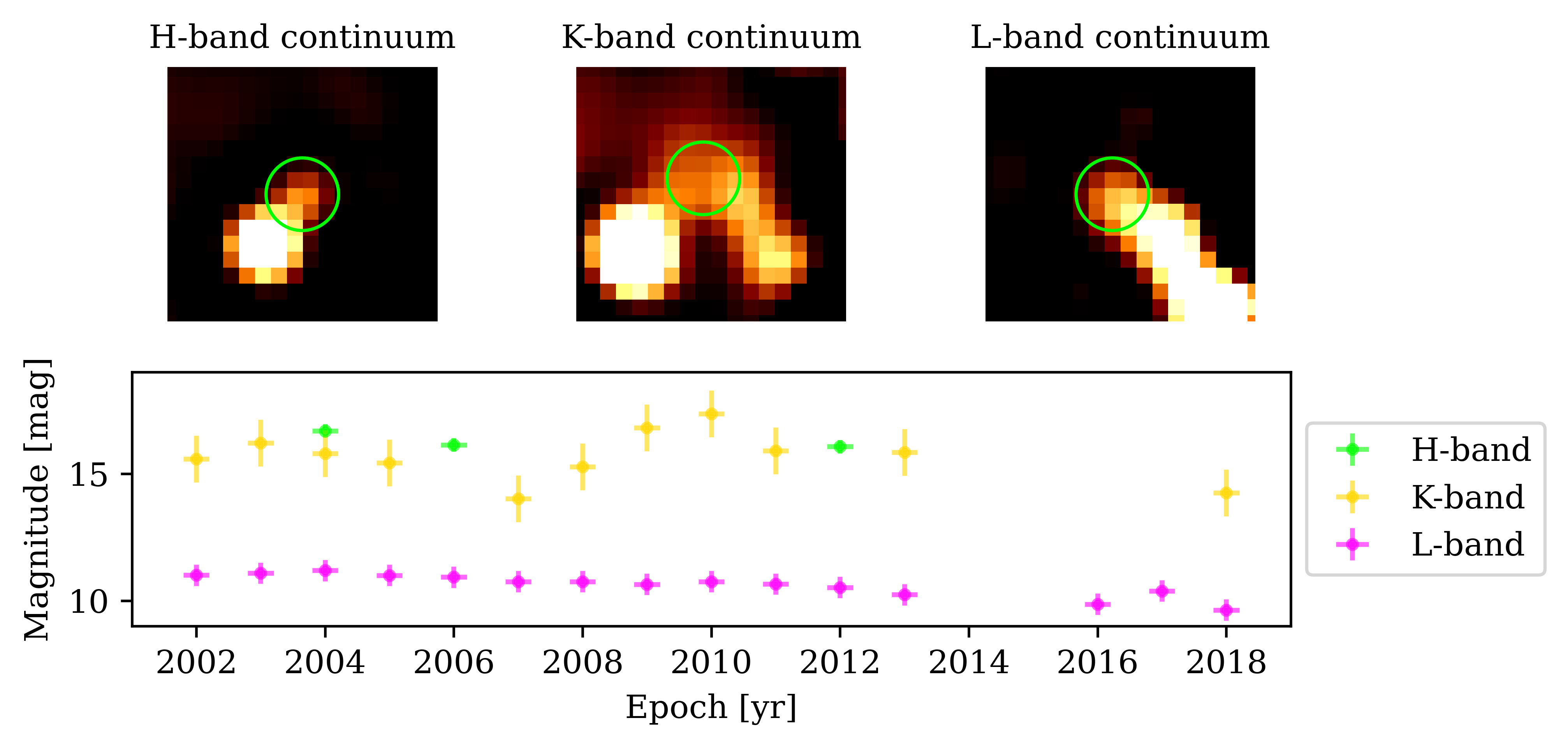}
	\caption{Dusty source 28 observed in the H-, K-, and L-band. The source is located at the northern tip of an elongated L-band feature.}
\label{fig:ident_ds28}
\end{figure*}

The source DS28 is located at the northern tip of an apparent elongated L-band feature (Fig. \ref{fig:ident_ds28}) consisting of DS31 and DS33. We detect DS28 without confusion in several bands and epochs. The estimated magnitude is listed in Table \ref{tab:ds28_mag}.

\begin{table*}
\centering
\begin{tabular}{|cccc|}
\hline
\hline
       & mean & median & std  \\
H-band & 16.14 & 16.30 & 0.27 \\
K-band & 15.80 & 15.68 & 0.92 \\
L-band & 10.76 & 10.63 & 0.42 \\
\hline
\hline
\end{tabular}
\caption{Estimated magnitudes for DS28 using multi-wavelength observations carried out with NACO between 2002 and 2018.}
\label{tab:ds28_mag}
\end{table*}

\subsection{DS29}

\begin{figure*}[htbp!]
	\centering
	\includegraphics[width=1.\textwidth]{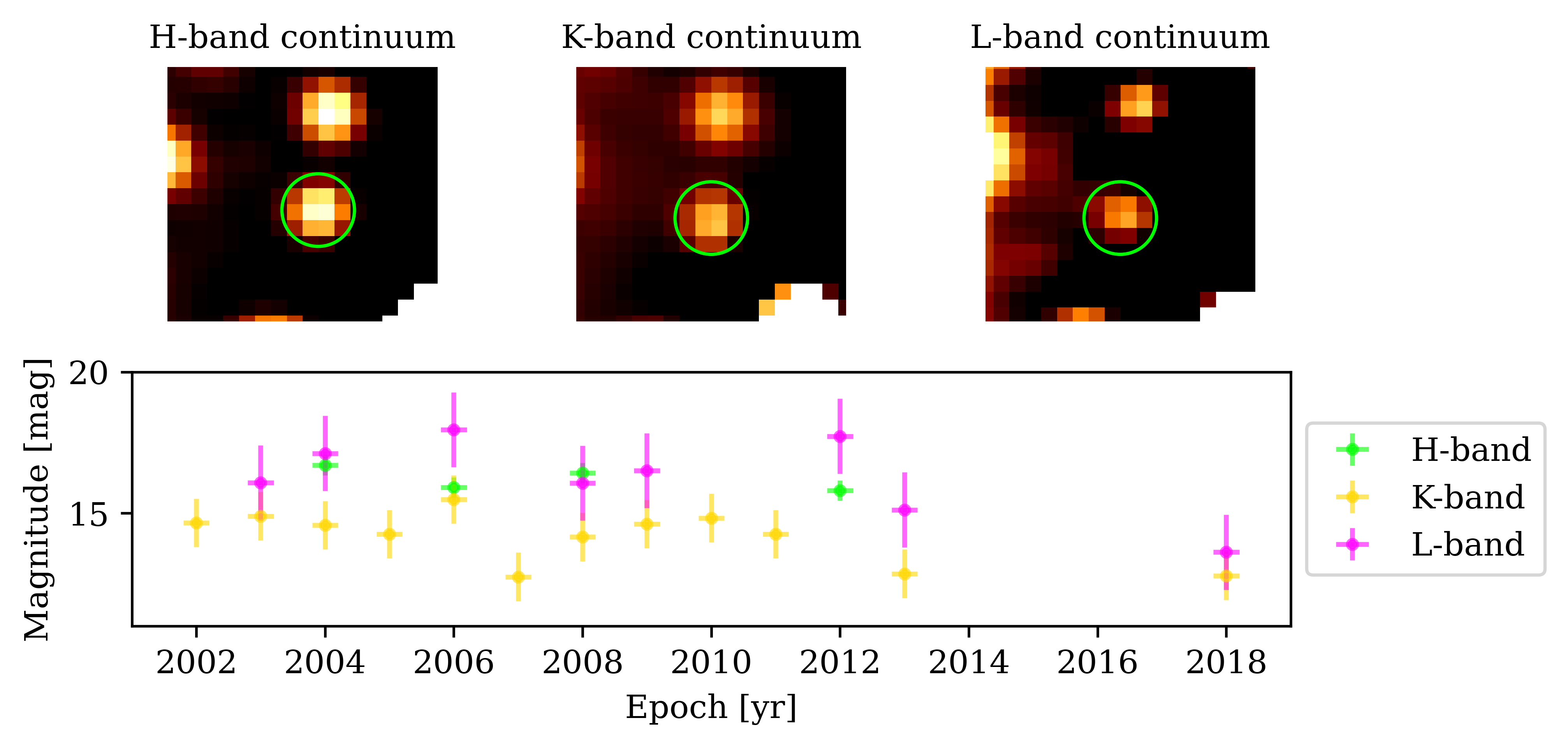}
	\caption{Identification of the bright source DS29 in the H-, K-, and L-band. The source is located south of DS25 (Fig. \ref{fig:ident_ds25}) and north of X3 \citep{peissker2023b}.}
\label{fig:ident_ds29}
\end{figure*}

The source DS29 is located on a north-south axis with DS24 (Fig. \ref{fig:ident_ds24}) and DS25 (Fig. \ref{fig:ident_ds25}). Like the former two sources, DS29 is located a the edge of the IRS 13 cluster and north of X3 \citep{peissker2023b}. We detect DS29 without confusion (Fig. \ref{fig:ident_ds29}) and indicate the related magnitude in Table \ref{fig:ident_ds29}.

\begin{table*}
\centering
\begin{tabular}{|cccc|}
\hline
\hline
       & mean & median & std  \\
H-band & 16.17 & 16.21 & 0.36 \\
K-band & 14.42 & 14.18 & 0.86 \\
L-band & 16.29 & 16.27 & 1.33 \\
\hline
\hline
\end{tabular}
\caption{Estimated magnitudes for DS29 using multi-wavelength observations carried out with NACO between 2002 and 2018.}
\label{tab:ds29_mag}
\end{table*}

\subsection{DS30}

\begin{figure*}[htbp!]
	\centering
	\includegraphics[width=1.\textwidth]{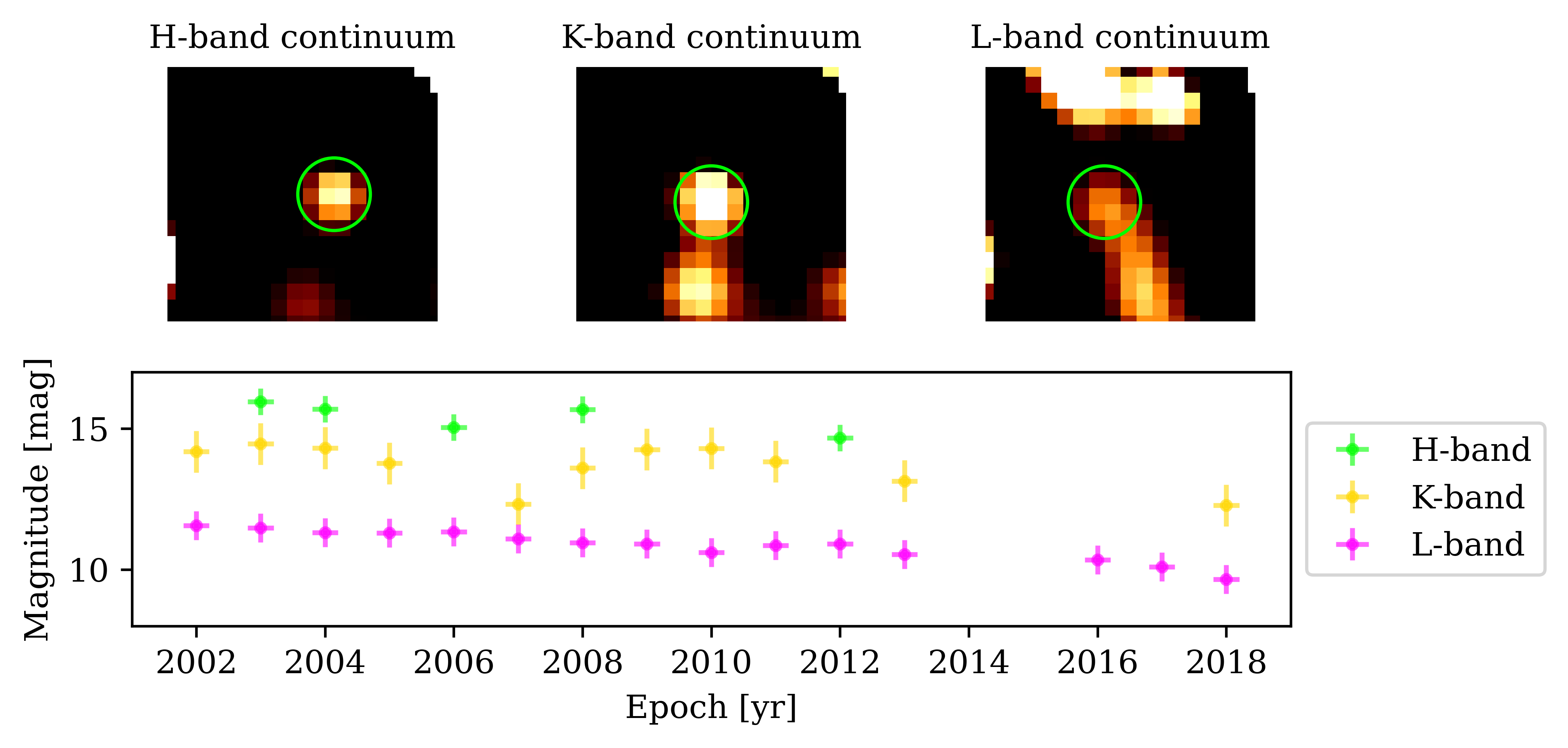}
	\caption{Dusty source 30 located close to the E-star E1.}
\label{fig:ident_ds30}
\end{figure*}

Like DS26 (Fig. \ref{fig:ident_ds26}), the source DS30 is located close to the E-star E1. Due to the brightness of DS30, we detect the source without confusion is various epochs (Fig. \ref{fig:ident_ds30}). In Table \ref{tab:ds30_mag}, we list the related magnitude.

\begin{table*}
\centering
\begin{tabular}{|cccc|}
\hline
\hline
       & mean & median & std  \\
H-band & 15.67 & 15.40 & 0.47 \\
K-band & 13.83 & 13.67 & 0.74 \\
L-band & 10.92 & 10.86 & 0.51 \\
\hline
\hline
\end{tabular}
\caption{Estimated magnitudes for DS30 using multi-wavelength observations carried out with NACO between 2002 and 2018.}
\label{tab:ds30_mag}
\end{table*}

\subsection{DS31}

\begin{figure*}[htbp!]
	\centering
	\includegraphics[width=1.\textwidth]{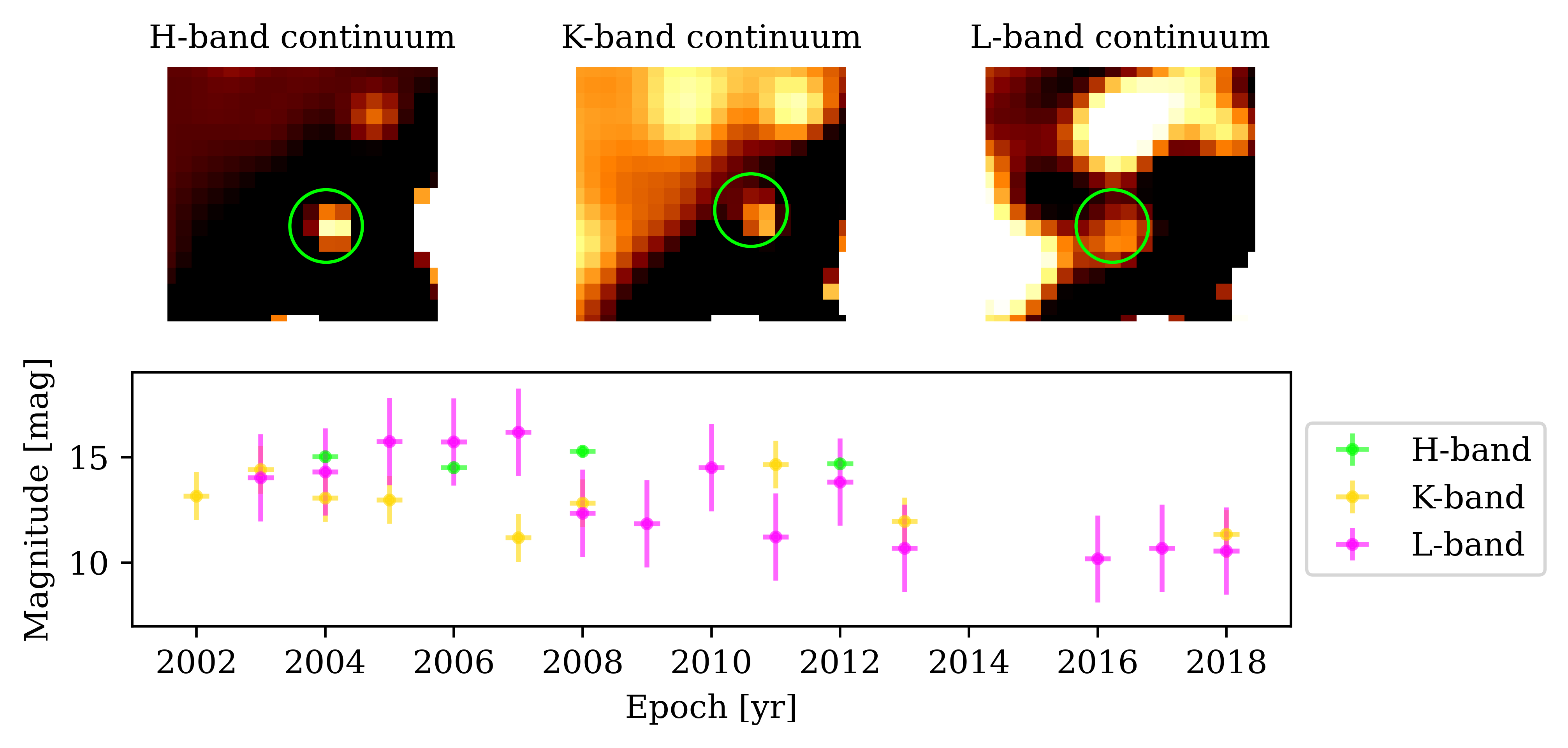}
	\caption{Detection of DS31 in the H-, K-, and L-band suffering from the dominant imprint of the PSF wings of surrounding stars.}
\label{fig:ident_ds31}
\end{figure*}

The source DS31 suffers, like, for example, DS26 (Fig. \ref{fig:ident_ds26}), from the toxic imprint of the core stars of the IRS 13 cluster. This is reflected by a variation of the magnitude in various bands as indicated in Fig. \ref{fig:ident_ds31}. The significance of the detection is lowered since the variation of the individual magnitude values is partially outside of the range of the standard deviation. The related magnitudes are listed in Table \ref{tab:ds31_mag}.

\begin{table*}
\centering
\begin{tabular}{|cccc|}
\hline
\hline
       & mean & median & std  \\
H-band & 14.84 & 14.85 & 0.29 \\
K-band & 12.97 & 12.83 & 1.13 \\
L-band & 13.08 & 12.98 & 2.06 \\
\hline
\hline
\end{tabular}
\caption{Estimated magnitudes for DS31 using multi-wavelength observations carried out with NACO between 2002 and 2018.}
\label{tab:ds31_mag}
\end{table*}

\subsection{DS32}

\begin{figure*}[htbp!]
	\centering
	\includegraphics[width=1.\textwidth]{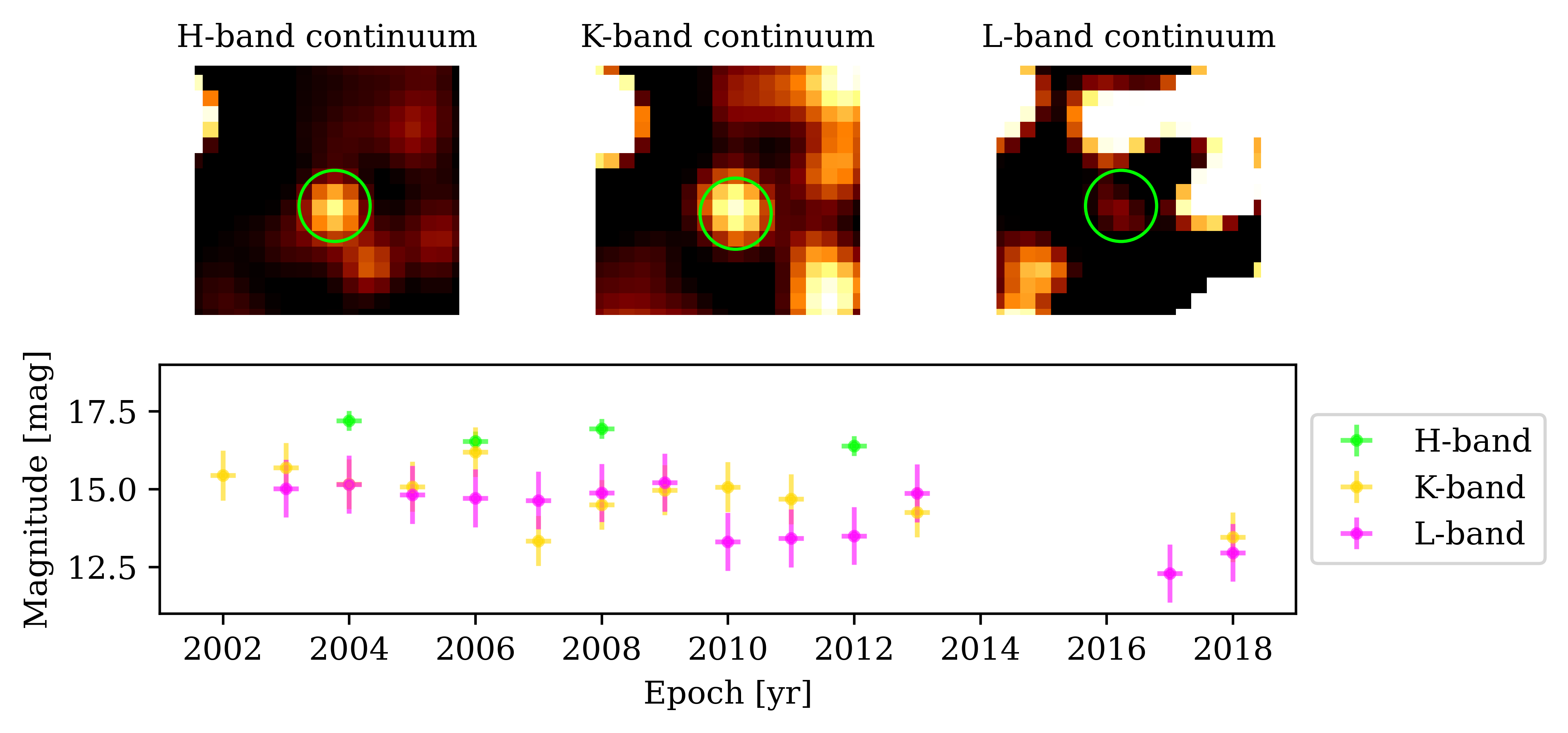}
	\caption{Observation of DS32 in various band close the bright dusty sources discussed in this work.}
\label{fig:ident_ds32}
\end{figure*}

The source DS32 is located close to the brighter dusty sources of the IRS 13 cluster. Although the chance of confusion is increased due to the amount of close by stars, we detect DS32 in all epochs between 2002 and 2018 (Fig. \ref{fig:ident_ds32}). The related magnitude of DS32 is listed in Table \ref{tab:ds32_mag}.

\begin{table*}
\centering
\begin{tabular}{|cccc|}
\hline
\hline
       & mean & median & std  \\
H-band & 16.74 & 16.76 & 0.32 \\
K-band & 15.02 & 14.82 & 0.80 \\
L-band & 14.71 & 14.21 & 0.93 \\
\hline
\hline
\end{tabular}
\caption{Estimated magnitudes for DS32 using multi-wavelength observations carried out with NACO between 2002 and 2018.}
\label{tab:ds32_mag}
\end{table*}

\subsection{DS33}

\begin{figure*}[htbp!]
	\centering
	\includegraphics[width=1.\textwidth]{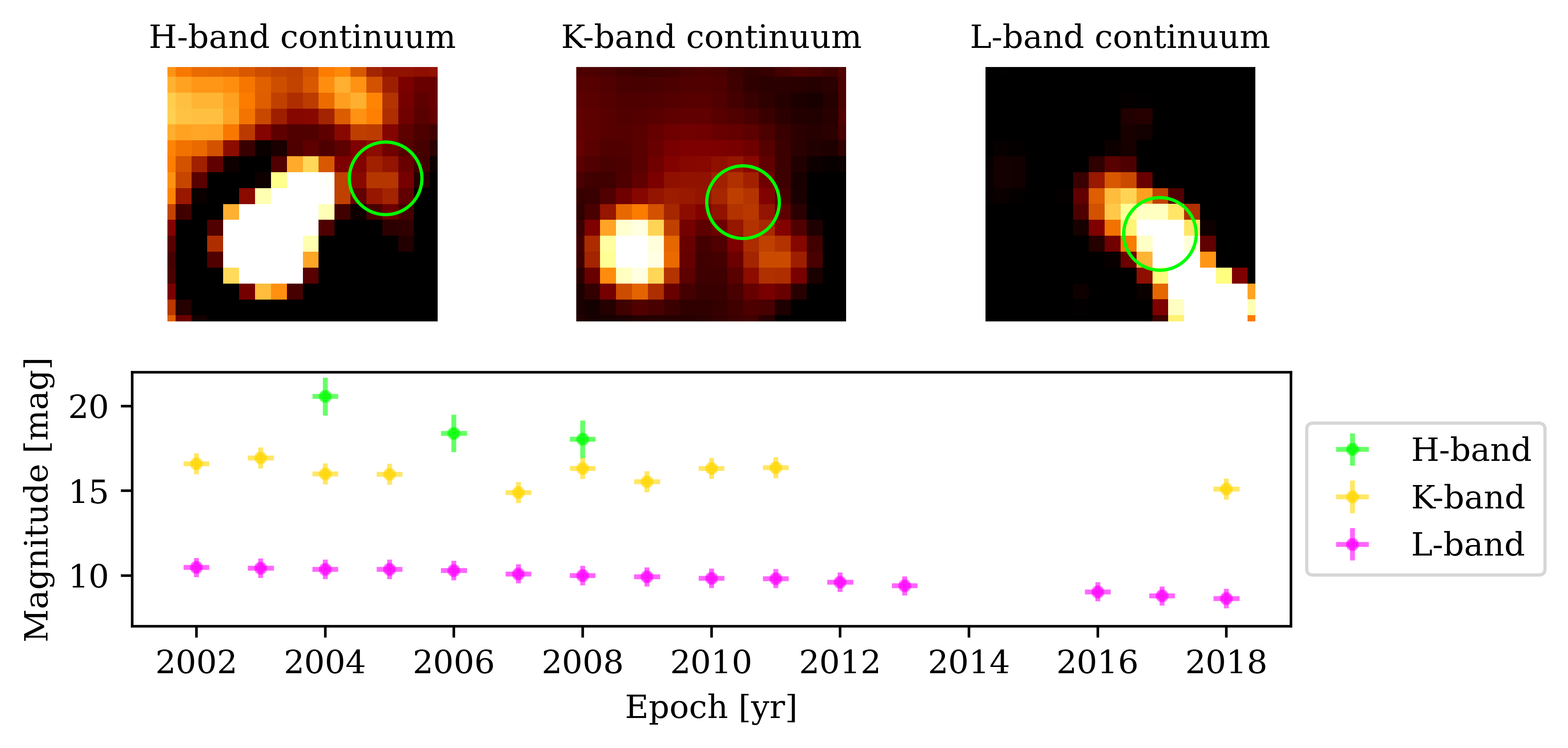}
	\caption{Identfication of DS33 in the elongated L-band feature consisting of several stars.}
\label{fig:ident_ds33}
\end{figure*}

As mentioned before, the location of DS33 can be associated with an elongated L-band feature (DS28, Fig. \ref{fig:ident_ds28}). However, this feature is most likely a chance association because we identify the individual components as DS sources (Fig. \ref{fig:ident_ds33}). Please consult Table \ref{tab:ds33_mag} for the related NIR and MIR magnitudes.

\begin{table*}
\centering
\begin{tabular}{|cccc|}
\hline
\hline
       & mean & median & std  \\
H-band & 18.39 & 19.00 & 1.11 \\
K-band & 16.16 & 16.01 & 0.62 \\
L-band & 9.92 & 9.80 & 0.57 \\
\hline
\hline
\end{tabular}
\caption{Estimated magnitudes for DS33 using multi-wavelength observations carried out with NACO between 2002 and 2018.}
\label{tab:ds33_mag}
\end{table*}

\end{document}